\documentclass[twocolumn]{jpsj3}
\usepackage{txfonts}
\usepackage{bm}
\usepackage{color}
\usepackage[dvipdfmx]{graphicx}
\usepackage{dcolumn}
\usepackage{mathrsfs}
\title{Field-Induced Switching of Ferro-Quadrupole Order Parameter in PrTi$_{2}$Al$_{20}$}

\author{Takanori  Taniguchi\thanks{Electronic address: taka.taniguchi@imr.tohoku.ac.jp, Present address: Institute for Materials Research, Tohoku University, Sendai 980-8577, Japan}$^{1}$, Kazumasa  Hattori\thanks{Electronic address: hattori@tmu.ac.jp}$^{2}$, Makoto  Yoshida\thanks{Present address: Max Planck Institute for Solid State Research, Heisenbergstrasse 1, 70569 Stuttgart, Germany}$^{1}$, Hikaru Takeda$^{1}$, Shota Nakamura\thanks{Present address: Department of Physical Science and Engineering, Graduate School of Engineering, Nagoya Institute of Technology, Nagoya 466-8555, Japan}$^{1}$, Toshiro Sakakibara$^{1}$, Masaki Tsujimoto\thanks{Present Address: Asahi Kasei Microdevide Corporation}$^{1}$, Akito Sakai$^{1}$, Yosuke Matsumoto\thanks{Present address: Max Planck Institute for Solid State Research, Heisenbergstrasse 1, 70569 Stuttgart, Germany}$^{1}$, Satoru Nakatsuji$^{1}$, and Masashi Takigawa\thanks{Electronic address: masashi@issp.u-tokyo.ac.jp}$^{1}$}
\inst{%
$^{1}$Institute for Solid State Physics, University of Tokyo, Kashiwa, Chiba, 277-8581, Japan.\\
$^{2}$Department of Physics, Tokyo Metropolitan University, Hachioji, Tokyo, 192-0937, Japan
}%

\date{\today}

\abst{We report magnetic-field-induced first-order phase transitions in the ferro-quadrupole (FQ) ordered state of PrTi$_{2}$Al$_{20}$, in which non-Kramers Pr$^{3+}$ ions with two 4$f$ electrons have a non-magnetic $\Gamma_3$ doublet ground state in the cubic $T_{d}$ crystalline electric field. For magnetic fields along [111], \textsuperscript{27}Al-NMR and magnetization experiments reveal $Q_z \propto 3z^2-r^2$ type FQ order below 2~K independent of field strength. Magnetic fields along [001] or [110], however, induce discontinuous switching of order parameters within the two dimensional space spanned by $Q_z$ and $Q_x \propto x^2-y^2$ at small field values less than a few tesla. A symmetry-based theoretical analysis shows that  the transitions can be caused by competition between the magnetic Zeeman interaction and anisotropy in the quadrupole-quadrupole interactions, if the latter dominates over the former in low fields and vice versa in high fields. Furthermore, striking violation of proportionality between NMR Knight shift and magnetic susceptibility is observed in the symmetry-broken FQ phases, indicating significant influence of FQ order on the hybridization between conduction and $f$ electrons, which in turn mediates the RKKY-type quadrupole interaction causing the FQ order. This feedback effect may be a specific feature of quadrupole orders not commonly observed in magnetic phase transitions and play a key role for inducing the discontinuous transitions. }

\begin{document}
\maketitle
\section{\label{sec:level1}Introduction}
A wide range of quantum phenomena caused by spin-orbit coupling combined with strong electronic correlation is one of the central issues of current condensed matter physics~\cite{Kim_Balents_2014, Shaffer_Kim_2016}. In $f$-electron materials containing rare earth or actinide elements, strong spin-orbit coupling combined with a highly symmetric crystalline electric field (CEF) often leads to a degenerate ground state with multiple degrees of freedom. The concept of multipoles developed in classical electromagnetism has been used to classify such diverse electronic degrees of freedom according to the symmetry of angular distribution of the charge and magnetization densities.~\cite{Kuramoto_2009, Kusunose_2008}. 

In fact, in addition to conventional order of magnetic dipole, orders of electric quadrupole, magnetic octupole, and even higher-rank mulitipoles have been investigated in various $f$-electron materials~\cite{Kuramoto_2009, Santini_2009, Onimaru_2016}. Furthermore, hybridization between the conduction and $f$ states ($c$-$f$ hybridization) in such materials leads to multipolar Kondo effects, which could induce non-Fermi liquid behavior and exotic superconductivity~\cite{Cox_1987, Hoshino_2014, Tsuruta_2015, Matsubayashi_PRL}. Unlike conventional magnetic orders, high-rank mulitipole orders are difficult to observe directly by microscopic experimental probes and often recognized only by thermodynamic anomalies. Nevertheless, progress in experimental methods and analysis such as resonant X-ray\cite{Mannix_2005,Wilkins_2006} and neutron\cite{Kuwahara_2007,Onimaru_2005} scattering and nuclear magnetic resonance (NMR)\cite{Takigawa_1983,Sakai_1997,Shina_1998,Tokunaga_2006,Sakai_2005,Kikuchi_2007,Sakai_2007} have enabled identification of multipole orders in a number of materials~\cite{Kuramoto_2009,Santini_2009}. 

A major fraction of research on heavy fermions has been devoted to materials containing Ce$^{3+}$ or Yb$^{3+}$, which have odd number of $f$-electrons and the ground states always have finite dipole moments due to the Kramers theorem. Symmetry breaking in most of these compounds occurs in magnetic dipole sectors and it is relatively rare that only high-rank multipoles are involved in ordering\cite{Tayama_1997,Kuwahara_2007,Takigawa_1983,Yamauchi_1999,Tanaka_2002}. In contrast, situation is different for materials containing non-Kramers ions such as Pr$^{3+}$ or U$^{4+}$, for which the CEF ground state can be non-magnetic and high-rank multipoles play the primary role. The incommensurate quadrupole order in PrPb$_{3}$\cite{Onimaru_2005}, the antiferro-order of highly symmetric electric multipole in Pr(Fe,Ru)$_{4}$P$_{12}$\cite{Kikuchi_2007,Sakai_2007,Sato_2007,Iwasa_2005,Takimoto_2006,Harima_2008,Kiss_2008} and the long standing mystery of \textit{hidden order} in URu$_{2}$Si$_{2}$\cite{Harima_2010} are notable examples. 

In this respect, the series of Pr-based cage compounds Pr$T_{2}X_{20}$ ($T$ = Ir, Rh, $X$ = Zn; $T$ = V, Ti, $X$ = Al), abbreviated as Pr~1-2-20, provide an ideal playground to explore multipole physics\cite{Onimaru_2016}. In these compounds, the Pr$^{3+}$ ions with $(4f)^2$ configuration form a diamond structure in the cubic crystal with the space group \textit{Fd$\bar{3}$m}.  Each Pr$^{3+}$ ion is surrounded by a Frank-Kasper cage consisting of 16 \textit{X} atoms\cite{Kangas,Niemann}, providing a stage for $c$-$f$ hybridization and CEF with $T_d$ point group symmetry at Pr sites.  The total angular momentum $J=4$ multiplet of the Hund's rule manifold ($L=5$, $S=1$) of the $(4f)^2$ configuration is split by the CEF into $\Gamma_{1}$ singlet, $\Gamma_{3}$ doublet, $\Gamma_{4}$ and $\Gamma_{5}$ triplets, of which the $\Gamma_{3}$ is the common ground state for this series of compounds\cite{Onimaru_2016}. 

The wave function of the $\Gamma_{3}$ doublet can be written by using the eigenstates $\left|m\right\rangle$ of the $z$-component of the total angular momentum $J_z$ ($m = -4, -3, \dots, 4$) as
\begin{eqnarray}
\left|\Gamma_{3}^{u}\right\rangle =\frac{\sqrt{21}}{6}\left( \left|+4\right\rangle +\left|-4\right\rangle \right) -\frac{\sqrt{15}}{6}\left|0\right\rangle
\label{eq:G3u},
\end{eqnarray}
\begin{eqnarray}
\left|\Gamma_{3}^{v}\right\rangle =\frac{1}{\sqrt{2}}\left( \left|+2\right\rangle +\left|-2\right\rangle \right)
\label{eq:G3v}.
\end{eqnarray}
The $\Gamma_{3}$ doublet is non-magnetic, i.e. all matrix elements of the dipole $\boldsymbol{J}$ are zero within the $\Gamma_{3}$ doublet,
\begin{eqnarray}
\boldsymbol{J} =\left(\begin{array}{cc}
0 & 0\\
0 & 0
\end{array}\right)
\label{eq:Jmatrix}. 
\end{eqnarray}
On the other hand, the $\Gamma_{3}$ doublet has two active quadrupole moments 
\begin{equation}
Q_{z}=\frac{1}{8}\left(3J_{z}^{2}-\boldsymbol{J}^{2}\right), \ \ \ Q_{x}=\frac{\sqrt{3}}{8}\left(J_{x}^{2}-J_{y}^{2}\right)
\label{eq:Qz},
\end{equation}
which also form a basis of $\Gamma_{3}$ representation of $T_d$ point group, and one octupole moment
\begin{equation}
T_{xyz} = \frac{\sqrt{15}}{6} \overline{J_{x}J_{y}J_{z}} ,
\label{eq:Txyz}
\end{equation}
transforming as $\Gamma_{2}$, where the overline represents the fully symmetrized product. The matrix elements of these multipoles are given in the basis of $\left\{ \left|\Gamma_{3}^{u}\right\rangle ,\left|\Gamma_{3}^{v}\right\rangle \right\} $ as 
\begin{eqnarray}
Q_{z} =\left(\begin{array}{cc}
1 & 0\\
0 & -1
\end{array}\right), \ \ 
Q_{x} =\left(\begin{array}{cc}
0 & -1\\
-1 & 0
\end{array}\right)
\label{eq:Qmatrix}, 
\end{eqnarray}
\begin{eqnarray}
T_{xyz}=\frac{18}{\sqrt{5}}\left(\begin{array}{cc}
0 & i\\
-i & 0
\end{array}\right)
\label{eq:Txyzmatrix}.
\end{eqnarray}

All members of the Pr(Ir, Rh, V, Ti)$_{2}$(Zn, Al)$_{20}$ family show a multipole order and superconductivity along with several features indicative of quadrupolar Kondo effects~\cite{Onimaru_2016}. The Zn-based materials, PrIr$_{2}$Zn$_{20}$ and PrRh$_{2}$Zn$_{20}$ show antiferro-quadrupole (AFQ) order at very low temperatures, 0.11~K\cite{Onimaru_2011} and 0.06~K\cite{Onimaru_2012}, respectively. Since the intersite quadrupole exchange interactions in such intermetallic compounds are primarily mediated by conduction electrons through the Ruderman-Kittel-Kasuya-Yosida (RKKY) mechanism, the very low ordering temperatures indicate weak $c$-$f$ hybridization. PrV$_{2}$Al$_{20}$ shows intriguing double phase transitions at 0.75~K and 0.65~K\cite{Tsujimoto_2014}, which might be due to separate ordering of quadrupole and octupole moments\cite{Freyer_2018,Lee_2018,Y_B_Kim_2019}, in addition to the superconducting transition at 0.05~K~\cite{Tsujimoto_2014}. It also exhibits heavy fermion behavior with highly enhanced effective mass and anomalous temperature dependence ($\sim T^{1/2}$) of the resistivity\cite{Sakai_2011,Tsujimoto_2014}. 

In this paper, we focus on PrTi$_{2}$Al$_{20}$, which is the only member of the Pr~1-2-20 family exhibiting ferro-quadrupole (FQ) order with the highest ordering temperature of 2~K\cite{Sakai_2011,Koseki,Ito}. The energies of excited CEF levels are determined by inelastic neutron scattering experiments as $\Gamma_{4}$(54 K)-$\Gamma_{5}$(108 K)-$\Gamma_{1}$(156 K)~\cite{Sato_2012}. Since these energies are much larger than the energy scales of quadrupole interaction and magnetic fields in typical experiments, active degrees of freedom of $\Gamma_{3}$ doublet should be responsible for all phenomena at low temperatures. The resistivity show $-\ln T$ behavior at high temperatures with a peak near 50~K due to Kondo effects from the excited magnetic triplets. However, it follows $T^2$ Fermi liquid behavior below 20~K\cite{Sakai_2011,Sakai_super}. Moderately strong $c$-$f$ hybridization in PrTi$_{2}$Al$_{20}$ can be inferred from the Kondo resonance peak observed in the photoemission spectroscopy\cite{Matsunami} and modest enhancements of physical quantities such as the Seebeck coefficient\cite{Machida,Kuwai}, the effective mass obtained from de Haas-van Alphen measurements\cite{Nagashima}, and the transferred hyperfine coupling constant obtained by Al-NMR\cite{Tokunaga,Taniguchi_Proc}. Evidence for the FQ order of $Q_z$ quadrupole moment is provided by the neutron\cite{Sato_2012} and NMR\cite{Taniguchi_JPSJ} experiments.  

Applying high pressure to PrTi$_{2}$Al$_{20}$ causes profound change of behavior. The superconducting transition temperature, which is 0.2~K at ambient pressure\cite{Sakai_super}, increases steeply under pressure above 7~GPa, reaching 1.1~K near 9~GPa\cite{Matsubayashi_PRL,Matsubayashi_Proc}. The enhanced superconductivity is correlated with suppression of the FQ order, enhanced effective mass and development of non-Fermi liquid behavior of resistivity, suggesting close relation between quadrupole fluctuations and superconductivity. 

 In the following, we present results of comprehensive NMR and magnetization measurements that determine the magnetic field vs. temperature phase diagram of the FQ order in PrTi$_{2}$Al$_{20}$. NMR is a powerful local probe that allows us to detect subtle breaking of local symmetry at specific sites by observing splitting of resonance lines. This method has been proven to be useful to identify the symmetry of multipole order parameters in a number of materials, for example, CeB$_{6}$\cite{Takigawa_1983,Sakai_1997,Shina_1998}, NpO$_{2}$\cite{Tokunaga_2006,Sakai_2005}, and PrFe$_{4}$P$_{12}$\cite{Kikuchi_2007,Sakai_2007}. In our previous NMR experiments on PrTi$_{2}$Al$_{20}$ with magnetic field along [111], the splitting of NMR lines from Al nuclei is attributed to the field-induced dipole moment perpendicular to [111] in the presence of FQ order of $Q_z$\cite{Taniguchi_JPSJ}. 

\begin{figure*}[t]
\begin{center}
\includegraphics[scale=0.95]{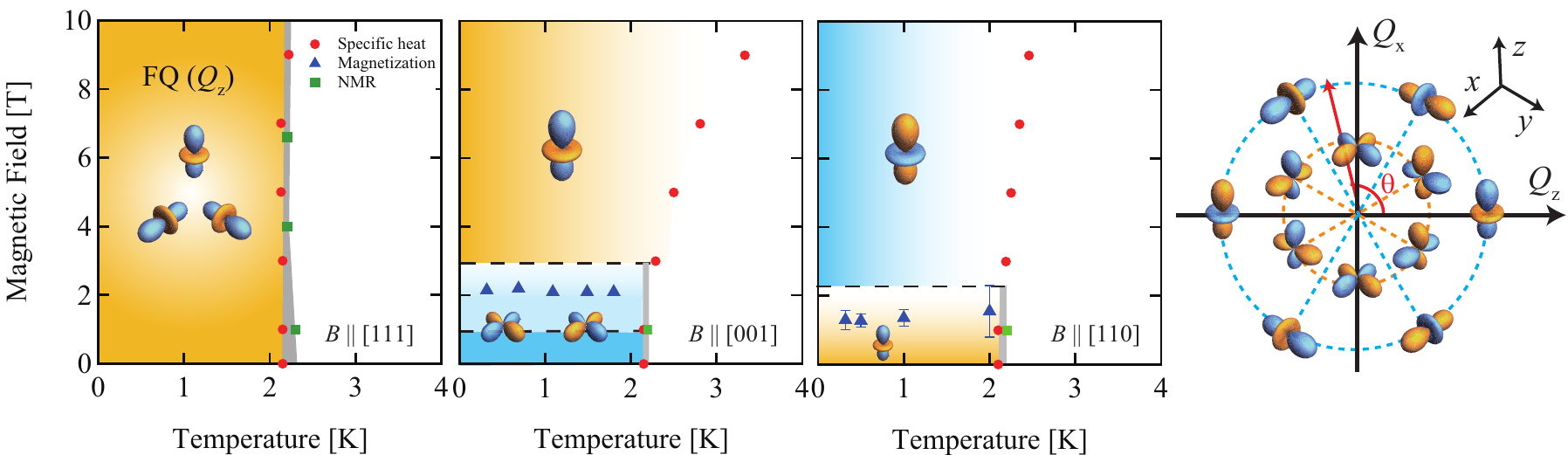}
\caption{\label{fig:phase_diagram} (Color online)  The temperature vs. magnetic-field phase diagrams of the ferro-quadrupole order in PrTi\textsubscript{2}Al\textsubscript{20} for $\boldsymbol{B}_{\rm ext}$ \textbar{}\textbar{} {[}111{]}, {[}001{]}, and {[}110{]}. The order parameters are indicated by graphical symbols, each of which corresponds to a specific angle $\theta$ in the order-parameter space as indicated in the rightmost panel and described in the text. The red dots indicate peak temperatures of $d(C/T)/dT$, $C$ being the specific heat\cite{Sakai_2011,Sakai_2012_proc}, which are identified either as the transition temperature or crossover. The green squares show either onset of the splitting of NMR lines (for ${\bm B}_{\rm ext} \parallel [111]$) or a kink in the temperature dependence of the Knight shift (for ${\bm B}_{\rm ext} \parallel [001]$ and [110]). The dashed lines for ${\bm B}_{\rm ext} \parallel [001]$ and ${\bm B}_{\rm ext} \parallel [110]$ indicate the region of coexistence of high- and low-field phases. The blue triangles show the representative transition fields taken from the peak (for ${\bm B}_{\rm ext} \parallel [001]$) or the average field of the two kinks (for ${\bm B}_{\rm ext} \parallel [110]$) in $dM/dB$.   
}
\end{center}
\end{figure*}

In this paper, we apply general symmetry arguments to analyze the NMR spectra and use mean-field calculation to examine the effects of FQ order on magnetization for various field directions. Our main results are represented by the phase diagram in Fig.~\ref{fig:phase_diagram}, the details of which are discussed in Sect.\ref{sec:level3_5}. Here, following Hattori and Tsunetsugu~\cite{Hattori_2014}, the quadrupole order parameter is expressed as a two dimensional vector in the space spanned by $Q_z$ and $Q_x$,
\begin{equation}
{\bm Q} = \left( \langle Q_z \rangle, \langle Q_x \rangle \right) = Q \left( \cos \theta, \sin \theta \right) 
\label{eq:OrderParam}.
\end{equation}
The symmetry of the FQ order is then specified by a single parameter $\theta$. One can see from Eq.~(\ref{eq:Qz}) that 2$\pi$/3-rotation in ${\bm Q}$-space is equivalent to \textit{C}\textsubscript{3} rotation along {[}111{]} ($J_x \rightarrow J_z \rightarrow J_y$), therefore, corresponds to an equivalent domain under cubic symmetry. For example, $\theta = 0, 2\pi/3$, and $-2\pi/3$ ($\theta = \pi/2$, $7\pi/6$, and $11\pi/6$) correspond to the orders of $3J_{z}^{2}-\boldsymbol{J}^{2}$, $3J_{x}^{2}-\boldsymbol{J}^{2}$, and $3J_{y}^{2}-\boldsymbol{J}^{2}$ ($J_{x}^{2} - J_{y}^{2}$, $J_{y}^{2} - J_{z}^{2}$, and $J_{z}^{2} - J_{x}^{2}$), respectively, with positive values of the order parameters. (See the rightmost panel of Fig.~\ref{fig:phase_diagram}.)     
In this work, we found that a FQ order of $Q_{z}$ quadrupole moments occurs for magnetic fields along the [111] direction. For the fields along [001] or [110], however, the FQ order parameter jumps discontinuously at very low fields less than a few tesla. The results can be explained by the competition between magnetic Zeeman and anisotropic quadrupole interactions, provided that the latter is dominant at low fields but get suppressed at high fields. We also observed that proportionality between NMR Knight shift and magnetic susceptibility breaks down violently in the low-field FQ phases, indicating significant influence of FQ order on the $c$-$f$ hybridization. Since the FQ order is driven by the quadrupole exchange interactions, which are in turn mediated by the $c$-$f$ hybridization, our results suggest a feedback effect, which may play an important role in the discontinuous field-induced transitions.     

The rest of the paper is organized as follows. Sample characteristics and experimental procedures are described in Sect.~\ref{sec:level2}. In Sect.~\ref{sec:level3}, results of magnetization and NMR measurements are presented for the magnetic field along [111], [001], and [110]. Most probable order parameters (possible values of $\theta$) are proposed on the basis of symmetry considerations of the NMR spectra and mean-field analysis of the magnetic susceptibility within a conventional model including isotropic quadrupole interaction.  In Sect.~\ref{sec:level4}, we develop a Landau theory and present mean-field analysis of an unconventional phenomenological model including field-dependent anisotropic quadrupole interaction to reproduce the experimental phase diagram and determine the order parameters more precisely. Sect.~\ref{sec:level5} is devoted to the concluding remarks. 

\section{\label{sec:level2}Experimental Procedures}
Single crystals of PrTi\textsubscript{2}Al\textsubscript{20} were synthesized by the Al flux method as described in Ref.~\citen{Sakai_2011}.
For NMR measurements, one crystal was shaped into a thin plate of the size $2.1\times1.1\times0.07$ mm\textsuperscript{3} with the {[}111{]} direction normal to the plate.
We chose a thin plate geometry because it reduces distribution of the demagnetizing field, thereby resulting in narrower NMR lines, and gives a better signal-to-noise ratio since the rf-penetration depth is much smaller than the thickness of the crystal. The large value of the residual resistivity ratio (RRR $\sim$ 170) ensures high quality of the crystal.
The NMR spectra were obtained by summing the Fourier transform of spin-echo signal recorded at equally spaced rf-frequencies with a fixed magnetic field. 
The orientation of the crystal with respect to the magnetic field was precisely controlled by a double-axis goniometer, typically within 0.2$^\circ$. 

There are three crystallographically inequivalent Al sites, Al(1), Al(2), and Al(3), in the crystal structure of PrTi$_{2}$Al$_{20}$ at the Wyckoff positions 16\textit{c}, 48\textit{f}, and 96\textit{g} in the space group \textit{Fd$\bar{3}$m}, respectively. Each Pr ion is surrounded by four Al(1) and twelve Al(3) atoms. In this paper, we focus on the NMR results at Al(3) sites. The crystallographically equivalent twelve Al(3) atoms split into several inequivalent sites by external magnetic fields as we discuss below.
Since \textsuperscript{27}Al nuclei have spin $I=5/2$, the NMR spectrum for each site consists of five equally-spaced resonance lines at the frequencies
\begin{eqnarray}
f_{k}=\gamma\left|\boldsymbol{B}_{\rm ext}+\boldsymbol{B}_{\rm hf}\right|+k\nu_{q} , \;\left(k=-2, -1, 0, 1, 2\right)
\label{eq:NMRfreq},
\end{eqnarray}
where $\gamma=11.09407$ MHz/T is the nuclear gyromagnetic ratio of \textsuperscript{27}Al, $\boldsymbol{B}_{\rm ext}$ the external magnetic field, $\boldsymbol{B}_{\rm hf}$ the magnetic hyperfine field produced by field-induced local magnetization from surrounding electrons, dominantly from 4$f$ electrons, $\nu_q$ the nuclear quadrupole splitting, which is proportional to the electric-field gradient (EFG) $\partial^2 V/\partial \xi^2$ at the nucleus ($V$ is the electrostatic potential and $\xi$ is the direction of the external field), and $k$ specifies the transition between nuclear spin levels, $\left|I_z=k+1/2\right\rangle\leftrightarrow\left|I_z=k-1/2\right\rangle$, for each of the five resonance lines~\cite{Abragam_1961,NMR_intro}. Complete site-assignment of all NMR lines in PrTi$_{2}$Al$_{20}$ for high-symmetry field directions is reported in Ref.~\citen{Taniguchi_Proc}. 

Note that both the frequency of the center line ($k = 0$) and the average frequencies of the first ($k = \pm 1$) and second ($k = \pm 2$) satellite lines are not affected by the quadrupole effects, allowing us to determine the Knight shift, which is defined by 
\begin{equation}
\left|\boldsymbol{B}_{\rm ext}+\boldsymbol{B}_{\rm hf}\right| = (1 + K) B_{\rm ext} .
\label{eq:Knight shift}
\end{equation}
Since $B_{\rm ext} \equiv \left|\boldsymbol{B}_{\rm ext}\right| \gg \left|\boldsymbol{B}_{\rm hf}\right|$ in paramagnetic states, only the component of $\boldsymbol{B}_{\rm hf}$ parallel to $\boldsymbol{B}_{\rm ext}$ contributes to the Knight shift,
\begin{equation}
K = \frac{\boldsymbol{B}_{\rm ext}\cdot\boldsymbol{B}_{\rm hf}}{B_{\rm ext}^{2}} .
\label{eq:Knight shift2}
\end{equation}
The quadrupole splitting $\nu_q$ can be determined from the spacing between the satellite lines.    

The magnetization ($M$) was measured on another single crystal with the weight of 1.40 mg. A Qunatum Design SQUID magnetometer was used at high temperatures above 1.8~K. At low temperatures ($0.3-3$~K), we used a Faraday magnetometer with capacitive detection~\cite{Sakakibara}. In order to remove the torque contribution $\boldsymbol{M}\times\boldsymbol{B}_{\rm ext}$, the difference of the signal with and without the field gradient ($dB_{z}/dz=5$~T/m) was taken.     

\section{\label{sec:level3}Experimental Results and Analysis}
\subsection{\label{sec:level3_1}General behavior of magnetization}

\begin{figure}
\begin{center}
\includegraphics[width=7cm,clip]{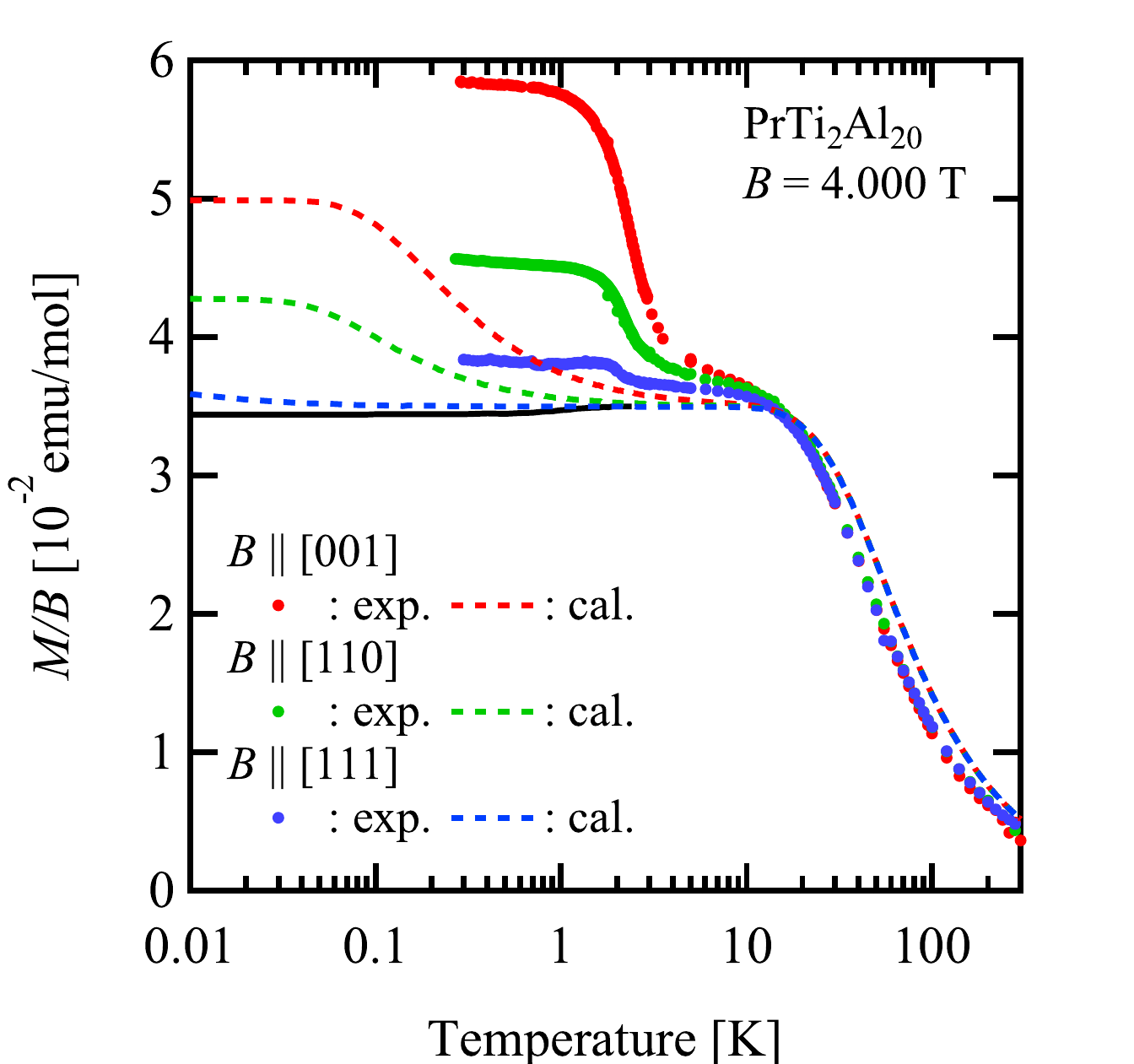}
\caption{\label{fig:mag_para} (Color online) Temperature dependence of the susceptibility $M/B$ for $B_{\rm ext} = 4.000$~T (dots) along {[}111{]} (blue), {[}001{]} (red), and {[}110{]} (green) directions. The dashed lines show the calculated results without quadrupole interaction based on Eq.~(\ref{eq:H_para}). The solid-black line indicates the calculated results assuming a ferro-quadrupole order of $Q_{z}$ ($\langle Q_{z} \rangle > 0$) or $Q_{x}$ in Eqs.~(\ref{eq:H_order}) and (\ref{eq:quadrupole_111}) for ${\bm B}_{\rm ext} \parallel [111]$. The results for two different order parameters are indistinguishable in this plot.   
}
\end{center}
\end{figure}

\begin{table}
\begin{center}
\caption{\label{tab:para}The energy splitting of the $\Gamma_{3}$ doublet ($\Delta E$ in unit of Kelvin)
and the magnetic moments of the ground state ($\mu_{\rm g}$ in unit of Bohr magneton) obtained by diagonalizing $\text{\ensuremath{\mathscr{H}}}_{\rm CEF}+\text{\ensuremath{\mathscr{H}}}_{\rm Z}$ for the magnetic field of 4~T along  {[}111{]},  {[}001{]}, and  {[}110{]}. 
}
\begin{tabular}{cccccc} 
\hline
\multicolumn{2}{c}{$\boldsymbol{B}_{\textrm{ext}}$ \textbar\textbar{} {[}111{]}} & \multicolumn{2}{c}{$\boldsymbol{B}_{\textrm{ext}}$ \textbar\textbar{} {[}001{]}} & \multicolumn{2}{c}{$\boldsymbol{B}_{\textrm{ext}}$ \textbar\textbar{} {[}110{]}}\tabularnewline
\hline 
$\Delta E$ & $\mu_{\textrm{g}}$ & $\Delta E$ & $\mu_{\textrm{g}}$ & $\Delta E$ & $\mu_{\textrm{g}}$\tabularnewline
0.03 & 0.33 & 0.49 & 0.49 & 0.24 & 0.40\tabularnewline 
\hline
\end{tabular}
\end{center}
\end{table}

We first discuss general features of temperature dependence and anisotropy of the magnetic susceptibility. Figure~\ref{fig:mag_para} shows temperature dependence of the magentic susceptibility $M/B$ for the magnetic field $B_{\rm ext}$ of 4~T parallel to {[}111{]}, {[}001{]}, and {[}110{]}.
At high temperatures, the susceptibility increases with decreasing temperature down to 20~K for all field directions and tends to saturate at a constant value at lower temperatures, which is consistent with the earlier report \cite{Sakai_2011}. The origin of the behavior above 20~K is the Curie-Weiss contribution from thermally excited magnetic states, $\Gamma_{4}$ and $\Gamma_{5}$, while the saturation at lower temperatures indicates the van Vleck paramagnetism of the nonmagnetic $\Gamma_{3}$ CEF ground state.
In contrast to the nearly isotropic behavior above 10~K, striking anisotropy develops below 10~K. The $M/B$ increases rapidly for $\boldsymbol{B}_{\rm ext}$ \textbar{}\textbar{} {[}001{]} while changes only slightly for $\boldsymbol{B}_{\rm ext}$ \textbar{}\textbar{} {[}111{]}. Such anisotropy can be partly attributed to the splitting of $\Gamma_{3}$ doublet by magnetic fields as explained below.  

For quantitative understanding, we calculate the temperature dependence of $M/H$ in a mean-field approximation by diagonalizing the following Hamiltonian within the $J$=4 multiplet of Pr$^{3+}$, 
\begin{eqnarray}
\text{\ensuremath{\mathscr{H}}}_{\rm m}=\text{\ensuremath{\mathscr{H}}}_{\rm CEF}+\text{\ensuremath{\mathscr{H}}}_{\rm Z}+\text{\ensuremath{\mathscr{H}}}_{\rm D}
\label{eq:H_para}.
\end{eqnarray}
$\text{\ensuremath{\mathscr{H}}}_{\rm CEF}$ is the CEF potential with the $T_{d}$ point group symmetry\cite{Hattori_2014}
\begin{eqnarray}
\text{\ensuremath{\mathscr{H}}}_{\rm CEF}=\epsilon_{2}\left|\boldsymbol{Q}\right|^{2}-\epsilon_{3}\left\{ Q_{z}^{3}-3\overline{Q_{z}Q_{x}^{2}}\right\}
\label{eq:CEF},
\end{eqnarray}
where the bar represents a cyclic permutation of three operators: $3\overline{AB^{2}}=AB^{2}+BAB+B^{2}A$.  Equation~(\ref{eq:CEF}) is equivalent to the conventional representation of CEF Hamiltonian in terms of the Stevens operator\cite{Stevens}. This expression is suitable for $\Gamma_{3}$ quadrupole systems since $Q_{z}$ and $Q_{x}$ are basis of the $\Gamma_{3}$ representation of the $T_{d}$ group and Eq.~(\ref{eq:CEF}) explicitly shows the anisotropy in the order parameter space, which will be discussed in Sect.~\ref{sec:level4}\cite{Hattori_2014}.
$\text{\ensuremath{\mathscr{H}}}_{\rm Z}$ is the Zeeman interaction
\begin{eqnarray}
\text{\ensuremath{\mathscr{H}}}_{\rm Z}=-g_{J}\mu_{B}\boldsymbol{J}\cdotp\boldsymbol{B}_{\rm ext}
\label{eq:Zeeman},
\end{eqnarray}
and $\text{\ensuremath{\mathscr{H}}}_{\rm D}$ is the mean-field Hamiltonian for the exchange between magnetic dipoles,
\begin{eqnarray}
\text{\ensuremath{\mathscr{H}}}_{\rm D}=-\lambda_{d}\left\langle \boldsymbol{J}\right\rangle \cdotp\boldsymbol{J}
\label{eq:dipole},
\end{eqnarray}
where $\langle A \rangle$ indicates the thermal average of $A$. 
The CEF parameters are determined from the inerastic neutron scattering measurements as $\epsilon_{2}=23$~K and $\epsilon_{3}=8.0$~K \cite{Sato_2012,Hattori_2014}.
The Lande's \textit{g}-factor is gievn as $g_{J}=4/5$ for Pr$^{3+}$ and $\lambda_{d}$ can be estemated from the Weiss temperature of the susceptibility $\Theta_{w}=-40$~K as $\lambda_{d}=3g_{J}^{2}\Theta_{w}/zn_{\rm eff}^{2}=-1.63$ K, where $z=4$ is the number of nearest neighbors, and $n_{\rm eff}\mu_{B}=3.43\mu_{B}$ is the effective magnetic moment~\cite{Sakai_2011}.

The susceptibility $M/B \equiv N_{A} g_{J} \mu_{B}\left\langle \boldsymbol{J}\right\rangle\cdotp\boldsymbol{B}_{\rm ext}/B_{\rm ext}^2$ is calculated self-consistently and shown in Fig.~\ref{fig:mag_para} by the dashed lines. 
The observed susceptibilities are reproduced reasonably well by the calculations above 20 K, indicating appropriate choice of the CEF parameters. However, small systematic deviation is present, which may be caused by Kondo effects, since interactions between $f$-electrons and conduction electrons are not explicitly considered in Eq.~(\ref{eq:H_para}).  

The pronounced anisotropy at lower temperature can be also captured qualitatively by the calculation. This is due to anisotropic mixing between the $\Gamma_{3}$ ground doublet and the $\Gamma_{4}$ and $\Gamma_{5}$ excited triplets by the Zeeman interaction Eq.~(\ref{eq:Zeeman}), which splits the $\Gamma_{3}$ doublet. We show in Table~\ref{tab:para} the energy splitting of the $\Gamma_{3}$ doublet ($\Delta E$) and the magnetic moments of the ground states ($\mu_{\rm g}$). Both ($\Delta E$) and $\mu_{\rm g}$ are largest for $\boldsymbol{B}_{\rm ext}$ \textbar{}\textbar{} {[}001{]} and smallest for $\boldsymbol{B}_{\rm ext}$ \textbar{}\textbar{} {[}111{]} in agreement with the anisotropy of $M/B$. The experimental results and the calculation, however, show quantitative discrepancy. The anisotropy of the observed susceptibility develops rapidly below 2~K in contrast to the gradual growth only below 1~K in the calculated results. To resolve this, we have to consider the quadrupole-quadrupole interaction and quadrupole order as we discuss below.   

Effects of quadrupole order can be taken into account by adding the quadrupole-quadrupole interaction in a mean-field approximation to the Hamiltonian of Eq. (\ref{eq:H_para}), 
\begin{eqnarray}
\text{\ensuremath{\mathscr{H}}}_{\rm tot}=\text{\ensuremath{\mathscr{H}}}_{\rm CEF}+\text{\ensuremath{\mathscr{H}}}_{\rm Z}+\text{\ensuremath{\mathscr{H}}}_{\rm D}+\text{\ensuremath{\mathscr{H}}}_{\rm Q}
\label{eq:H_order},
\end{eqnarray}
\begin{eqnarray}
\text{\ensuremath{\mathscr{H}}}_{\rm Q}=-\lambda\left\langle Q_{z}\right\rangle Q_{z}-\lambda\left\langle Q_{x}\right\rangle Q_{x}
\label{eq:quadrupole_111}.
\end{eqnarray}
Generally, the quadrupole coupling should be expressed as a 2$\times$2 matrix in ${\bm Q}$-space. However, it can be shown on the basis of the symmetry of diamond structure that the coupling is isotropic at zero field. This is because Eq.~(\ref{eq:quadrupole_111}) is the only invariant combination obtained from the product of two $\Gamma_{3}$ representations.  In magnetic fields, the quadrupole interaction can be anisotropic in ${\bm Q}$ space as will be discussed in detail in Sect.~\ref{sec:level4}. 

The value of $\lambda$ is set to $\lambda=0.96$~K to reproduce the transition temperature $T_{\rm Q}=2.2$~K at zero-field. In this section, we consider the cases where either $\langle Q_z \rangle$ or $\langle Q_x \rangle$ is non-zero. Since the Hamiltonian is not symmetric with respect to the sign change of $\langle Q_z \rangle$ due to the third-order term in Eq.~(\ref{eq:CEF}), the cases for $\langle Q_z \rangle > 0$ ($\theta$ = 0) and $\langle Q_z \rangle < 0$ ($\theta$ = $\pi$) must be considered separately. In contrast, the Hamiltonian is symmetric for the sign change of $\langle Q_x \rangle$, therefore it is sufficient to consider only positive $\langle Q_x \rangle$ ($\theta$ = $\pi/2$). The magnetic susceptibility in FQ ordered phase is calculated by diagonalizing ${\mathscr{H}}_{\rm tot}$ [Eq.~(\ref{eq:H_order})] for various values of $\langle {\bm J} \rangle \parallel {\bm B}_{\rm ext}$ and ${\bm Q}$ with the constraint $\theta =$ 0, $\pi/2$, or $\pi$ [Eq.~(\ref{eq:OrderParam})] to find a self-consistent solution. Strictly speaking, solutions for $\theta$ = $\pi/2$ and $\pi$ are not stable against small changes in $\theta$. Here we try to capture overall feature of $M/B$ for a given value of $\theta$. The fully self-consistent results are discussed in Sect.~\ref{sec:level4}. 

\subsection{\label{sec:level3_2}FQ order for $\boldsymbol{B}_{\rm ext}$ \textbar{}\textbar{} {$\mathrm{[111]}$}}
\subsubsection{\label{sec:level3_2_1}Magnetization behavior for $\boldsymbol{B}_{\rm ext}$ \textbar{}\textbar{} {$\mathrm{[111]}$}}

\begin{figure}[b]
\begin{center}
\includegraphics[width=9cm,clip]{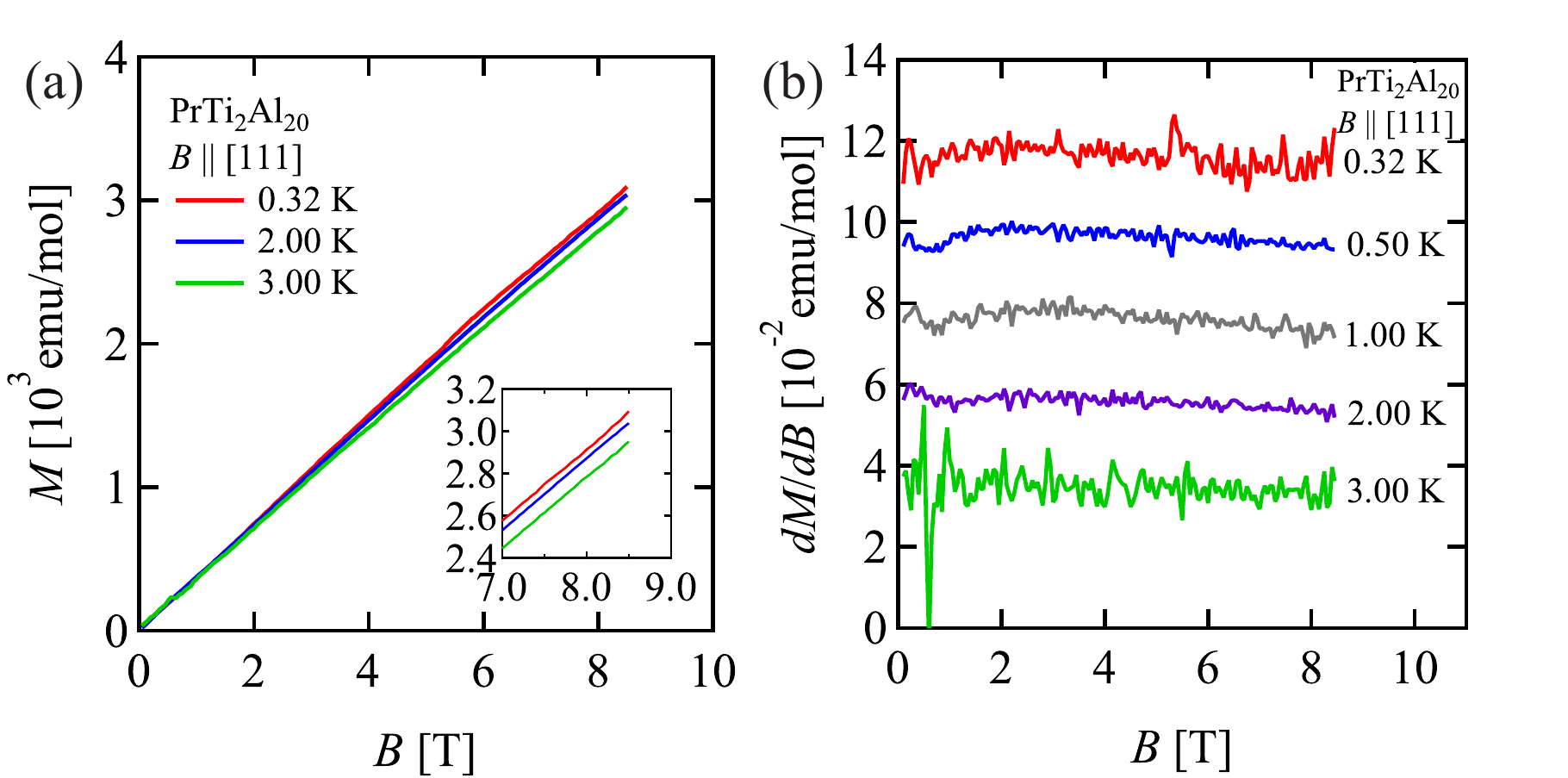}
\caption{\label{fig:mag_111} (Color online) 
Field dependence of (a) the magnetization and (b) the differential susceptibility for $\boldsymbol{B}_{\rm ext}$  \textbar{}\textbar{} {[}111{]}. The plots in (b) are vertically shifted consecutively by 0.02 emu/mole. 
}
\end{center}
\end{figure}

The results of mean-field calculation of the susceptibility for $\boldsymbol{B}_{\rm ext}$ \textbar{}\textbar{} {$\mathrm{[111]}$} with FQ orders of $Q_z$ ($\theta$ = 0) and $Q_x$  ($\theta$ = $\pi/2$) are shown in Fig.~\ref{fig:mag_para} by the solid-black line. These two cases are indistinguishable in this plot. The calculated susceptibility is nearly constant below 10~K without any anomaly at $T_{Q}$, qualitatively consistent with the experimental result. However, we should remark that the experimental susceptibility shows slight increase below 10~K and a small step at $T_{Q}$, which are not reproduced by the calculation. This feature may be caused by modestly strong $c$-$f$ hybridization and beyond the applicability of models with purely localized $f$ electrons.      
Figures~\ref{fig:mag_111}(a) and \ref{fig:mag_111}(b) show the field dependence of the magnetization and differential susceptibility ($dM/dB$) for $\boldsymbol{B}_{\rm ext}$ \textbar{}\textbar{} {$\mathrm{[111]}$} at several temperatures. The magnetization is proportional to the magnetic field in a good approximation both below and above $T_{\rm Q}$ without any anomaly.

\subsubsection{\label{sec:level3_2_2}NMR line splitting for $\boldsymbol{B}_{\rm ext}$ \textbar{}\textbar{} {$\mathrm{[111]}$}}

\begin{figure}
\begin{center}
\includegraphics[width=8.5cm,clip]{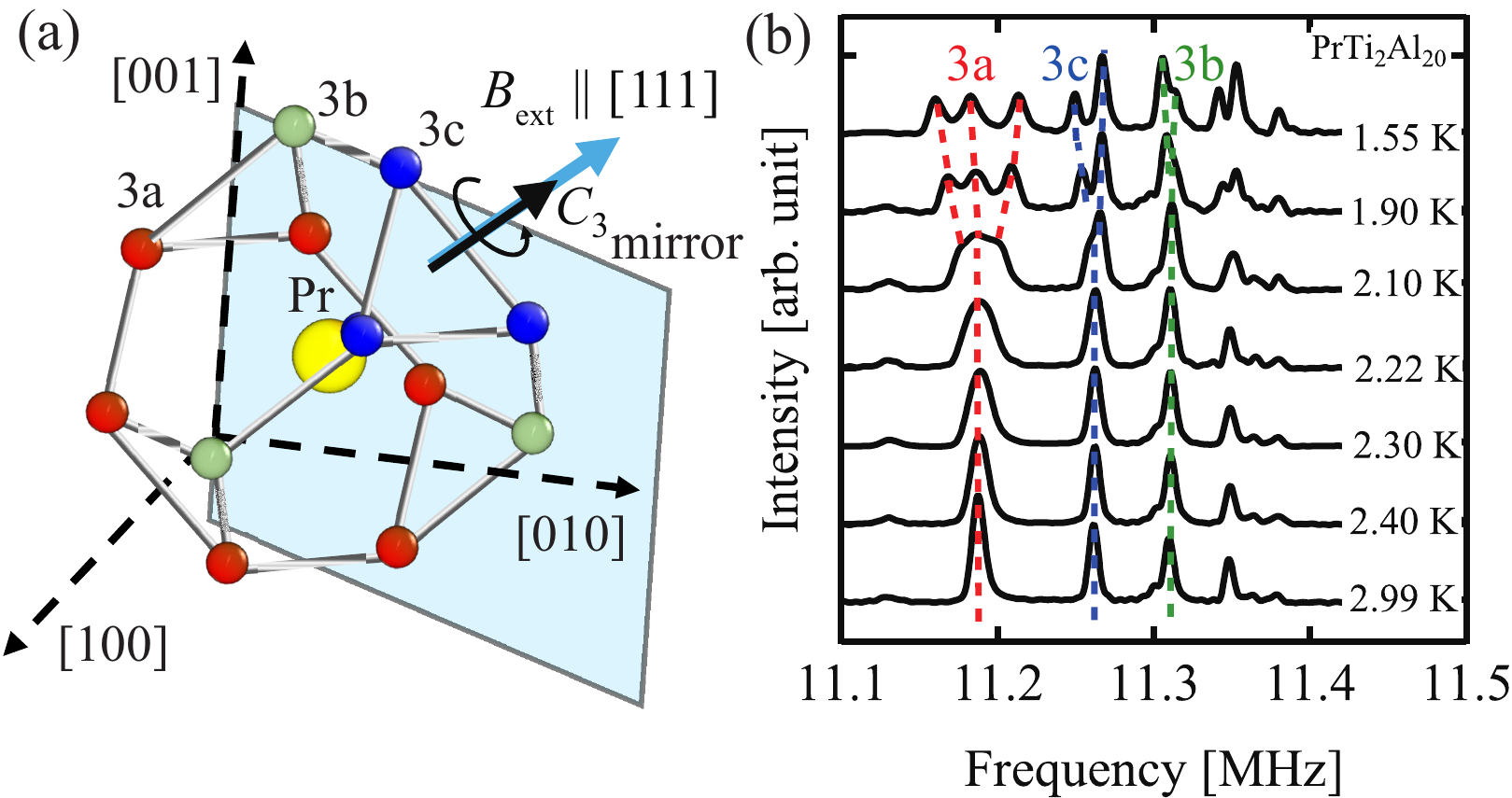}
\caption{\label{fig:sp111_1T} (Color online) (a) Twelve Al(3) sites forming a cage surrounding a Pr site. Under magnetic fields along  {[}111{]}, they split into three inequivalent sites, six 3a (red), three 3b (green), and three 3c (blue) sites. (b)Temperature dependence of NMR spectrum of the center line [$k=0$ in Eq.~(\ref{eq:NMRfreq})] at Al(3) sites for $B_{\rm ext} = 1.0062$~T along  {[}111{]}. Each NMR line from 3a, 3b, and 3c sites further split in the quadrupole ordered phase below $T_{Q}=2.2$~K.  
}
\end{center}
\end{figure}

\begin{figure}
\begin{center}
\includegraphics[width=8.5cm,clip]{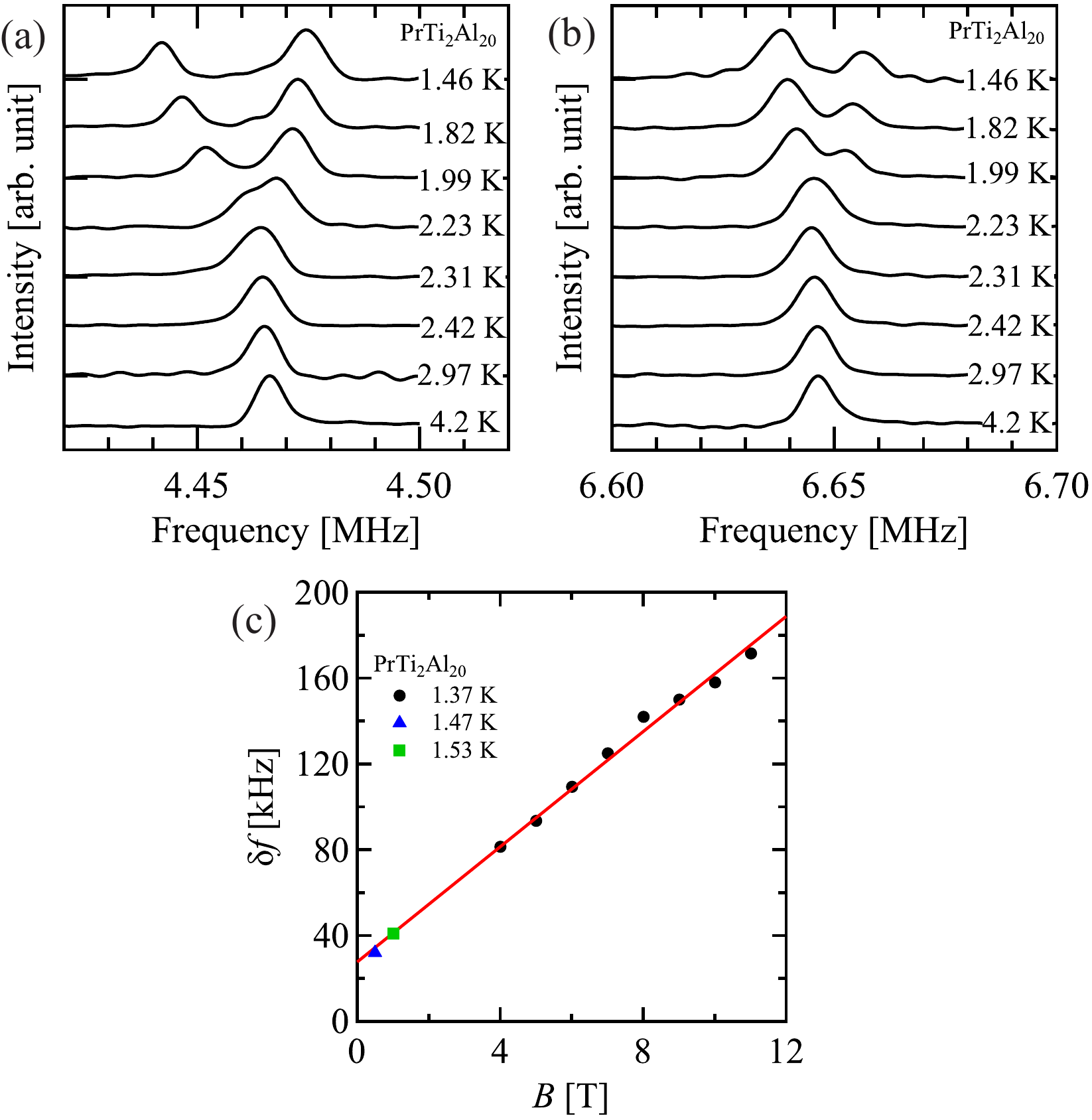}
\caption{\label{fig:sp111_0p5T} (Color online) Temperature dependence of the second satellite NMR spectra for (a) $k=-2$ and (b) $k=+2$ [Eq.~(\ref{eq:NMRfreq})] at 3c sites for $B_{\rm ext} = 0.4993$~T along {[}111{]}. (c) Field dependence of the splitting ($\delta f$) of the $k=-2$ line for $\boldsymbol{B}_{\rm ext}$ \textbar{}\textbar{} {$\mathrm{[111]}$}. The data above 4~T are taken from Ref.~\citen{Taniguchi_JPSJ}.
}
\end{center}
\end{figure}

Among three different Al sites, our NMR measurements focus on Al(3) sites, which have a mirror symmetry. Figure~\ref{fig:sp111_1T}(a) shows a cage composed of twelve Al(3) sites surrounding a Pr ion. These sites are crystallographically equivalent. However, their NMR spectra can be different since application of a magnetic field generally reduces the symmetry. For a field along {[}111{]}, Al(3) sites split into three groups of sites denoted by 3a, 3b, and 3c in Fig.~\ref{fig:sp111_1T}(a) with the population ratio of 2:1:1~\cite{Taniguchi_Proc}. Those sites belonging to the same group can be interchanged either by \textit{C}\textsubscript{3} rotation along {[}111{]} or $(1\bar{1}0)$ mirror, which leaves the field direction invariant. Therefore, they give the identical NMR spectra above $T_{\rm Q}$.

Figure~\ref{fig:sp111_1T}(b) shows temperature dependence of the NMR spectrum of the center line [$k=0$ in Eq.~(\ref{eq:NMRfreq})] at Al(3) sites for $B_{\rm ext} = 1.0062$~T along  {[}111{]}. Each line from 3a, 3b, and 3c sites split into three, two, and two lines below $T_{Q}=2.2$~K with the intensity ratio of approximately 1:1:1, 1:2, and 1:2, respectively.
Since the three 3b sites and the three 3c sites on a single cage are interchanged by \textit{C}\textsubscript{3} rotation, the line splitting provides the direct evidence for the loss of \textit{C}\textsubscript{3} symmetry in the ordered phase as already discussed in Ref.~\citen{Taniguchi_JPSJ}. 

Figures~\ref{fig:sp111_0p5T}(a) and \ref{fig:sp111_0p5T}(b) show the pair of second satellite spectra ($k=\pm2$) at 3c sites for $B_{\rm ext} = 0.4993$~T along {[}111{]}. Both of them split into two lines below $T_{\rm Q}=2.2$~K. While the higher frequency peak of the split line is more intense for $k=-2$ [Fig.~\ref{fig:sp111_0p5T}(a)], the lower frequency peak is stronger for $k=+2$ [Fig.~\ref{fig:sp111_0p5T}(b)]. This indicates that the splitting is primarily caused by differentiation in EFG [$\nu_q$ in Eq.~(\ref{eq:NMRfreq})] rather than in magnetic hyperfine field [$\boldsymbol{B}_{\rm hf}$  in Eq.~(\ref{eq:NMRfreq})] among the three 3c sites. The spacing between the split peaks of the $k=-2$ line increases linearly with external field as shown in Fig.~\ref{fig:sp111_0p5T}(c), where the data above 4~T are taken from Ref.~\citen{Taniguchi_JPSJ}. This behavior is due to increase of differentiation in $\boldsymbol{B}_{\rm hf}$ produced by the component of field-induced magnetic dipole perpendicular to $\boldsymbol{B}_{\rm ext}$, which becomes dominant at high fields\cite{Taniguchi_JPSJ}. The splitting extrapolates to a finite value $\sim27$~kHz at zero field, where $\boldsymbol{B}_{\rm hf}$ = 0. Thus, the differentiation in $\nu_q$ caused by quadrupole order gives dominant contribution to the splitting at low fields. 

\subsubsection{\label{sec:level3_2_3}Order parameters for $\boldsymbol{B}_{\rm ext}$ \textbar{}\textbar{} {$\mathrm{[111]}$}}

\begin{figure}
\begin{center}
\includegraphics[width=8.5cm,clip]{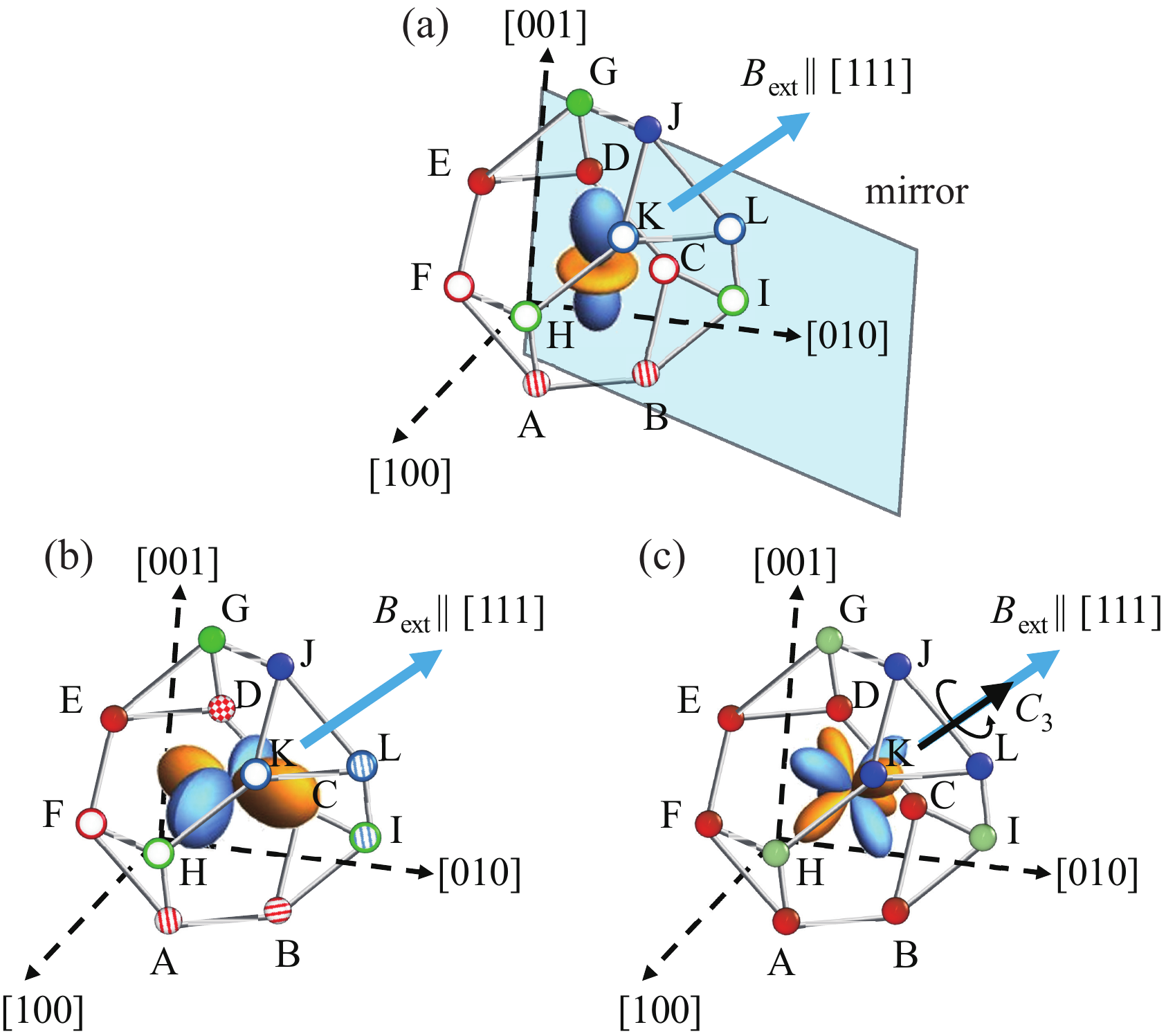}
\caption{\label{fig:op111} (Color online) Symmetry reduction of Al(3) cage due to order of (a) \textit{Q}\textsubscript{z}, (b) \textit{Q}\textsubscript{x}, and (c) \textit{T}\textsubscript{xyz} for $\boldsymbol{B}_{\rm ext}$ \textbar{}\textbar{} {$\mathrm{[111]}$}. In each panel, twelve Al(3) sites labeled as A-L are marked by different colors to distinguish 3a (red), 3b (green), and 3c (blue) sites. Further splitting of these sites due to mulitipole order is indicated by different filling patterns of the circle. 
}
\end{center}
\end{figure}

Here, we show that our NMR results uniquely determine the order parameter for $\boldsymbol{B}_{\rm ext}$ \textbar{}\textbar{} {$\mathrm{[111]}$} based on general symmetry arguments. In the following, we consider possible breaking of local symmetry at Al(3) sites caused by either ferro- or antiferro-order of $Q_{z}$ or $Q_{x}$. We also consider orders of $T_{xyz}$ octupole, which is odd in time reversal, therefore, couples directly to nuclear spins and produces magnetic hyperfine fields\cite{Sakai_1997}. Symmetry of such coupling is analyzed in Appendix.   

Let us first consider ferro-quadrupole (FQ) order of $Q_{z}$. As shown in Fig.~\ref{fig:op111}(a), an order of $Q_{z}$ breaks \textit{C}\textsubscript{3} symmetry along the external-field direction but preserves mirror symmetry with respect to the $(1\overline{1}0)$ plane. Therefore, both 3b and 3c sites split into two groups with 2:1 ratio, while 3a sites split into three groups with equal population. This splitting is displayed schematically in Fig.~\ref{fig:op111}(a) by different filling patterns of the circles representing Al(3) sites. In the case of antiferro-quadrupole (AFQ) order of $Q_{z}$, Pr sites are divided into two sub-lattices (A and B) with different values of the quadrupole moments, $\left\langle Q_{z}\right\rangle_{A}  \sim -\left\langle Q_{z}\right\rangle_{B}$. Since no symmetry operation causes sign reversal of $\left\langle Q_{z}\right\rangle$, the number of inequivalent sites should be doubled from the case of FQ order.  

We next consider FQ order of $Q_{x}$ [Fig.~\ref{fig:op111}(b)]. In this case, not only \textit{C}\textsubscript{3} but also mirror symmetry is broken, leaving no symmetry preserved. Therefore, all Al(3) sites become inequivalent. In the case of AFQ order of $Q_{x}$, by noting that the sign of $\langle Q_{x} \rangle \propto \langle J_{x}^{2} - J_{y}^{2} \rangle$ is reversed by mirror operation with respect to $(1\overline{1}0)$, we conclude that the number of inequivalent sites does not change from the case of FQ order.
 
Finally we consider order of $T_{xyz}$ octupole [Fig.~\ref{fig:op111}(c)]. The magnetic hyperfine field produced by $T_{xyz}$ moments is discussed in Appendix. From the results in Tables~\ref{tab:Hf_oct}, we conclude none of NMR lines from 3a, 3b, and 3c sites split by ferro-octupole (FO) order of $T_{xyz}$. In the case of antiferro-octupole (AFO) order of $T_{xyz}$, since the direction of the hyperfine field is reversed for the two sub-lattices, the number of NMR lines should be doubled for all sites.

The number of NMR lines expected for various types of multipole orders is summarized in Table~\ref{tab:split_111}. Only the case of FQ order of $Q_{z}$ is compatible with the experimental observation of three, two, and two lines for 3a, 3b, and 3c sites. This is also supported by the behavior of the susceptibility. We therefore conclude that FQ order of $Q_{z}$ occurs for $\boldsymbol{B}_{\rm ext}$ \textbar{}\textbar{} {$\mathrm{[111]}$}. Because \textit{C}\textsubscript{3} symmetry is preserved for this field direction, there should exist three domains of equivalent order parameters, $3J_{x}^{2}-\boldsymbol{J}^{2}$, $3J_{y}^{2}-\boldsymbol{J}^{2}$, and $3J_{z}^{2}-\boldsymbol{J}^{2}$, which correspond to $\theta = 2\pi/3$, $-2\pi/3$, and 0 in Eq.~(\ref{eq:OrderParam}), respectively.    
\begin{table}
\begin{center}
\caption{\label{tab:split_111}Expected number of split NMR lines from 3a, 3b, and 3c sites in various multipole ordered phases for $\boldsymbol{B}_{\rm ext}$ \textbar{}\textbar{} {$\mathrm{[111]}$}. 
}
\begin{tabular}{ccccccc}
\hline 
 & \multicolumn{2}{c}{$Q_{z}$} & \multicolumn{2}{c}{$Q_{x}$} & \multicolumn{2}{c}{$T_{xyz}$}\tabularnewline
\cline{2-7} 
site & F & AF & F & AF & F & AF\tabularnewline
\hline 
3a & 3 & 6 & 6 & 6 & 1 & 2\tabularnewline 
3b & 2 & 4 & 3 & 3 & 1 & 2\tabularnewline 
3c & 2 & 4 & 3 & 3 & 1 & 2\tabularnewline
\hline 
\end{tabular}
\end{center}
\end{table}

\subsection{\label{sec:level3_3}FQ order and the field-induced transition for $\boldsymbol{B}_{\rm ext}$ \textbar{}\textbar{} {$\mathrm{[001]}$}}
\subsubsection{\label{sec:level3_3_1}Magnetization behavior for $\boldsymbol{B}_{\rm ext}$ \textbar{}\textbar{} {$\mathrm{[001]}$}}

\begin{figure}
\begin{center}
\includegraphics[width=9cm,clip]{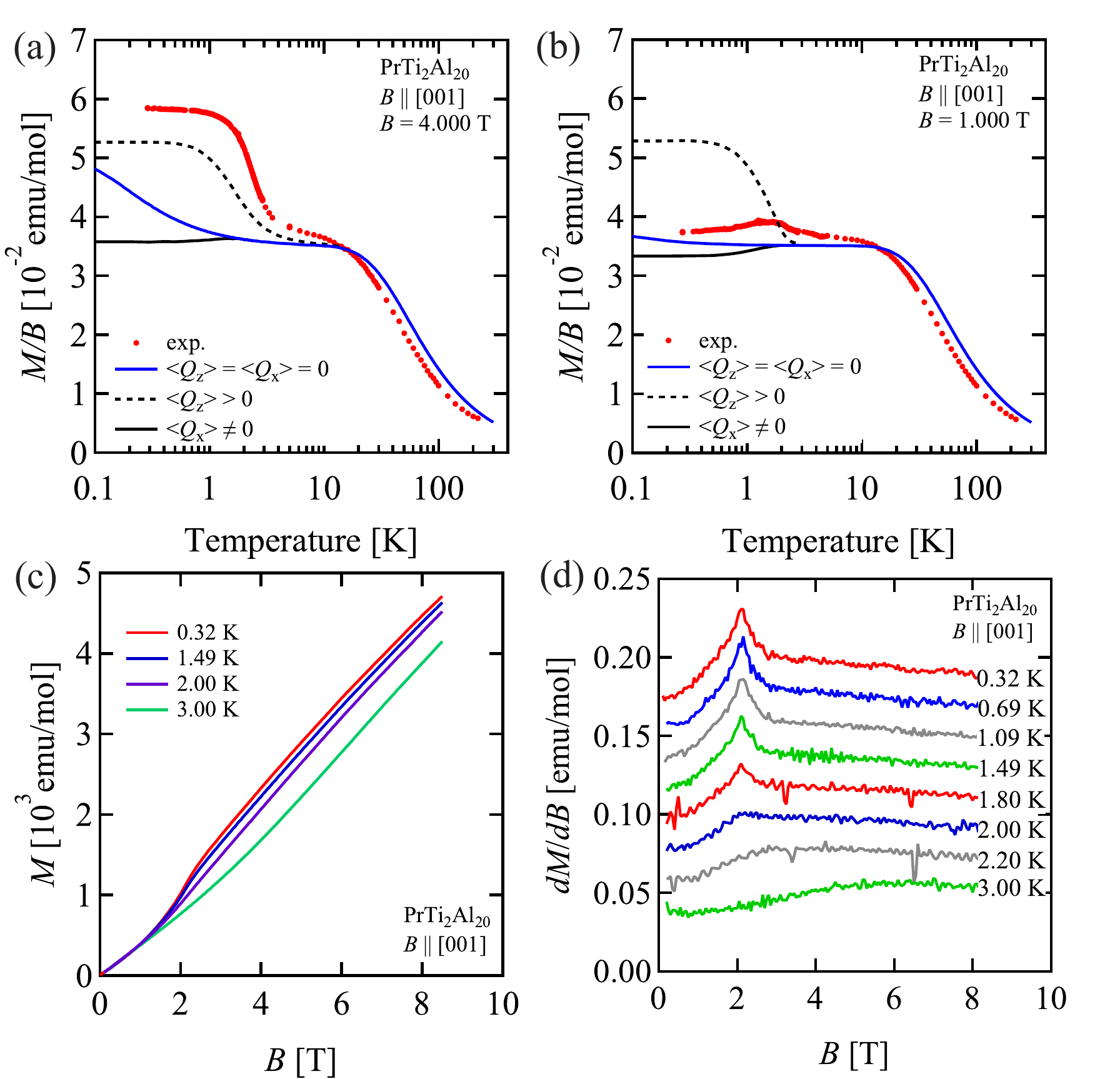}
\caption{\label{fig:mag001} (Color online) (a)(b):Temperature dependence of the magnetic susceptibility $M/B$ for the field of (a) 4~T and (b) 1~T along {[}001{]} (red dots). The blue lines show the results of calculation without quadrupole interaction based on Eq. (\ref{eq:H_para}). The solid- (dashed-) black line indicates the results assuming the FQ order of $Q_{x}$ (FQ order of $Q_{z}$ with $\langle Q_{z} \rangle >0$, which is actually a crossover) in Eqs.~(\ref{eq:H_order}) and (\ref{eq:quadrupole_111}). (c)(d): Field dependence of (c) the magnetization and (d) the differential susceptibility for $\boldsymbol{B}_{\rm ext}$  \textbar{}\textbar{} {[}001{]}. The plots in (d) are vertically shifted consecutively by 0.02 emu/mole. 
}
\end{center}
\end{figure}

Figure~\ref{fig:mag001}(a) shows temperature dependence of the magnetic susceptibility $M/B$ at $B_{\rm ext} = 4$~T along {[}001{]} (red dots, also shown in Fig.~\ref{fig:mag_para}). The dashed-black line shows the result of calculation assuming $Q_{z}$-type FQ order ($\theta$ = 0) in Eqs.~(\ref{eq:H_order}) and (\ref{eq:quadrupole_111}). This agrees with the experimental data much better than the calculation without quadrupole interaction [Eq. (\ref{eq:H_para})] (the blue line, also shown in Fig.~\ref{fig:mag_para}). Assumption of $Q_{x}$-type FQ order ( $\theta$ = $\pi/2$, solid-black line), on the other hand, fails to reproduce the rapid increase of $M/B$ below about 3~K. As mentioned in Sect.~\ref{sec:level3_1}, the increase of $M/B$ at low temperatures for $\boldsymbol{B}_{\rm ext}$ \textbar{}\textbar{} {$\mathrm{[001]}$} is attributed to the splitting of $\Gamma_{3}$ doublet by the Zeeman interaction [Eq.~(\ref{eq:Zeeman})]. A similar splitting is also caused by $Q_{z}$-type FQ order. This is understood by examining the matrix elements of $Q_{z}$ and $J_{z}$ within the $\Gamma_{3}$ doublet perturbed by the Zeeman interaction, 
\begin{eqnarray}
Q_{z} = \left(\begin{array}{cc}
1.05 & 0\\
0 & -1.00
\end{array}\right), \ \ 
J_{z} = \left(\begin{array}{cc}
0.61 & 0\\
0 & 0.16
\end{array}\right)
\label{eq:QzJz_matrix_001}, 
\end{eqnarray}
where the eigenstates of ${\mathscr{H}}_{\rm CEF}+{\mathscr{H}}_{\rm Z}$ for $B_{\rm ext} = 4$~T along {[}001{]} are taken as the basis. 
Thus, magnetic fields along {[}001{]} work cooperatively with $Q_{z}$ type FQ interaction. This is why $M/B$ increase rapidly below about 3~K, which is close to $T_{Q}$ at zero field. One may also say that $J_{z}$ dipole and $Q_{z}$ quadrupole have the same symmetry under magnetic fields along {[}001{]}. Let us emphasize that the rapid variation of $M/B$ is not a phase transition but a crossover, since a finite $\langle Q_{z} \rangle$ is induced by the field already above $T_{Q}$. 

Surprisingly, the susceptibility shows completely different behavior at a lower field of 1~T as shown in Fig.~\ref{fig:mag001}(b). The susceptibility increases slightly below 10~K down to $\sim$2~K, then decreases slowly below 1~K. This behavior is in contradiction to the case with finite $\langle Q_{z} \rangle > 0$ but agrees qualitatively with the calculation assuming FQ order of $Q_{x}$. Note that $Q_{x}$ has different symmetry from $J_{z}$, therefore, does not couple to the external fields along {[}001{]}. These results suggest a change of order parameter as the field varies between 1 and 4~ T. 

Indeed, the magnetization and the differential susceptibility $dM/dB$ data provide clear evidence for a phase transition in magnetic fields. As shown in Figs.~\ref{fig:mag001}(c) and (d), \textit{M} is proportional to $B$ above 3 K with no anomaly in $dM/dB$. However, a clear step in $M$ and the corresponding peak in $dM/dB$ develop at $B_{\rm ext}=2$~T below 2.2~K, strongly indicating a field-induced phase transition below $T_{Q}$. 

\subsubsection{\label{sec:level3_3_2}NMR evidence of the field-induced transition for $\boldsymbol{B}_{\rm ext}$ \textbar{}\textbar{} {$\mathrm{[001]}$} and the Knight shift anomaly}

\begin{figure}
\begin{center}
\includegraphics[width=9cm]{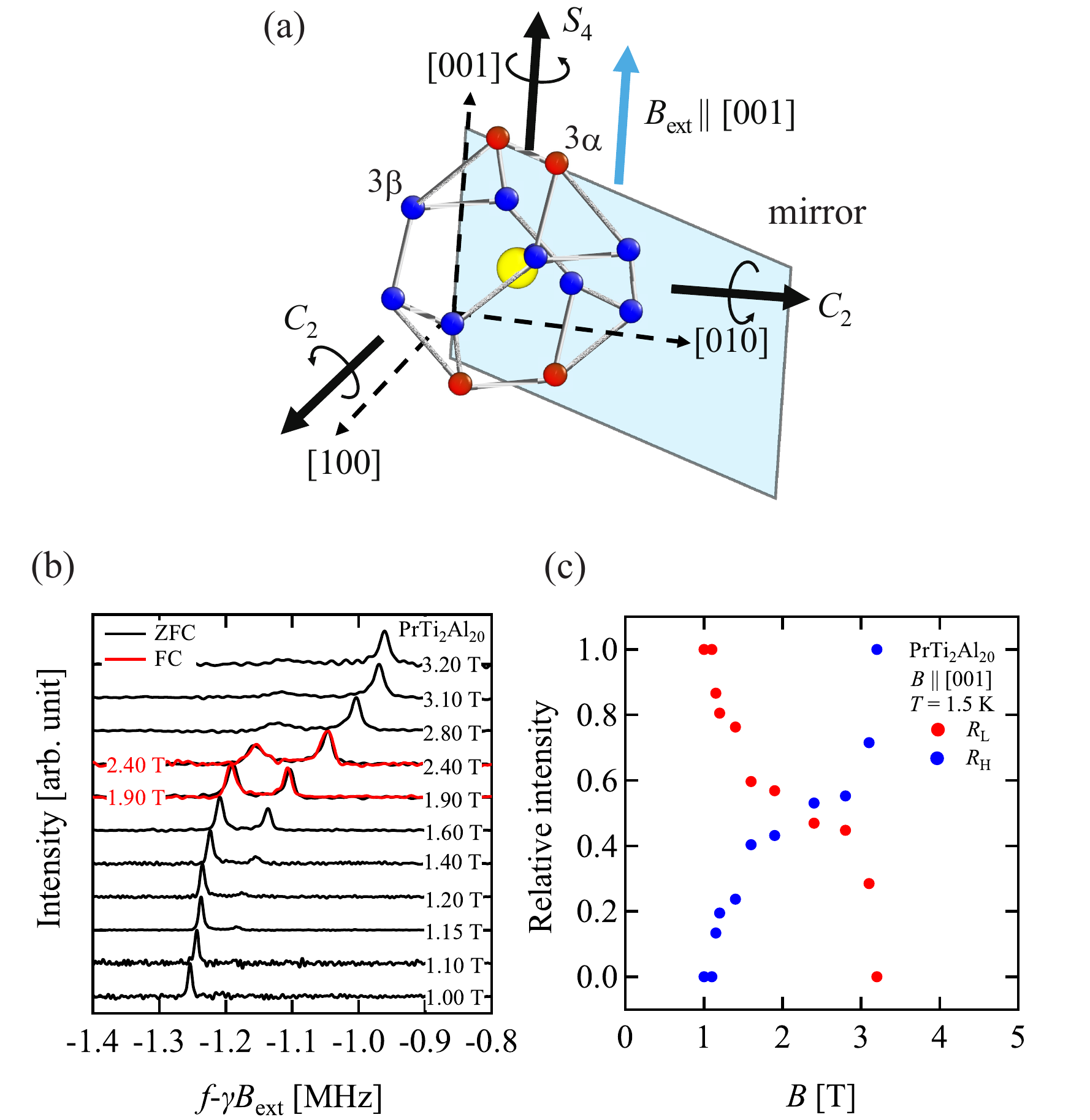}
\caption{\label{fig:sp001_1p5K} (Color online) (a) Twelve Al(3) sites forming a cage surrounding a Pr site. Under magnetic fields along  {[}001{]}, they split into two inequivalent sites, four 3$\alpha$ (red) and eight 3$\beta$ (blue) sites. 
(b) Field dependence of the NMR spectrum (plotted as a function of shift from the nuclear Zeeman frequency) of the low-frequency second satellite line ($k=-2$) at 3$\alpha$ site at 1.5~K. For $B_{\rm ext}=1.9$ and 2.4~T, two spectra obtained under field-cooled (red) and zero-field cooled (black) conditions are shown with no discernible difference. (c) Field dependence of the normalized intensities, $R_{L} = I_{L}/\left(I_{H}+I_{L}\right)$, $R_{H} = I_{H}/\left(I_{H}+I_{L}\right)$, where $I_{L}$ ($I_{H}$) are the intensity of the NMR line from the low field (high field) phase.
}
\end{center}
\end{figure}

While magnetic fields along {[}001{]} reduce the cubic symmetry by breaking $C_{3}$ symmetry, \textit{S}\textsubscript{4} rotoreflection along {[}001{]} and mirror with respect to $(1\bar{1}0)$ and (110) planes are preserved~\cite{mirror_note}. The twelve Al(3) sites on a cage then split into two groups, four 3$\alpha$ and eight 3$\beta$ sites as shown in Fig.~\ref{fig:sp001_1p5K}(a)~\cite{Taniguchi_Proc}.
Figure~\ref{fig:sp001_1p5K}(b) shows field dependence of the NMR spectrum of the low-frequency second satellite line ($k=-2$) at 3$\alpha$ site at 1.5~K. While the spectrum consists of a single line at 1.0~T, a second line appears at a higher frequency above 1.15~T and grows with increasing field. At the same time, the first low-frequency line gradually disappears with increasing field. No discernible difference is observed between field-cooled and zero-field-cooled conditions. The relative intensities of the two lines are plotted against $B_{\rm ext}$ in Fig.~\ref{fig:sp001_1p5K}(c). These results establish a field-induced first order phase transition near 2~T for $\boldsymbol{B}_{\rm ext}$ \textbar{}\textbar{} {$\mathrm{[001]}$} with a finite range of magnetic field where two phases coexist. 

\begin{figure}
\begin{center}
\includegraphics[width=8.5cm,clip]{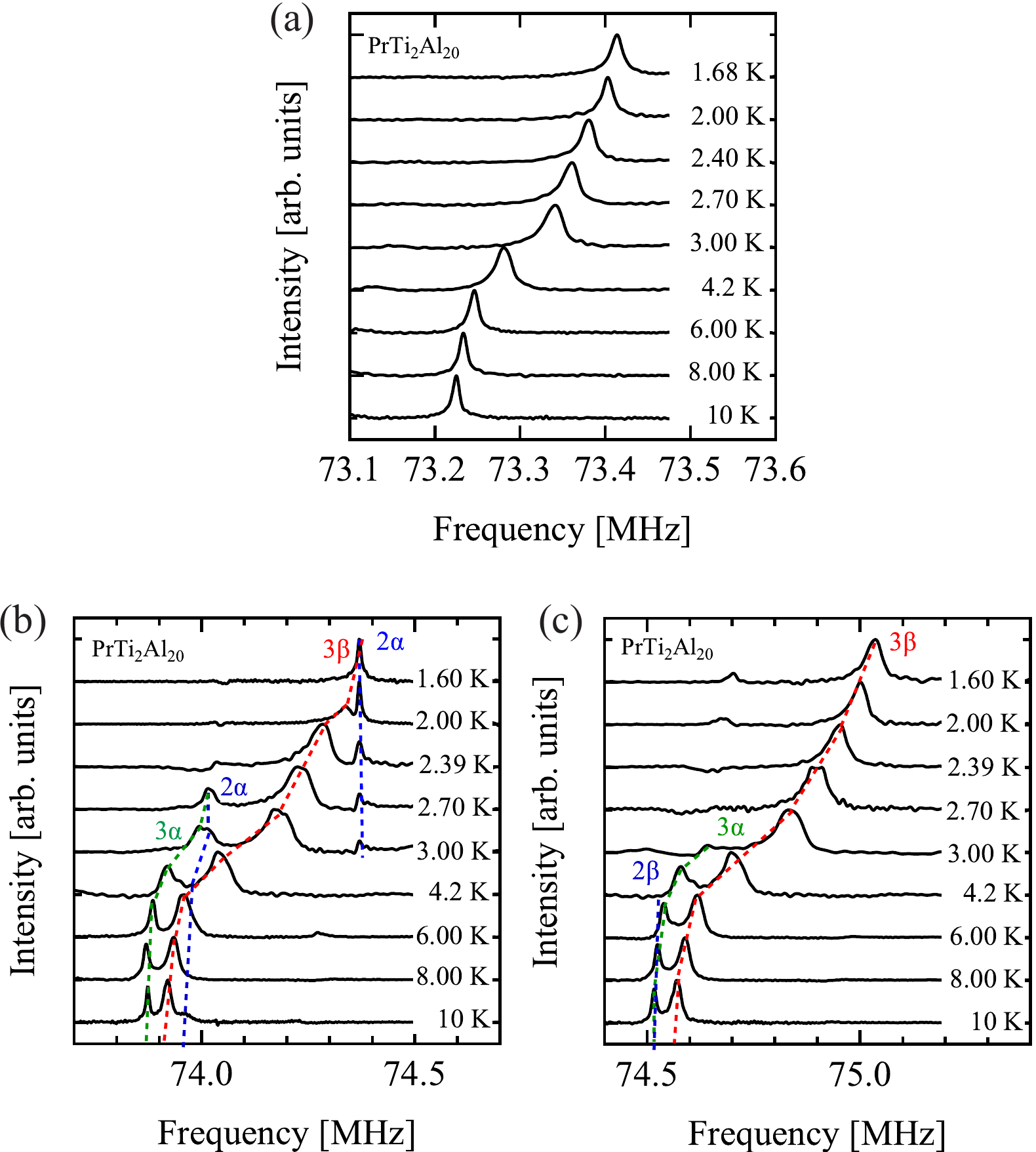}
\caption{\label{fig:sp001_6p615T} (Color online)  Temperature dependence of the NMR spectra for $B_{\rm ext}=6.615$~T along {[}001{]}: (a) the low-frequency first satellite line ($k=-1$) at 3$\alpha$ site, (b) the low-frequency ($k=-1$) and (c) the high-frequency ($k=1$) first satellite lines at 3$\beta$ site (marked by the dashed red lines), which overlap with lines from other sites at some temperatures (dashed blue and green lines). The 2$\alpha$ and 2$\beta$ represent Al(2) sites as defined in Ref.~\citen{Taniguchi_Proc}.
}
\end{center}
\end{figure}

We now examine temperature dependence of the NMR spectra in the high- and low-field phases separately. In Fig.~\ref{fig:sp001_6p615T}, we show temperature dependence of selected NMR lines of both 3$\alpha$ (a) and 3$\beta$  (b, c) sites in the high-field phase ($B_{\rm ext}=6.615$~T). In all cases, the peak frequency increases with decreasing temperature: first gradually from 10~K down to 6~K then rapidly at lower temperatures without any splitting. Since both the $k=-1$ and $k=+1$ lines of 3$\beta$ site move in the same direction, this behavior is attributed to the changes in $\boldsymbol{B}_{\rm hf}$ rather than in $\nu_{q}$ in  Eq.~(\ref{eq:NMRfreq}).
\begin{figure}
\includegraphics[width=8.5cm,clip]{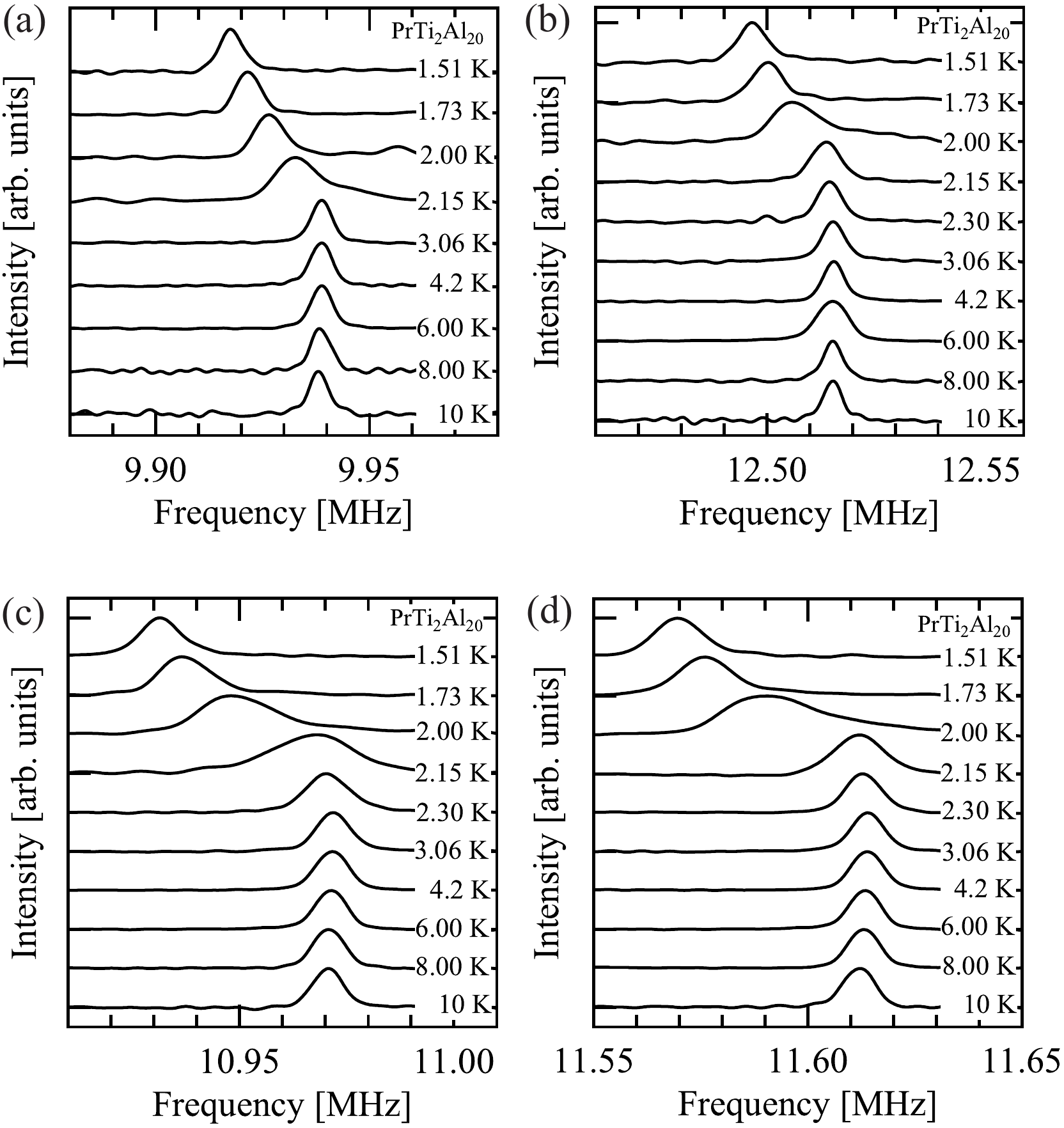}
\caption{\label{fig:sp001_1p0T} Temperature dependence of the NMR spectra for $B_{\rm ext}=1.007$~T along {[}001{]}: (a) the low-frequency ($k=-2$) and (b) high-frequency ($k=2$) second satellite lines at 3$\alpha$ site, (c) the low-frequency ($k=-1$) and (d) high-frequency ($k=1$) first satellite lines at 3$\beta$ site.
}
\end{figure}
The low-field phase, however, shows opposite behavior as shown in Fig.~\ref{fig:sp001_1p0T}. For $B_{\rm ext}=1.007$~T, all lines from 3$\alpha$ (a, b) and 3$\beta$ (c, d) sites move to lower frequencies below 2.2~K. Again the origin of this shift is the changes in $\boldsymbol{B}_{\rm hf}$. 

\begin{figure}
\begin{center}
\includegraphics[width=7cm,clip]{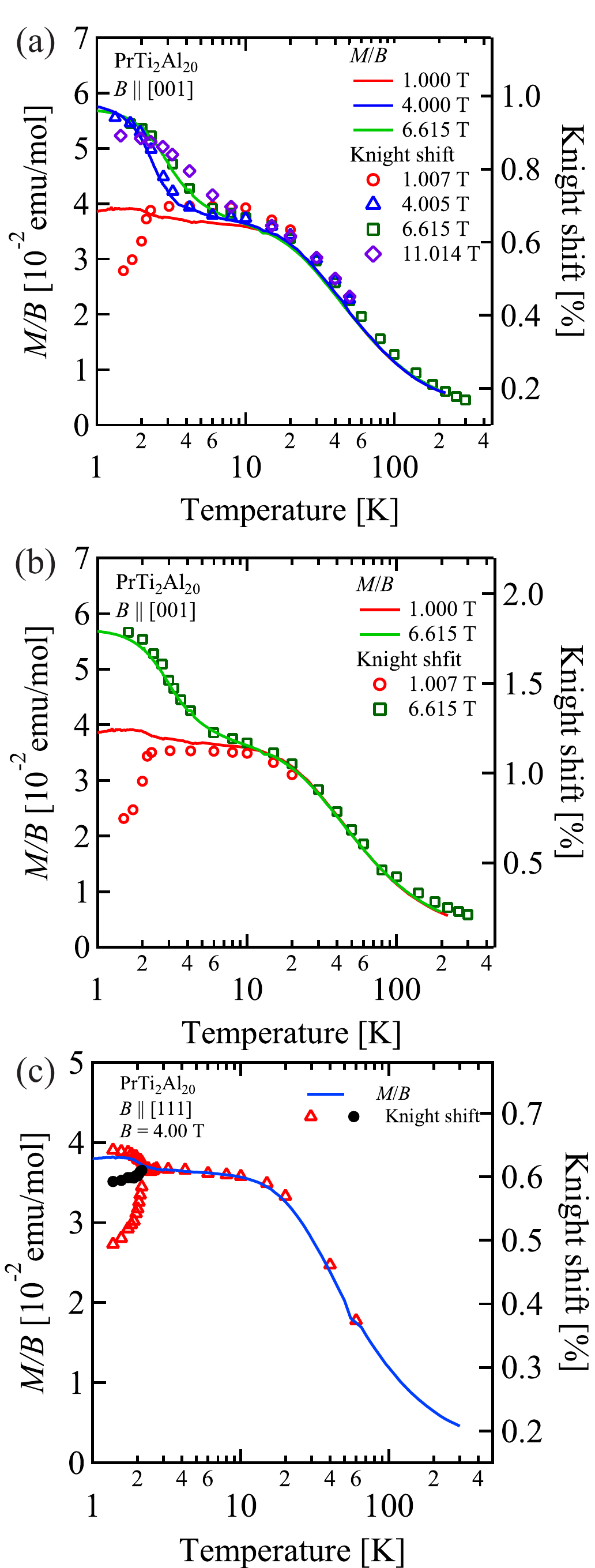}
\caption{\label{fig:Kx001} (Color online) Temperature dependence of the Knight shift at (a) 3$\alpha$ and (b) 3$\beta$ sites for several values of magnetic field along {[}001{]}. The magnetic susceptibility $M/B$ for the same field values are also shown for comparison. (c) Temperature dependences oxief the susceptibility and the Knight shift at 3c sites for the field of 4~T along [111].   The Knight shift (red triangle) split into two values in the FQ ordered phase. The weighted average of the Knight shift is shown by the black dots.  
}
\end{center}
\end{figure}

\begin{figure}
\begin{center}
\includegraphics[width=7cm,clip]{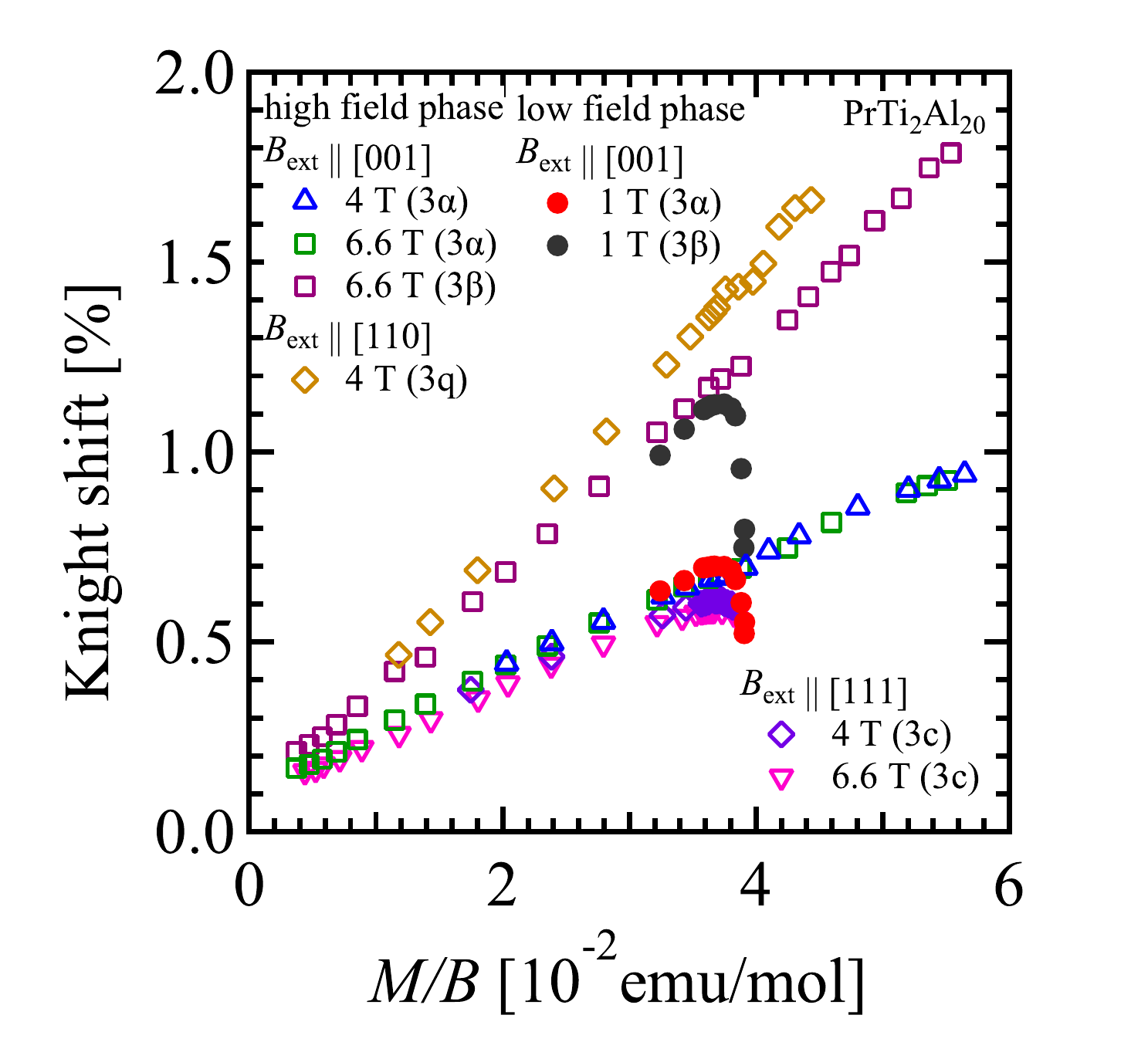}
\caption{\label{fig:K_chi} (Color online) The Knight shift is plotted against susceptibility with temperature as an implicit parameter ($K$-$\chi$ plots) for various directions and values of magnetic fields and at different Al sites indicated in the parentheses. The slope of these plots determines the value of the hyperfine coupling constant $A_{\xi,\xi}$ in Eq.(\ref{eq:Kchi}). The proportionality of Eq.(\ref{eq:Kchi}) holds in all cases except in the low temperature phases with a symmetry breaking FQ order, which appear for $B_{\rm ext}$ = 1~T along [001] and for $\boldsymbol{B}_{\rm ext} \parallel [111]$. 
}
\end{center} 
\end{figure}

Figures~\ref{fig:Kx001}(a) and \ref{fig:Kx001}(b) show temperature dependence of the Knight shift at 3$\alpha$ and 3$\beta$ sites, respectively, for various values of magnetic field along {[}001{]} obtained from the peak frequency of NMR lines via Eqs.~(\ref{eq:NMRfreq}) and (\ref{eq:Knight shift}). Also shown are the susceptibility data ($M/B$) measured at the same field values.
The magnetic hyperfine field acting on nuclei can be generally expressed in terms of magnetic dipole, i.e. the magnetization $\boldsymbol{M}$ and octupole moment $T = \langle T_{xyz} \rangle$,
\begin{equation}
\boldsymbol{B}_{\rm hf} = \hat{A}\cdot\boldsymbol{M}+\boldsymbol{n}T ,
\label{eq:Bhf}
\end{equation}
where $\hat{A}$ is the hyperfine coupling tensor between a nuclear spin and the magnetic dipole of 4$f$ electrons and $\boldsymbol{n}$ is the hyperfine field per unit  $T_{xyz}$ octupole moment discussed in Appendix. In the absence of octupole order, which is indeed the case for PrTi$_{2}$Al$_{20}$ as we discuss below, the Knight shift can be expressed from Eq.~(\ref{eq:Knight shift2}), assuming $\boldsymbol{M} \parallel \boldsymbol{B}_{\rm ext}$, as 
\begin{equation}
K = A_{\xi\xi} \frac{M}{B_{\rm ext}} , 
\label{eq:Kchi}
\end{equation}
where $\xi$ represents direction of the external field.

Like other rare earth intermetallic compounds, the hyperfine couping in PrTi$_{2}$Al$_{20}$ arises dominantly from the transferred hyperfine mechanism\cite{Taniguchi_Proc}, where the dipole moments of 4$f$ electrons induce spin polarization of conduction electrons via the second order process of $c$-$f$ hybridization and the latter then provides a local field to nuclear spins. Because of the large energy denominator of this process, of the order of the Coulomb energy between 4$f$ electrons, the hyperfine coupling tensor is usually independent of temperature. Therefore, $K$ and $M/B$ are expected to show the same temperature dependence.     

As shown in Figs.~\ref{fig:Kx001}(a) and \ref{fig:Kx001}(b), the Knight shift and $M/B$ indeed follow the same temperature dependence at high fields above 4~T. Their increase at low temperatures becomes more gradual at higher magnetic fields, consistent with the expected crossover behavior. On the other hand, they show a remarkable divorce in low fields. At 1~T, the Knight shifts at both 3$\alpha$ and 3$\beta$ sites show a steep reduction upon entering into the FQ ordered phase below $T_{Q}$ = 2.2~K, which is a clear signature for a phase transition. In contrast, $M/B$ stays nearly constant below $T_{Q}$ as already discussed in Sect.~\ref{sec:level3_3_1}. This means that the temperature dependence of Knight shift below $T_{Q}$ is caused by the reduction of the hyperfine coupling, i.e. the $c$-$f$ hybridization, not of the magnetization. 

Similar violation of the proportionality between Knight shift and susceptibility in the FQ ordered phase is observed also for ${\bm B}_{\rm ext} \parallel [111]$ as shown in Fig.~\ref{fig:Kx001}(c). In this case, the Knight shift of 3c sites split into two values below $T_{Q}$ with the population ratio of 1:2 as discussed in Sect.~\ref{sec:level3_2_2}. The weighted average of the split Knight shift then should follow the temperature dependence of the susceptibility if the hyperfine coupling is unchanged by the FQ order. However, this is not the case: while the susceptibility increases slightly below $T_{Q}$, the average Knight shift decreases at lower temperatures, indicating that the transferred hyperfine coupling is affected by the FQ order. 

The remarkable divorce between the Knight shift and susceptibility can be seen also in the so-called $K$-$\chi$ plots displayed in Fig.~\ref{fig:K_chi}. It is interesting to note that the proportionality of Eq.(\ref{eq:Kchi}) holds in high fields above 4~T both for ${\bm B}_{\rm ext} \parallel [001]$ and ${\bm B}_{\rm ext} \parallel [110]$, where low temperature states are reached by crossover. (The NMR spectra for ${\bm B}_{\rm ext} \parallel [110]$ are discussed in Sect.~\ref{sec:level3_4}.) On the other hand, at low fields, the proportionality breaks suddenly upon entering into symmetry breaking FQ ordered phases, corresponding to the cases for ${\bm B}_{\rm ext} \parallel [001]$ at 1~T and for ${\bm B}_{\rm ext} \parallel [111]$. From these results, we conclude that symmetry-breaking FQ orders in PrTi$_{2}$Al$_{20}$ strongly change the $c$-$f$ hybridization.   
 
Breaking of the $K$-$\chi$ proportionality has been observed in a number of heavy fermion materials below a certain crossover temperature\cite{Curro_2009}. Although the microscopic mechanism is not fully understood, the results are analyzed by a phenomenological two-fluid model composed of local moments of $f$-electrons and itinerant quasiparticles generated by the Kondo screening of local moments by conduction electrons\cite{Curro_2004}. Compared with the complex phenomenology of magnetic Kondo lattices, our case of FQ order appears to be simpler. Since the spatial charge distribution of 4$f$ electrons is modified by quadrupole order, it is physically quite natural that a quadrupole order affects the $c$-$f$ hybridization. Nevertheless, to our knowledge this is the first observation of clear violation of $K$-$\chi$ proportionality in the quadrupole ordered phase. The reason may be the relatively weak $c$-$f$ hybridization in most of the materials showing quadrupole order studied so far compared with the Pr~1-2-20 system. Furthermore, a change in the $c$-$f$ hybridization should influence the RKKY interaction, which causes the FQ order. Thus, our results strongly suggest a feedback mechanism between the quadrupole order and the RKKY interaction, which may either accelerate or destabilize the quadrupole order. 

 We remark that octupoles other than $T_{xyz}$, ${\bm T}^{\alpha}$ and ${\bm T}^{\beta}$ defined in Refs.~\citen{Shiina_1997} and \citen{Kuramoto_2009}, contribute in general to the hyperfine field in Eq.~(\ref{eq:Bhf}) in the FQ ordered states, even though they are not active multipoles of $\Gamma_{3}$. This is because these octupole moments are induced by external fields via coupling to the ordered quadrupole moments. In our case of the low-field phase for $\boldsymbol{B}_{\rm ext} \parallel [001]$, finite $\langle T_{z}^{\beta} \rangle$ can be induced by the fields in the presence of $Q_{x}$ order\cite{Shiina_1997, Kuramoto_2009}. However, as will be discussed in detail in Sect.~\ref{sec:level3_3_3}, such an octupole moment should cause NMR line splitting, which is not observed. Therefore, we conclude that the contribution from $\langle T_{z}^{\beta} \rangle$ is small and cannot account for the breaking of $K$-$\chi$ proportionality.

\subsubsection{\label{sec:level3_3_3}Order parameters for $\boldsymbol{B}_{\rm ext}$ \textbar{}\textbar{} {$\mathrm{[001]}$}}

\begin{table}
\begin{center}
\caption{\label{tab:split_001}Expected number of split NMR lines from 3$\alpha$ and 3$\beta$ sites in various multipole ordered phases for $\boldsymbol{B}_{\rm ext}$ \textbar{}\textbar{} {$\mathrm{[001]}$}. 
}
\begin{tabular}{ccccccc}
\hline  
& \multicolumn{2}{c}{$Q_{z}$} & \multicolumn{2}{c}{$Q_{x}$} & \multicolumn{2}{c}{$T_{xyz}$}\tabularnewline
\cline{2-7} 
site & F & AF & F & AF & F & AF\tabularnewline
\hline 
3$\alpha$ & 1 & 2 & 1 & 1 & 2 & 2\tabularnewline
3$\beta$ & 1 & 2 & 2 & 2 & 2 & 2\tabularnewline
\hline 
\end{tabular}
\end{center}
\end{table}

\begin{figure}
\begin{center}
\includegraphics[width=8.5cm,clip]{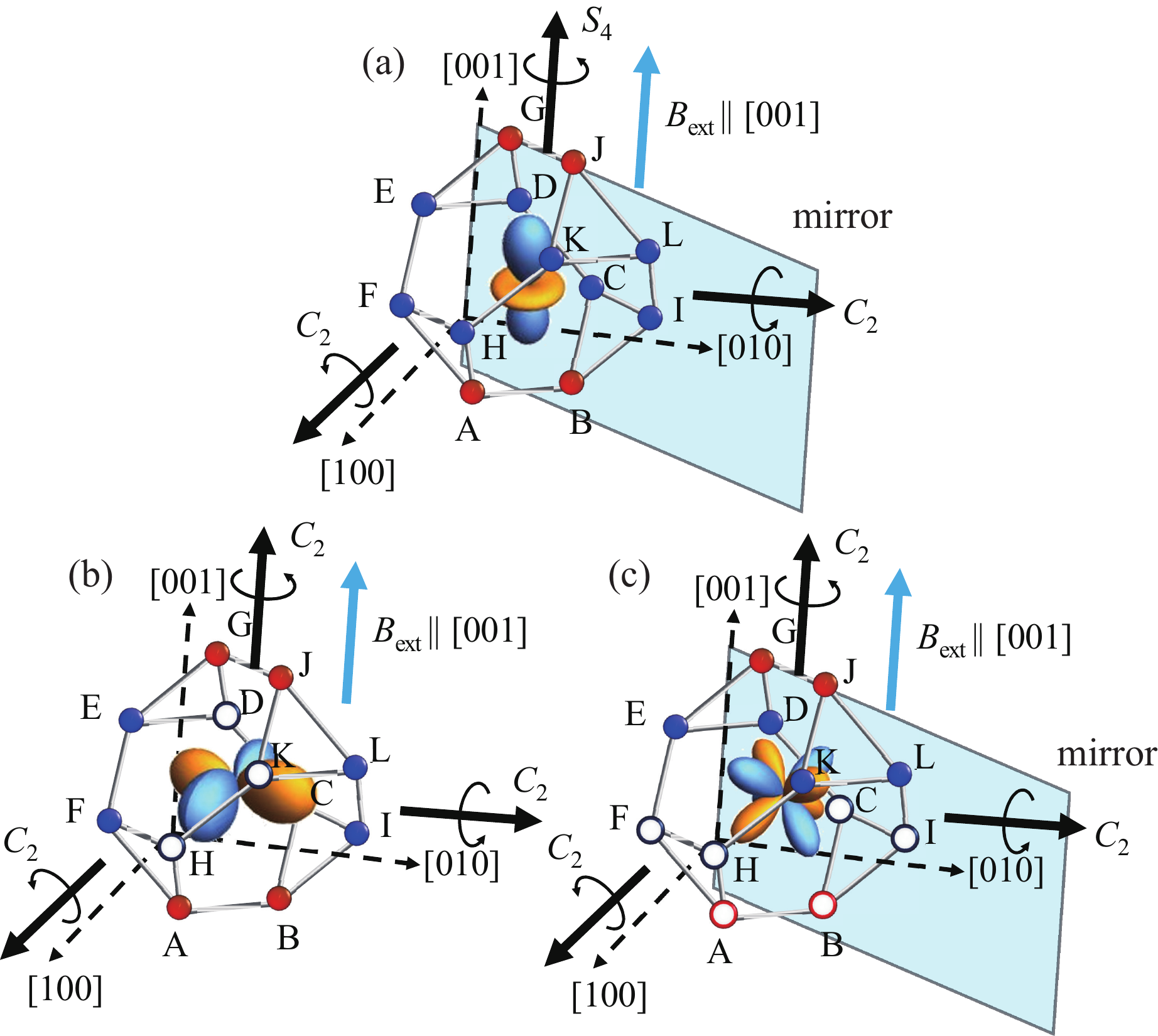}
\caption{\label{fig:op001} (Color online) Symmetry reduction of an Al(3) cage due to order of (a) \textit{Q}\textsubscript{z}, (b) \textit{Q}\textsubscript{x}, and (c) \textit{T}\textsubscript{xyz} for $\boldsymbol{B}_{\rm ext}$ \textbar{}\textbar{} {$\mathrm{[001]}$}. In each panel, twelve Al(3) sites labeled as A-L are marked by different colors to distinguish 3$\alpha$ (red) and 3$\beta$ (blue) sites. Further splitting of these sites due to mulitipole order is indicated by different filling patterns of the circle. 
}
\end{center}
\end{figure}
Following the analysis presented in Sect.~\ref{sec:level3_2_3}, we consider possible reduction of the point symmetry at Al(3) sites due to multipole order for magnetic fields along {[}001{]}. As already mentioned in Sect.~\ref{sec:level3_3_1}, appearance of finite $\langle Q_{z} \rangle$ causes no symmetry reduction [Fig.~\ref{fig:op001}(a)]. Therefore, FQ order of $Q_{z}$, which is actually a crossover, does not cause splitting and AFQ order of $Q_{z}$ causes two-fold splitting for both 3$\alpha$ and 3$\beta$ sites.  

In the case of FQ order of $Q_{x}$, mirror symmetry with respect to $(1\bar{1}0)$ and $S_{4}$ rotoreflection along {[}001{]} are broken but $C_{2}$ rotation along {[}001{]} is preserved. In addition, $C_{2}$ along {[}100{]} and {[}010{]} are also preserved, since reversal of the magnetic-field direction by $C_{2}$ does not change NMR spectra. Therefore, while 3$\alpha$ sites do not split, 3$\beta$ sites split into two groups: (C, D, H, K) and (E, F, I, L) as shown in Fig.~\ref{fig:op001}(b). In the case of AFQ order of $Q_{x}$, since the sign reversal of $Q_{x}$ is equivalent to the mirror (see Sect.~\ref{sec:level3_2_3}), which exchanges the two groups of 3$\beta$ sites, the number of NMR lines remains the same as in the case of FQ order. 

Finally in the case of FO order of $T_{xyz}$, from the results in Table~\ref{tab:Hf_oct} we conclude that both 3$\alpha$ and 3$\beta$ sites are divided into two groups with opposite sign of hyperfine fields. Therefore, the number of NMR lines of both $\alpha$ and 3$\beta$ sites should be doubled either for FO or AFO order of $T_{xyz}$. 

The results of our analysis are summarized in Table~\ref{tab:split_001}. At high fields, both the susceptibility and the NMR results indicate a crossover to $\langle Q_{z} \rangle > 0$ states ($\theta$ = 0) at low temperatures. For the low-field phase, however, conclusion is not straightforward. Although absence of line splitting in NMR is only compatible with the FQ state of $Q_{z}$, such a state is reached only by a crossover, therefore, the phase transition as a function of neither temperature nor magnetic field can be explained. Furthermore, the susceptibility data strongly suggests $Q_{x}$-type ($\theta$ = $\pi/2$) order. A possible solution may be the splitting of the NMR line at 3$\beta$ sites by $Q_{x}$ order being too small to be resolved in the experimental spectra. In the case of $Q_{x}$ order, the line splitting is caused by field-induced octupole of $T_{\beta z} \propto \overline{J_{z}J_{x}^2} - \overline{J_{y}^2J_{z}}$, not by induced dipole~\cite{Shiina_1997}. Since $T_{\beta z}$ is not an active multipole of $\Gamma_{3}$ doublet, the field-induced moment of $T_{\beta z}$ is generated only by hybridization with excited CEF states, which would be indeed very small. Therefore, the order parameter of the low-field phase for $\boldsymbol{B}_{\rm ext}$ \textbar{}\textbar{} {$\mathrm{[001]}$} should be dominantly of the $Q_{x}$ type ($\theta \sim \pm \pi/2$). 

We should remark that $\theta$ may change with field in the low-field phase and the precise value is difficult to determine experimmentally. Since there are expected to be three equivalent domains of $Q_{z}$ with $\theta$ = 0 and $\pm 2\pi/3$ at zero field, our results suggest that an infinitesimal small field along [001] will select the two domains with $\theta = \pm 2\pi/3$, which have a large weight of $Q_{x}$ component. This is indeed what is predicted by our theoretical analysis presented in Sect.~\ref{sec:level4}.

\subsection{\label{sec:level3_4}FQ order and the field-induced transition for $\boldsymbol{B}_{\rm ext}$ \textbar{}\textbar{} {$\mathrm{[110]}$}}
\subsubsection{\label{sec:level3_4_1}Magnetization behavior for $\boldsymbol{B}_{\rm ext}$ \textbar{}\textbar{} {$\mathrm{[110]}$}}

\begin{figure}
\begin{center}
\includegraphics[width=9cm,clip]{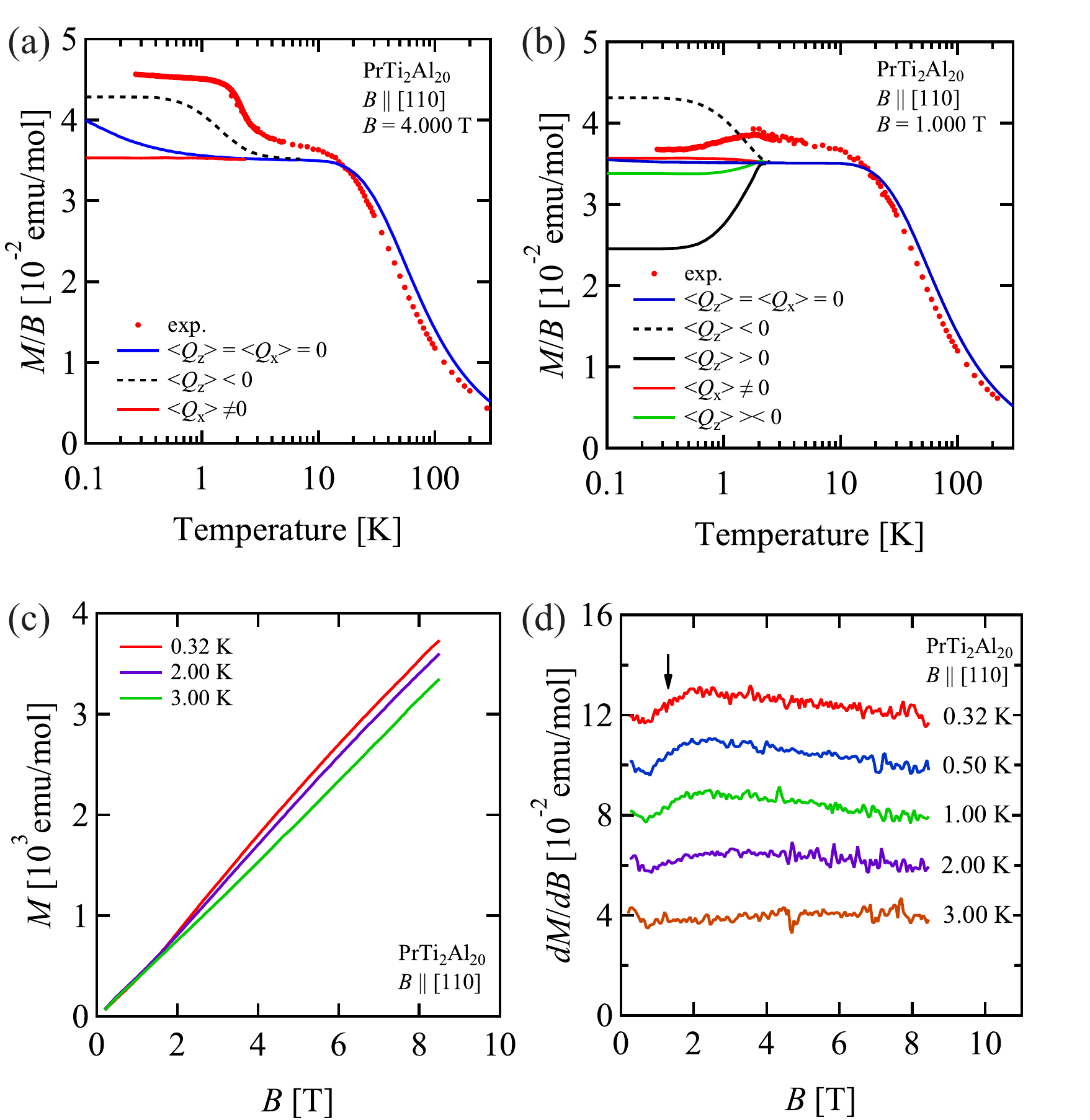}
\caption{\label{fig:mag110}  (Color online) (a)(b):Temperature dependence of the magnetic susceptibility $M/B$ for the field of (a) 4~T and (b) 1~T along {[}110{]} (red dots). The blue line shows the result of calculation without quadrupole interaction based on Eq. (\ref{eq:H_para}). The red, dashed-black, and solid-black [only in (b)] lines indicate the results assuming FQ order of $Q_{x}$, $Q_{z}$ with $\langle Q_{z} \rangle <0$, and $Q_{z}$ with $\langle Q_{z} \rangle >0$, respectively. The green line in (b) represents the case where the two phases with $\langle Q_{z}\rangle <0$ and $\langle Q_{z}\rangle >0$ coexist with 1:1 ratio. (c)(d):Field dependence of (c) the magnetization and (d) the differential susceptibility for $\boldsymbol{B}_{\rm ext}$  \textbar{}\textbar{} {[}110{]}. The plots in (d) are vertically shifted consecutively by 0.02 emu/mole. The black arrow indicates the transition field determined by the average of two kinks.
}
\end{center}
\end{figure}
Figure~\ref{fig:mag110}(a) shows the temperature dependence of the magnetic susceptibility $M/B$ at $B_{\rm ext}$ = 4~T along {[}110{]} (red dots, also shown in Fig.~\ref{fig:mag_para}). The dashed-black line shows the result of calculation assuming $Q_{z}$-type FQ order in Eqs.~(\ref{eq:H_order}) and (\ref{eq:quadrupole_111}), which agrees with the experimental result much better than the calculation without quadrupole interaction (the blue line, also shown in Fig.~\ref{fig:mag_para}). However, unlike the case for $\boldsymbol{B}_{\rm ext}$ \textbar{}\textbar{} {[}001{]}, we have to choose negative values of $\left\langle Q_{z}\right\rangle$ to reproduce the experimental result. This is understood again by examining the matrix elements of $Q_{z}$ and $J_{z}$ within the Zeeman split $\Gamma_{3}$ doublet (eigenstates of ${\mathscr{H}}_{\rm CEF}+{\mathscr{H}}_{\rm Z}$ for $B_{\rm ext} = 4$~T along {[}110{]}),  
\begin{eqnarray}
Q_{z} =\left(\begin{array}{cc}
-1.00 & 0\\
0 & 0.98
\end{array}\right) \ \ 
\frac{J_{x}+J_{y}}{\sqrt{2}} = \left(\begin{array}{cc}
0.50 & 0\\
0 & 0.27
\end{array}\right)
\label{eq:QzJz_matrix_110}.
\end{eqnarray}
Thus, the magnetic fields along {[}110{]} induce finite $\langle Q_{z} \rangle <0$ and work cooperatively with FQ interaction with negative values of $\langle Q_{z} \rangle$. However, the CEF energy is higher for states with $\langle Q_{z} \rangle <0$ than that with $\langle Q_{z} \rangle >0$, owing to the third-order term in Eq.~(\ref{eq:CEF}). Then, in order to minimize the CEF energy, the order parameter may contain small admixture of $Q_{x}$ component in addition to the major $Q_{z}$ component, unless the magnetic field is sufficiently strong. This point is discussed in detail in Sect.~\ref{sec:level4_4}.  If this occurs, since the CEF energy is independent of the sign of $\langle Q_{x} \rangle$, the low temperature phase breaks an Ising-type symmetry. Therefore, there can be a phase transition unlike the crossover behavior for ${\bm B}_{\rm ext} \parallel [001]$. However, anomalies associated with such a transition could be too weak to be detected experimentally as discussed in Sect.~\ref{sec:level4_4}. The small $Q_{x}$ component should be suppressed in sufficiently strong fields, where a crossover behavior is expected.

Similarly to the case of $\boldsymbol{B}_{\rm ext}$ \textbar{}\textbar{} {[}001{]}, the susceptibility shows completely different behavior at low magnetic fields as shown in Fig.~\ref{fig:mag110}(b). The susceptibility at 1~T decreases slightly below 2~K,  which is incompatible with the case of negative $\left\langle Q_{z}\right\rangle$ (dashed-black line). Ferro-quadrupole order of positive $\left\langle Q_{z} \right\rangle$ (solid-black line) causes the susceptibility to decrease at low temperatures, however, the amount of reduction is much larger that in the experimental result. The FQ order of $Q_{x}$, on the other hand, causes almost no change of susceptibility. Since the NMR results that will be presented in Sect.~\ref{sec:level3_4_2} below indicate coexistence of two phases at $B_{\rm ext}=1$~T along {[}110{]}, we consider the case of 1:1 mixture of two phases with $\left\langle Q_{z} \right\rangle <0$ and $\left\langle Q_{z} \right\rangle >0$. The result shown by the green line appears to reproduce the experimental features quite well. 

Figure~\ref{fig:mag110}(c) and (d) show the field dependence of the magnetization $M$ and differential susceptibility $dM/dB$. While $M$ is proportional to $B$ with no anomaly in $dM/dB$ at 3~K, non-linear increase in $M$ and the corresponding two kinks in $dM/dB$ develop below 2~K. This behavior is similar to the case of $\boldsymbol{B}_{\rm ext}$ \textbar{}\textbar{} {[}001{]. However, anomaly in magnetization is substantially weaker and broader for $\boldsymbol{B}_{\rm ext}$ \textbar{}\textbar{} {[}110{].   

\subsubsection{\label{sec:level3_4_2}NMR evidence of the field-induced transition for $\boldsymbol{B}_{\rm ext}$ \textbar{}\textbar{} {$\mathrm{[110]}$}}

\begin{figure}
\begin{center}
\includegraphics[width=8.5cm,clip]{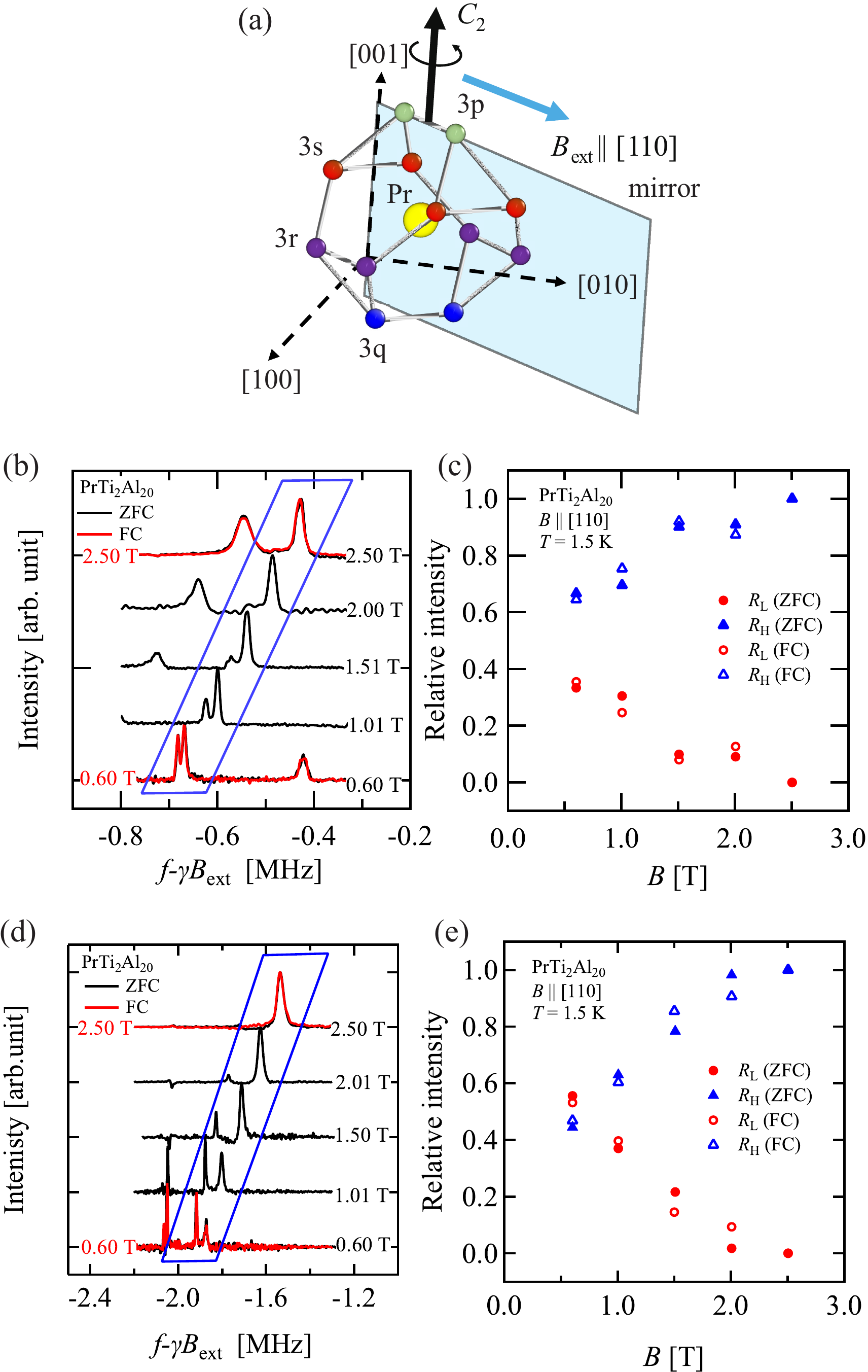}
\caption{\label{fig:sp110_1p5K} (Color online)  (a) Twelve Al(3) sites forming a cage surrounding a Pr site. Under magnetic fields along  {[}110{]}, they split into four inequivalent sites, two 3p (green), two 3q (blue), four 3r (purple) and four 3s (red) sites. 
(b) (d) Field dependence of the NMR spectra (plotted as a function of shift from the nuclear Zeeman frequency) of the low-frequency second satellite line ($k=-2$) at (b) 3p and (d) 3q sites at 1.5~K are surrounded by the blue frames. (c) (e) Field dependence of the normalized intensities, $R_{L} = I_{L}/\left(I_{H}+I_{L}\right)$, $R_{H} = I_{H}/\left(I_{H}+I_{L}\right)$, where $I_{L}$ ($I_{H}$) are the intensity of the NMR line from the low field (high field) phase at (c) 3p and (e) 3q sites. 
}
\end{center}
\end{figure}

\begin{figure}
\begin{center}
\includegraphics[width=8.5cm,clip]{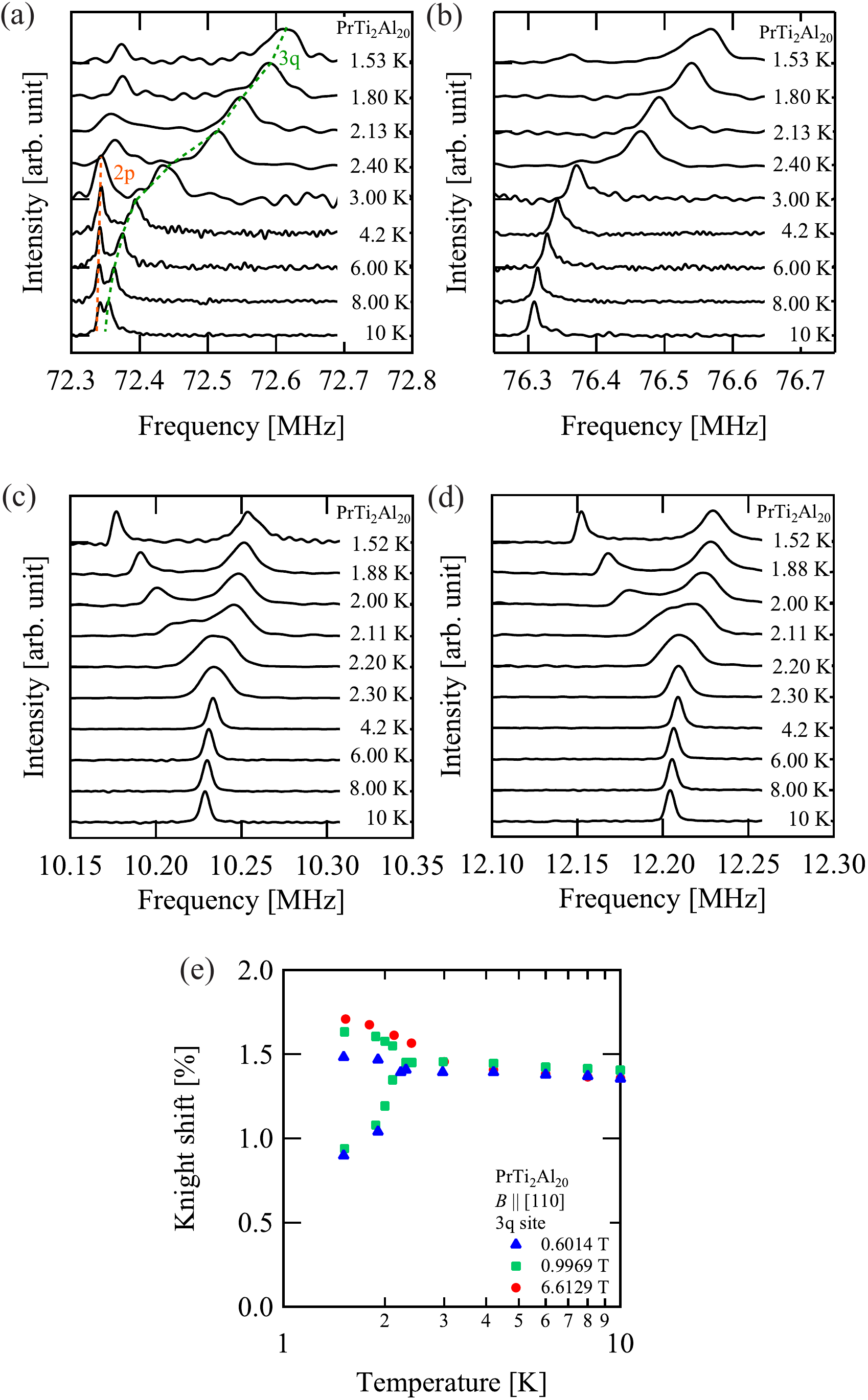}
\caption{\label{fig:sp110} (Color online) (a) (b) Temperature dependence of the NMR spectra of 3q sites for $B_{\rm ext}=6.613$~T along {[}110{]}: (a) the low-frequency ($k=-2$) and (b) the high-frequency ($k=2$) second satellite lines. In (a) NMR lines from Al(2) sites (red dashed line) and 3q sites (green dashed line) overlap at some temperatures.  (c) (d) Temperature dependence of the NMR spectra of 3q sites for $B_{\rm ext}=0.9969$~T along {[}110{]}: (c) the low frequency ($k=-1$) and (d) the high-frequency ($k=1$) satellite lines. (e) Temperature dependence of the Khight shift at 3q site for several values of magnetic field along {[}110{]}.
}
\end{center}
\end{figure}

While magnetic fields along {[}110{]} reduce the cubic symmetry, \textit{C}\textsubscript{2} rotation along {[}001{]} and mirror with respect to $(1\bar{1}0)$ are preserved. The twelve Al(3) sites on a cage then split into four groups, two 3p (green), two 3q (blue), four 3r (purple) and four 3s (red) sites as shown in Fig.~\ref{fig:sp110_1p5K}(a). Because of frequent overlap of NMR lines from a large number of inequivalent sites, we are able to follow complete temperature and field dependences of the spectra only for 3p and 3q sites.  

Figures~\ref{fig:sp110_1p5K}(b) and \ref{fig:sp110_1p5K}(d) show field dependence of NMR spectrum of the low-frequency second satellite line ($k=-2$) at (b) 3q and (d) 3p sites at 1.5~K. While the spectrum consists of a single line at 2.5~T, a second line appears at a lower frequency when the field is reduced below 2~T and grows in intensity with decreasing field. Intensity of the first high-frequency line gradually decreases with decreasing field but two lines still coexist at the lowest field of 0.6~T. Systematic difference is not observed between field-cooled and zero-field-cooled conditions. The relative intensities of the two lines are plotted against $B_{\rm ext}$ in Fig.~\ref{fig:sp110_1p5K}(c) for 3q sites and in Fig.~\ref{fig:sp110_1p5K}(e) for 3p sites. By comparing with the results for $\boldsymbol{B}_{\rm ext}$ \textbar{}\textbar{} {$\mathrm{[001]}$} in Fig.~\ref{fig:sp001_1p5K}(c), we conclude that magnetic fields along {[}110{]} induce first-order phase transition with a wide range of field for the coexistence of two phases. While magnetic fields above 2.5~T suppress the low-field phase completely, the high-field phase survives down to the lowest field (0.6~T) of our NMR measurements.

We next examine temperature dependence of the NMR spectra. In Figs.~\ref{fig:sp110}(a) and \ref{fig:sp110}(b), we show temperature dependence of the pair of second satellite NMR spectra ($k=\pm2$) of 3q site in the high-field phase ($B_{\rm ext}$ = 6.613~T). The peak frequency of both lines increases with decreasing temperature without splitting, indicating increase of $\boldsymbol{B}_{\rm hf}$. Figures~\ref{fig:sp110}(c) and \ref{fig:sp110}(d) show the pair of first satellite NMR spectra ($k=\pm1$) of 3q sites for $B_{\rm ext}$ = 0.9969~T. Both lines split into two lines below $T_{Q}$ = 2.2~K. This corresponds to separation into the high-field and low-field phases, consistent with the coexistence of two peaks at this field value at 1.5~K (see Fig.~\ref{fig:sp110_1p5K}). In both spectra, the higher-frequency (lower-frequency) peak corresponds to the high-field (low-field) phase. Since the $k=-1$ and $k=+1$ lines move in the same direction in both phases, the temperature variation of the NMR frequency is attributed to the change in $\boldsymbol{B}_{\rm hf}$. 

Temperature dependence of the Knight shifts at 3q sites obtained from these spectra are shown in Fig.~\ref{fig:sp110}(e) for different values of the field. For $B_{\rm ext}$ = 0.6014 and 0.9969~T, data from both high-field and low-field phases are presented. In the low-field phase, the Knight shift is independent of magnetic field and decreases suddenly below $T_{Q}$ = 2.2~K, indicating a phase transition. In the high-field phase, on the other hand, the Knight shift depend on magnetic field and increases gradually with decreasing temperature, consistent with the crossover behavior. Since magnetic susceptibility from low-field phase alone cannot be obtained, we are not able to examine whether the proportionality between Knight shift and susceptibility holds in the low-field phase for $\boldsymbol{B}_{\rm ext}$ \textbar{}\textbar{} {[}110{]}, 

\subsubsection{\label{sec:level3_4_3}Order parameters for $\boldsymbol{B}_{\rm ext}$ \textbar{}\textbar{} {$\mathrm{[110]}$}}

\begin{figure}
\begin{center}
\includegraphics[width=8.5cm,clip]{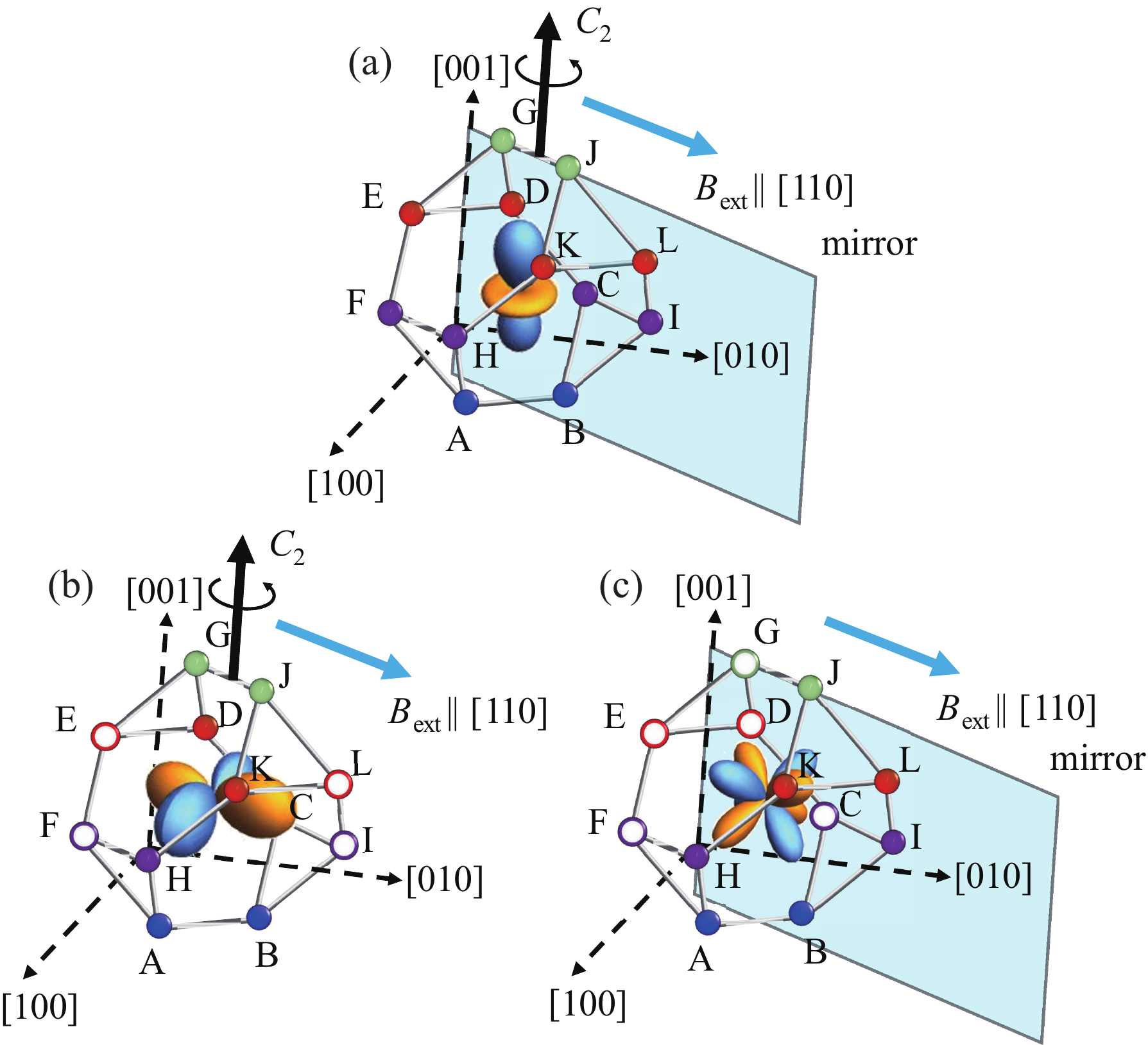}
\caption{\label{fig:op110} (Color online) Symmetry reduction of an Al(3) cage due to order of (a) \textit{Q}\textsubscript{z}, (b) \textit{Q}\textsubscript{x}, and (c) \textit{T}\textsubscript{xyz} for $\boldsymbol{B}_{\rm ext}$ \textbar{}\textbar{} {$\mathrm{[110]}$}. In each panel, twelve Al(3) sites labeled as A-L are marked by different colors to distinguish 3p (green), 3q (blue), 3r (purple) and 3s (red) sites. Further splitting of these sites due to mulitipole order is indicated by different filling patterns of the circle. 
}
\end{center}
\end{figure}

\begin{table}[t]
\begin{center}
\caption{\label{tab:split_110}Expected number of split NMR lines from 3p, 3q, 3r, and 3s sites in various multipole ordered phases for $\boldsymbol{B}_{\rm ext}$ \textbar{}\textbar{} {$\mathrm{[110]}$}. 
}
\begin{tabular}{ccccccc}
\hline 
 & \multicolumn{2}{c}{$Q_{z}$} & \multicolumn{2}{c}{$Q_{x}$} & \multicolumn{2}{c}{$T_{xyz}$}\tabularnewline
\cline{2-7} 
site & F & AF & F & AF & F & AF\tabularnewline
\hline
3p & 1 & 2 & 1 & 1 & 2 & 2\tabularnewline
3q & 1 & 2 & 1 & 1 & 1 & 1\tabularnewline
3r & 1 & 2 & 2 & 2 & 2 & 2\tabularnewline
3s & 1 & 2 & 2 & 2 & 2 & 2\tabularnewline
\hline 
\end{tabular}
\end{center}
\end{table}

Here, we consider possible reduction of the point symmetry at Al(3) sites due to multipole order for magnetic fields along {[}110{]}. As already mentioned, states with a finite value of $Q_{z}$ cause no symmetry reduction. Therefore, FQ order of $Q_{z}$, which can be actually a crossover if $\langle Q_{z} \rangle <0$, does not cause any splitting of NMR lines and AFQ order of $Q_{z}$ causes two-fold splitting for all sites [see Fig.~\ref{fig:op110}(a)].   

In the case of FQ order of $Q_{x}$, mirror symmetry with respect to $(1\bar{1}0)$ is broken while $C_{2}$ rotation along {[}001{]} is preserved. Then, while 3p and 3q sites do not split, each of 3r and 3s site splits into two groups as indicated in Fig.~\ref{fig:op110}(b). In the case of AFQ order of $Q_{x}$, since sign reversal of $Q_{x}$ is equivalent to the mirror operation, which interchanges the two groups of 3r sites and also the two groups of 3s sites, the number of NMR lines remains the same as in the case of FQ order. 

Finally, in the case of FO order of $T_{xyz}$, we conclude from Table~\ref{tab:Hf_oct} that while no hyperfine field is produced at 3q sites, all other sites are divided into two groups with opposite sign of hyperfine fields. Therefore, the number of NMR lines should be doubled either for FO or AFO order of $T_{xyz}$ except for 3q sites. 

The results of our analysis are summarized in Table~\ref{tab:split_110}. The NMR lines of both 3q and 3p sites show no splitting in either the low- or the high-field phase. From Table~\ref{tab:split_110}, the NMR results are compatible with FQ orders of $Q_{z}$ and $Q_{x}$. For the high-field phase, the crossover behavior in magnetic susceptibility and Knight shift strongly indicate FQ orders of $Q_{z}$ with $\langle Q_{z} \rangle <0$. For the low-field phase, weak temperature dependence of the susceptibility at $B_{\rm ext}=1$~T in the ordered phase [Fig.~\ref{fig:mag110}(b)] may appear to support the FQ order of $Q_{x}$. However, since the NMR results confirm coexistence of the two phases at 1~T, highly enhanced susceptibility of the high-field phase with negative $\langle Q_{z} \rangle$ at low temperatures must be compensated by strongly  reduced susceptibility of the low-field phase in order for the total susceptibility to be weakly temperature dependent. This is possible only if the low-field phase has FQ order of $Q_{z}$ with positive $\langle Q_{z} \rangle$.    

\subsection{\label{sec:level3_5}Phase diagram}

The complete magnetic field vs. temperature phase diagram of the ferro-quadrupole order in PrTi\textsubscript{2}Al\textsubscript{20} is displayed in Fig.~\ref{fig:phase_diagram}. The order parameters of individual phases are indicated by graphical symbols, representing either $Q_{z}$ or $Q_{x}$ type moment. These are defined by the polar anlge $\theta$ in the ${\bm Q}$-space as shown in the right end of Fig.~\ref{fig:phase_diagram}. The $Q_{z}$ type moments correspond to $\theta = n\pi/3$ ($n = 0, 1, \dots , 5$) and the $Q_{x}$ typer moments are associated with $\theta = (n+1/2)\pi/3$. In the following, we summarize the preceding discussions on the phase diagram. Assignment of order parameters is based on the presence or absence of NMR line splitting and qualitative features of temperature dependence of the susceptibility. Considering finite line width of the NMR spectra and the qualitative aspects of our analysis of the susceptibility, our assignment is not always precise enough to determine the exact values of $\theta$ but in some cases only indicates approximate area of most probable oder parameters.  

When the magnetic field is applied along [111] direction, ferro-quadrupole (FQ) order of $Q_{z} \propto 3J_{z}^{2}-\boldsymbol{J}^{2}$ ($\theta$ = 0) moment is established below $T_{Q}$ = 2.2~K. The transition temperature is determined from the onset of NMR line splitting, as well as the peak of $d(C/T)/dT$, where $C$ is the specific heat\cite{Sakai_2011,Sakai_2012_proc}. Both are nearly independent of magnetic field up to $\sim$~10~T. Because of the $C_3$ symmetry along [111], other two domains of FQ order, $-Q_{z}/2+\sqrt{3}Q_{x}/2\propto3J_{x}^{2}-\boldsymbol{J}^{2}$ ($\theta= 2\pi/3$) and $-Q_{z}/2-\sqrt{3}Q_{x}/2\propto3J_{y}^{2}-\boldsymbol{J}^{2}$ ($\theta=-2\pi/3$), are equally possible. 

For the field along [001], symmetry breaking phase transition occurs at $T_{Q}$ = 2.2~K only for low fields below about 2~T. The FQ moment is primarily of $Q_{x} \propto J_{x}^{2}-J_{y}^{2}$ type ($\theta\sim\pm\pi/2$). Since $\langle Q_{x} \rangle$ can take either sign, this should be a Ising-type phase transition. With increasing field, the FQ moment jumps to $\theta=0$ ($\langle Q_{z} \rangle >0$) in ${\bm Q}$-space. This discontinuous transition is accompanied by a stepwise increase of magnetization, leading to a peak in $dM/dB$ at 2~T. However, field dependence of the NMR spectra reveals that there is rather wide range of magnetic field (1-3~T) in which two phases coexist. Since finite $\langle Q_{z} \rangle >0$ is induced by the fields along [001] even in the para state, the high-field state is reached by a crossover from high temperatures without symmetry breaking.  

Similar phase diagram is obtained for the field along [110]. The low-field phase most likely has order of $Q_{z}$ with $\langle Q_{z} \rangle >0$. A stepwise increase in magnetizatioin and a peak in $dM/dB$ are also observed, but these features are much weaker and broader than the case for the fields along [001]. Indeed, NMR spectra show that two phases coexist in a very wide range of field below about 2.5~T down to at least 0.6~T, which is the lowest field of measurements. Actually, two phases may coexist even at zero field. The low-field phase is completely suppressed above 2.5~T. The FQ moment in the high field phase is primarily of $Q_{z}$ with $\langle Q_{z} \rangle <0$. Since finite $\langle Q_{z} \rangle <0$ is induced by the fields along [110] even in the para state, there should be no phase transition in temperature for sufficiently high magnetic fields.

\section{\label{sec:level4}Theory of Ferro Quadrupole Orders in ${\bf PrTi}_2{\bf Al}_{20}$}
In this section, we will show that emergence of the low-field phases for ${\bm B} \parallel$ [001] and [110] can be understood by introducing anisotropic quadrupole-quadrupole interactions induced by magnetic fields. For notational convenience, we will denote external field ${\bm B}_{\rm ext}$ simply as ${\bm B}$ throughout this section. We will start by introducing a Landau theory for FQ orders under magnetic fields for qualitative understanding of the essential aspect of the present FQ states. Then, we will demonstrate that a microscopic CEF model including competition between the Zeeman and the anisotropic quadrupole interactions can indeed reproduce the experimental phase diagram, provided that their relative dominance is reversed by increasing magnetic field.

\subsection{\label{sec:level4_1}Landau theory}
A Landau's free energy for interacting $\Gamma_3$ quadrupole ${\bm Q} = \left( Q_{z}, Q_{x} \right)$ moments under cubic  \textit{T}\textsubscript{d} symmetry has been investigated in the early stage of the study about 1-2-20 compounds\cite{Hattori_2014}.
Generalization including $\Gamma_2$ octupole $T_{xyz}$ moment has been recently investigated for analyzing PrV$_2$Al$_{20}$\cite{Lee_2018,Freyer_2018,Y_B_Kim_2019}.
Here, we focus on FQ orders and examine possible terms induced by magnetic fields. The Landau's free energy is given as 
\begin{align}
F=\frac{a}{2}|\bm Q|^2-\frac{b}{3}(Q_z^3-3Q_zQ_x^2) + \frac{c}{4}|\bm Q|^4+F_2+\cdots,\label{Free0}
\end{align}
where we have omitted a trivial constant term and $a$, $b$, and $c$ are constants. As in the conventional parameterization, $a=a_0(T-T_c^0)$, where $T_c^0$ is 
the ``transition temperature'' (which is not actual transition temperature owing to the presence of the third order $b$ term).
The last term $F_2$ is the one proportional to ${\bm B}^2$, which is the leading-order effect of the magnetic fields. Due to the cubic \textit{T}\textsubscript{d} symmetry, the form of $F_2$ is 
restricted, and for the lower orders in ${\bm Q}$ and ${\bm B}$, one finds 
\begin{align}
F_2&=F_Z+F_I,\label{eq:F2}\\
F_Z&=-\mu\left( h_z Q_{z}
+h_xQ_{x}\right),\label{eq:FZ}\\
F_I&=\frac{\lambda'}{2}\left[ h_z(Q_{z}^2-Q_x^2)
-2h_x Q_{z}Q_x\right]. \label{eq:FI}
\end{align}
Here, $\mu\equiv \alpha\left(g_J\mu_{B}\right)^{2}$, with $\alpha=[7/(3E_{\Gamma_4})-1/E_{\Gamma_5}]>0$ as represented by the CEF excited-state energy $\Gamma_{4,5}$\cite{Hattori_2014} and 
\begin{align}
h_z=2B_{z}^{2}-B_{x}^{2}-B_{y}^{2}, \quad 
h_x=\sqrt{3}\left(B_{x}^{2}-B_{y}^{2}\right). \label{defh}
\end{align}
$F_Z$ is the Zeeman coupling for the quadrupole moments, which can be derived by the second-order perturbation of $H_Z$ in Eq. (\ref{eq:Zeeman})\cite{Hattori_2014}. 

$F_I$ represents a symmetry-allowed anisotropic interaction induced by the fields, which is obtained as a cubic invariant constructed from the three $\Gamma_3$ quantities: $(h_z,h_x)$ and the two $\bm{Q}$'s. $F_I$ plays a key role in the following discussion, although its microscopic mechanism is not identified. Since a trivial constant term $\propto|\bm{B}|^2$ and an isotropic interaction $\propto|\bm{B}|^2|\bm{Q}|^2$ are not important in the following analysis, we have omitted them in Eq. (\ref{eq:F2}) for simplicity. If remembering the fact that the second-order coefficients in the Landau expansion are related to the susceptibility of the corresponding order parameters, one can easily realize that $F_I$ induces anisotropy of the transition temperature in the ${\bm Q}$ space. Although there are in principle $O\left(\boldsymbol{B}^{2}\boldsymbol{Q}^{n}\right)$ terms with $n\ge 3$, it is sufficient to include contribution with $n\le 2$ for the present purpose.

Before analyzing the effects of $F_I$, we briefly explain the nature of FQ order for $|{\bm B}|$ = 0. Introducing ${\bm Q}=Q(\cos\theta,\sin\theta)$, one can rewrite Eq. (\ref{Free0}) as follows.
\begin{align}
F=\frac{a}{2}Q^2-\frac{b}{3}Q^3\cos3\theta + \frac{c}{4}Q^4.\label{Free1}
\end{align}
Here, we have ignored the higher-order terms, which are irrelevant for the present discussions.
Minimizing Eq. (\ref{Free1}) with respect to $\theta$ and $Q$, one obtains first-order FQ transitions for $\theta=0, \pm 2\pi/3$ 
when $a \equiv a_c \equiv a_0(T_c-T_c^0) =2b^2/9c$,  
with a 
discontinuous jump of the order parameter $\Delta Q$:
\begin{align}
\Delta Q=\frac{b}{2c}\left(1+\sqrt{1-\frac{4c}{b^2}a_c}\right)=\frac{2b}{3c} \propto\frac{\eta T_c}{E_{\Gamma_1}},
\end{align}
where $E_{\Gamma_1}$ is the energy of the $\Gamma_1$ excited state and $\eta$ is related to the matrix element of the quadrupole operators between
 the $\Gamma_3$ and $\Gamma_1$ as will be introduced in Eqs. (\ref{def:newQz}) and (\ref{def:newQx}). 
 As will be discussed, $\Delta Q$ is very small and typically $0.01$ for PrTi$_2$Al$_{20}$. This is 
 indeed consistent with the experimental observation that detects no signiture of the first-order transition\cite{Taniguchi_JPSJ,Sakai_2011}. Note that the three choices of $\theta$ correspond to the three domains in the cubic  symmetry.

Now, let us consider cases for finite fields. Equation (\ref{Free0}) now leads to
\begin{align}
F=&-\mu\left( h_z Q_{z}
+h_xQ_{x}\right)\nonumber\\
&+\frac{a+\lambda'h_z}{2}Q_z^2+\frac{a-\lambda'h_z}{2}Q_x^2
-\lambda' h_xQ_zQ_x\nonumber\\
&-\frac{b}{3}(Q_z^3-3Q_zQ_x^2) + \frac{c}{4}|\bm Q|^4.\label{Free2}
\end{align}
When we concentrate on the cases for which ${\bm B}$ varies from [001] to [110] via [111], Eq. (\ref{Free2}) reduces further to 
\begin{align}
F=&-\mu h_z Q_{z}+\frac{a+\lambda'h_z}{2}Q_z^2+\frac{a-\lambda'h_z}{2}Q_x^2\nonumber\\
&-\frac{b}{3}(Q_z^3-3Q_zQ_x^2) + \frac{c}{4}|\bm Q|^4.\label{Free3}
\end{align}
In the absence of the $\lambda'$ term, 
FQ with $\theta=0$ is chosen among the three domains for ${\bm B}\parallel [001]$. In this case, the first-order transition line ends at some field strength with the critical end point. There is only a crossover at higher fields, since a finite $Q_z$ is induced even in the para state by the Zeeman term in Eq. (\ref{Free3}).
For ${\bm B}\parallel [110]$, the Zeeman term favors the two domains $\theta=\pm 2\pi/3$ with $Q_z<0$, since $h_z<0$ for this field direction. 
Thus, there are still (Ising like) phase transition under the fields. Increasing ${\bm B}$ leads to gradual changes (or first-order transition)
 in ${\bm Q}$ to $\theta=\pi$. When $\theta$ reaches $\pi$, 
the state is smoothly connected to the para state for $|{\bm B}|=0$. 
Nevertheless, there are no surprises for $\lambda'=0$ and this is the reason why
 FQ orders have not been paid attention in the early studies \cite{Hattori_2014}.

Now consider the cases for $\lambda'>0$.
It is clear that $\lambda'$ term favors $Q_x$ type FQ orders for ${\bm B}\parallel [001]$, while 
does $Q_z$ type for ${\bm B}\parallel [110]$, since the second-order coefficient in Eq. (\ref{Free3}) is shifted by $\pm \lambda'h_z/2$. Therefore, if $\lambda'$ term overwhelms the Zeeman term, there is a chance to realize $\theta\sim \pm 2\pi/3$ type FQ order, which has a large weight of $Q_{x}$ component, for an infinitesimal small field along [001] and $\theta=0$ type FQ order for ${\bm B}\parallel [110]$. 
This is in principle possible, since the both terms are $O({\bm B}^2)$. 
The fact that the both terms are proportional to ${\bm B}^2$ is unique and crucial property for quadrupolar systems.  
This is a clear contrast to the cases for systems with only spin degrees of freedom, where the Zeeman term is proportional 
to ${\bm B}$, while the anisotropic interactions generated by the magnetic fields (even if exist) are $O({\bm B}^2)$ because of the time-reversal symmetry. 
Thus, for spin systems, it never happens that the magnetic-field induced anisotropic interactions play important roles under small magnetic fields. There, the anisotropy is solely determined by the zero field parameters. 

When one carries out an elementary analysis about the free energy (\ref{Free3}) for small ${\bm B}\parallel [001]$ and at $T=T_c$, one finds that the anisotropic interaction wins for 
\begin{align}
\lambda'>\frac{3c\mu}{2b},\label{eq:cond1}
\end{align}
when an infinitesimal magnetic field is applied. Thus,
$\theta\sim \pm 2/3\pi$ FQ orders are realized, which are close to $Q_x$ type FQ orders. Increasing magnetic field should shift $\theta$ even closer to $\pm \pi/2$.
Note that these orders break the mirror symmetry $x\leftrightarrow y$, therefore, the transition does not become a crossover under the fields.
This can solve the puzzling experimental results for ${\bm B}\parallel [001]$. 
For ${\bm B}\parallel [110]$, when
\begin{align}
\lambda'>\frac{3c\mu}{b},\label{eq:cond2}
\end{align}
the anisotropic interaction overwhelms the Zeeman term at $T=T_c$ with an infinitesimal magnetic field and the $\theta=0$ solution is realized. 
As will be demonstrated in Sect. \ref{sec:level4_4}, the observed susceptibility under the fields can be qualitatively explained when Eq. (\ref{eq:cond2}) is satisfied.

So far, we have seen that the low-field phases, $\sim Q_x$ type FQ order for $\bm{B}\parallel [001]$ and positive $Q_z$ type FQ orders for $\bm{B}\parallel [110]$, can be stabilized by the field-induced anisotropic interaction (the $\lambda'$ term). 
As discussed in Sect.~\ref{sec:level3}, the experimental results show that these phases undergo first-order phase transitions at around 1-2 T. The high field states are those favored by the Zeeman interaction, which smears out phase transitions into crossover. This means that the inequality in Eqs. (\ref{eq:cond1}) and (\ref{eq:cond2}) should be reversed above 1-2 T. To reproduce the field-induced phase transitions, it is necessary to introduce further 
field dependence in $\lambda'$. Let us define this field-dependent factor as $f({\bm B})$, i.e., $\lambda'\to \lambda'f$. For simplicity, we assume $f=f(|{\bm B}|^2)$ with $f(0)=1$ and $f$ is a monotonically decreasing function of $|{\bm B}|^2$ at least for small fields we are interested in. As shown below, this simple form can qualitatively explain the experimentally observed phase diagram. In reality, $f$ may have discontinuity at the first-order transition, as a consequence of feedback from the changes in the order parameter as discussed in Sect.~\ref{sec:level3_3_2}. In any case, $f(|{\bm B}|^2)$ should become smaller in the high-field phase.
For rough estimate of the critical fields at the transition to conventional ($\lambda'$ = 0) states, we ignore the field dependence of the order parameters and other variables. Then, from Eq.~(\ref{eq:cond1}), the transition occurs for ${\bm B} \parallel [001]$ when	
\begin{align}
&f(|{\bm B}_c|^2)\sim \frac{3c\mu}{2b\lambda'}\equiv f_{[001]} 
\label{eq:cond_transition1}
\end{align}
and for ${\bm B} \parallel [110]$, 
\begin{align}
&f(|{\bm B}_c|^2)\sim\frac{3c\mu}{b\lambda'}\equiv f_{[110]}=2f_{[001]}  < 1 ,
\label{eq:cond_transition2} 
\end{align}
from Eq.~(\ref{eq:cond2}). Note that, to obtain the positive $Q_{z}$ type FQ order, it is necessary that $f_{[110]}  < 1$. Thus, since $f(|{\bm B}|^2)$ is monotonically decreasing function of $|{\bm B}|^2$, this simple argument explains
why the critical field for ${\bm B} \parallel [001]$ is larger than that for ${\bm B} \parallel [110]$ as observed in the present experiments.

In the following, instead of analyzing the free energy Eq. (\ref{Free3}) in details, we will discuss the CEF model in Sect.~\ref{sec:level4_4} to discuss the phase diagram for wider range of the magnetic fields.

\subsection{\label{sec:level4_4}CEF model}
In this subsection, we numerically calculate the CEF model with the anisotropic interaction under finite fields. 
Let us first consider the field effect on quadrupole-quadrupole 
interaction as a possible origin for the anisotropic interaction $F_I$ in Eq. (\ref{eq:FI}) introduced in Sect. \ref{sec:level4_1}.
Starting from a model of localized f-electrons in the 1-2-20 systems, the quadrupole-quadrupole interactions are mediated by conduction electrons, which is the RKKY interactions for the quadrupole sector. Existence of the small Fermi surface(s) would be helpful for enhancing the magnitude of the field dependence in the RKKY interactions, since the effective energy scale for the conduction electrons reduces. Actually, in the band-structure calculations of LaTi$_2$Al$_{20}$ \cite{Nagashima,Swatek_2018}, there are several small Fermi surfaces. This would be related to the generation of the anisotropic interactions we have been introduced in Sect. \ref{sec:level4_1}.

 We should remark that effects of high-rank multipoles such as octupoles are not considered in this paper. Interaction among high-rank multipole moments induced by field in the presence of quadrupole order may lead to effective anisotropy  \cite{Shiina_1997, Kuramoto_2009}. We have not considered such effects partly because these high-rank multipoles are inactive in $\Gamma_{3}$ and induced only through mixing with the excited CEF states, therefore, the effects would be minor. Also the effective anisotropy caused by high-rank multipoles should be again proportional to ${\bm B}^2$ in the low field limit, and is likely to result in the same form as Eq.~(\ref{eq:FI}). Thus, we considered the RKKY interaction with highly non-trivial field dependence as the mechanism for the field-induced transitions in this paper. Nevertheless, careful examination of the role of high-rank multipoles should be necessary, which is left for future studies.

When concerning the interactions between the nearest-neighbor pairs with a term corresponding to $F_I$ in Eq. (\ref{eq:FI}), the mean-field parts read 
\begin{align}
\text{\ensuremath{\mathscr{H}}}_{\rm Q}^{\prime}  &=
-\lambda\Big(
\langle Q_{z}\rangle Q_z + \langle Q_{x}\rangle Q_x
\Big)\nonumber\\
&+\frac{\lambda'f(|{\bm B}|^2) }{2}\Big[\Big(h_z \Big(\langle Q_{z}\rangle Q_{z}-\langle Q_{x}\rangle Q_{x}\Big)\nonumber\\
&-h_x\Big(\langle Q_{x}\rangle Q_{z}+\langle Q_{z}\rangle Q_{x}\Big)\Big].\ \ \ \ \ \ \ \ 
\label{eq:Haniso}
\end{align}
Here, we have set the coefficient  $1/2$ for $\lambda'$ term so that the parameterization matches the previous expressions Eq. (\ref{eq:FI}), and, 
 for $\lambda,\lambda'>0$ the interaction becomes ferroic one. This is suitable to the experimental results in finite fields. Taking into account the mean-field part [Eq. (\ref{eq:Haniso})], the Zeeman coupling [Eq. (\ref{eq:Zeeman})], CEF potential [Eq. (\ref{eq:CEF})], the dipole-dipole interactions [Eq. (\ref{eq:dipole})], and a small ${\bm T}^\alpha$ octupole-octupole interaction $\lambda_T$, 
\begin{align}
\text{\ensuremath{\mathscr{H}}}=&\text{\ensuremath{\mathscr{H}}}_{\rm CEF}-\Big( g_J\mu_B {\bm B}+\lambda_d\langle {\bm J}\rangle\Big) \cdot {\bm J}-\lambda_T \langle {\bm T}^\alpha\rangle \cdot {\bm T}^\alpha\nonumber\\
-& \left(\lambda-\frac{\lambda'f(|{\bm B}|^2)h_z}{2}\right)\langle Q_z\rangle Q_z
-\left(\lambda+\frac{\lambda'f(|{\bm B}|^2)h_z}{2}\right)\langle Q_x\rangle Q_x,\label{HcefMF}
\end{align}
 for ${\bm B}$ lying the plane including [001] and [110]. The interaction between the octupoles $\bm{T}^\alpha$ with the same symmetry as the dipole, $\lambda_T$, has been introduced in order to reproduce the low-temperature susceptibility in its magnitude, although the following discussion does not alter if $\lambda_T$ is set to zero and tune $\lambda_d$ instead. Since the CEF model does not include the many-body renormalization present in the $f$-electron system, the value of 
 	$\lambda_d$ is in general different from that estimated experimentally for high temperatures. The factor $f$ is  assumed to be a phenomenological form that suppresses the magnitude as the field increases:
\begin{eqnarray}
f(|{\bm B}|^2)\equiv \frac{1}{1+c_2|{\bm B}|^2+c_4|{\bm B}|^4},
\end{eqnarray}
where $c_2$ and $c_4$ are constants. 
In the following discussion, we fix $\lambda=2$ K, $\lambda_d=-0.8$ K, and $\lambda_T=0.0004$ K (note that the matrix elements of ${\bm T}^\alpha$ is much larger than those for ${\bm J}$), which, for $|{\bm B}|=0$, leads to a FQ transition at $T=T_c\simeq 2$ K.
For simplicity, the phenomenological parameters $c_2$ and $c_4$ are set to $c_2=0.8$ T$^{-2}$ and $c_4=0.8$ T$^{-4}$, respectively, as representative values among parameters that qualitatively reproduce the experimental results. Note that although $f>0$ for these parameters, the situation where $f<0$ is in principle possible. The point here is that, to reproduce the experimental results, the magnitude of $f$ should decrease as the field increases. 

In order to take into account the interactions in higher order $\Gamma_3$ multipole sectors, 
the $\Gamma_3$ operators in Eq. (\ref{HcefMF}) is parameterized as 
\begin{align}
	Q_z &=|\Gamma_3 u\rangle\langle \Gamma_3 u|-|\Gamma_3 v\rangle\langle \Gamma_3 v| \nonumber\\
	&+ \eta \left[ \frac{\sqrt{35}}{2}\left(
	|\Gamma_3 u\rangle\langle \Gamma_1| +{\rm H.c.}
	\right)+\cdots \right],\label{def:newQz}\\
	Q_x &=|\Gamma_3 u\rangle\langle \Gamma_3 v|+|\Gamma_3 v\rangle\langle \Gamma_3 u| \nonumber\\
	&+ \eta \left[ \frac{\sqrt{35}}{2}\left(
	|\Gamma_3 v\rangle\langle \Gamma_1| +{\rm H.c.}
	\right)+\cdots\right].\label{def:newQx}
\end{align}
Here, ``$\cdots$'' in Eqs. (\ref{def:newQz}) and (\ref{def:newQx}) 
represents terms involving only excited $\Gamma_4$ and $\Gamma_5$ states in the original quadrupole operators. 
Thus, $\eta$ is a parameter controlling the weight of excited states in the quadrupole operator, which influences the anisotropic parameter $b$ in eq. (\ref{Free0}). This corresponds to taking account for the hybridization between the quadrupole moments and the higher-order $\Gamma_3$ multipole moments. In this paper, we set $\eta$ to $\eta=0.33$ as a typical value.

\begin{figure}[t]
\begin{center}
    \includegraphics[width=0.55\textwidth]{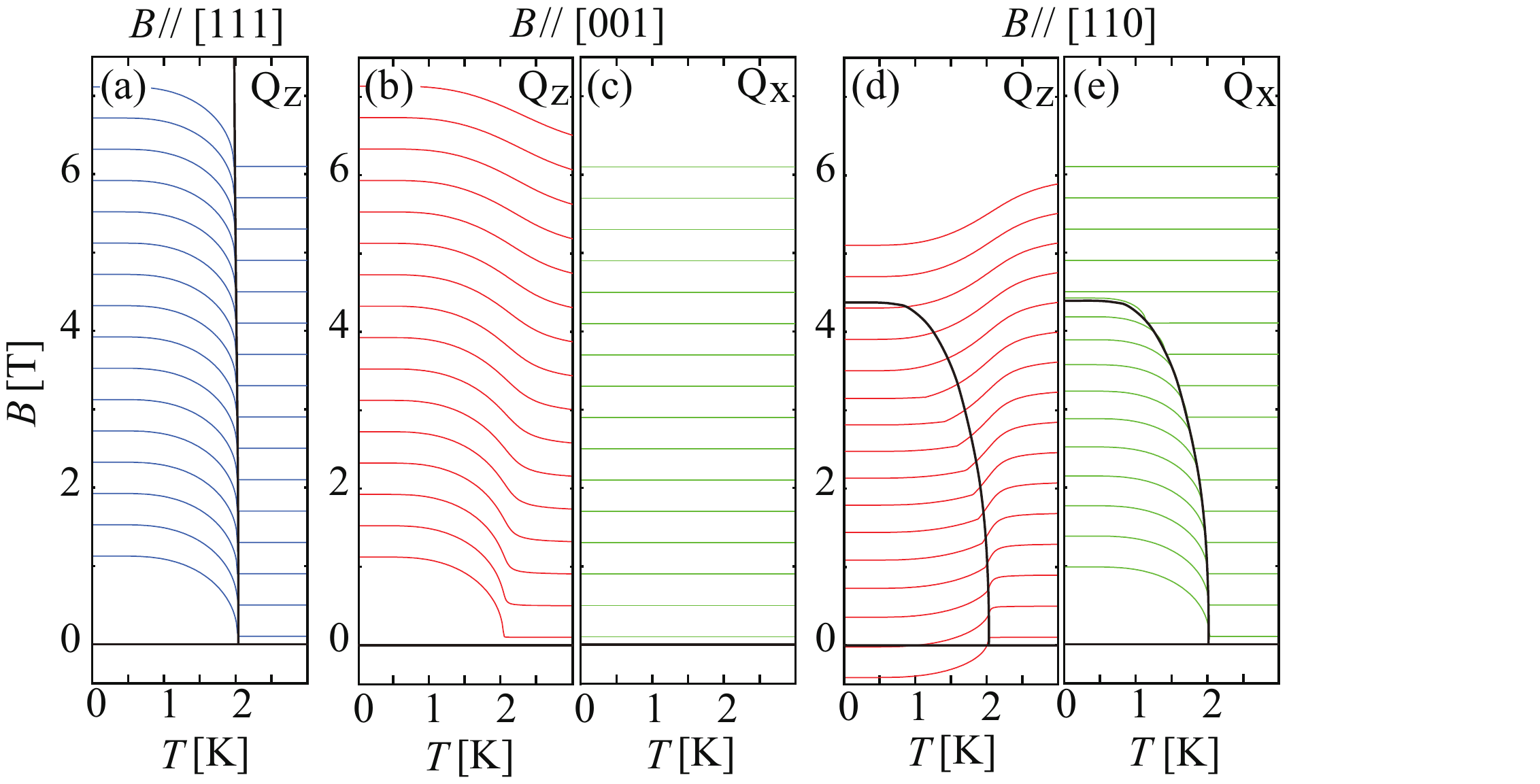}
\caption{\label{fig:phase_diagram0}(Color online) $B$-$T$ phase diagram and the temperature dependence of the quadrupole moments for $\lambda'=0$. $Q_z$ and $Q_x$ are plotted for every 0.4 T from 0.1 T to 6.1 T. The offset for each curve is equal to the value of the magnetic field. For $\boldsymbol{B}\parallel [111]$ in (a) and for 
$\boldsymbol{B}\parallel [110]$ in (c) and (d), $Q_{z}>0$ and $Q_{x}>0$ domains, respectively, are chosen.
}
\end{center}
\end{figure}

\begin{figure}[t]
    \includegraphics[width=0.5\textwidth]{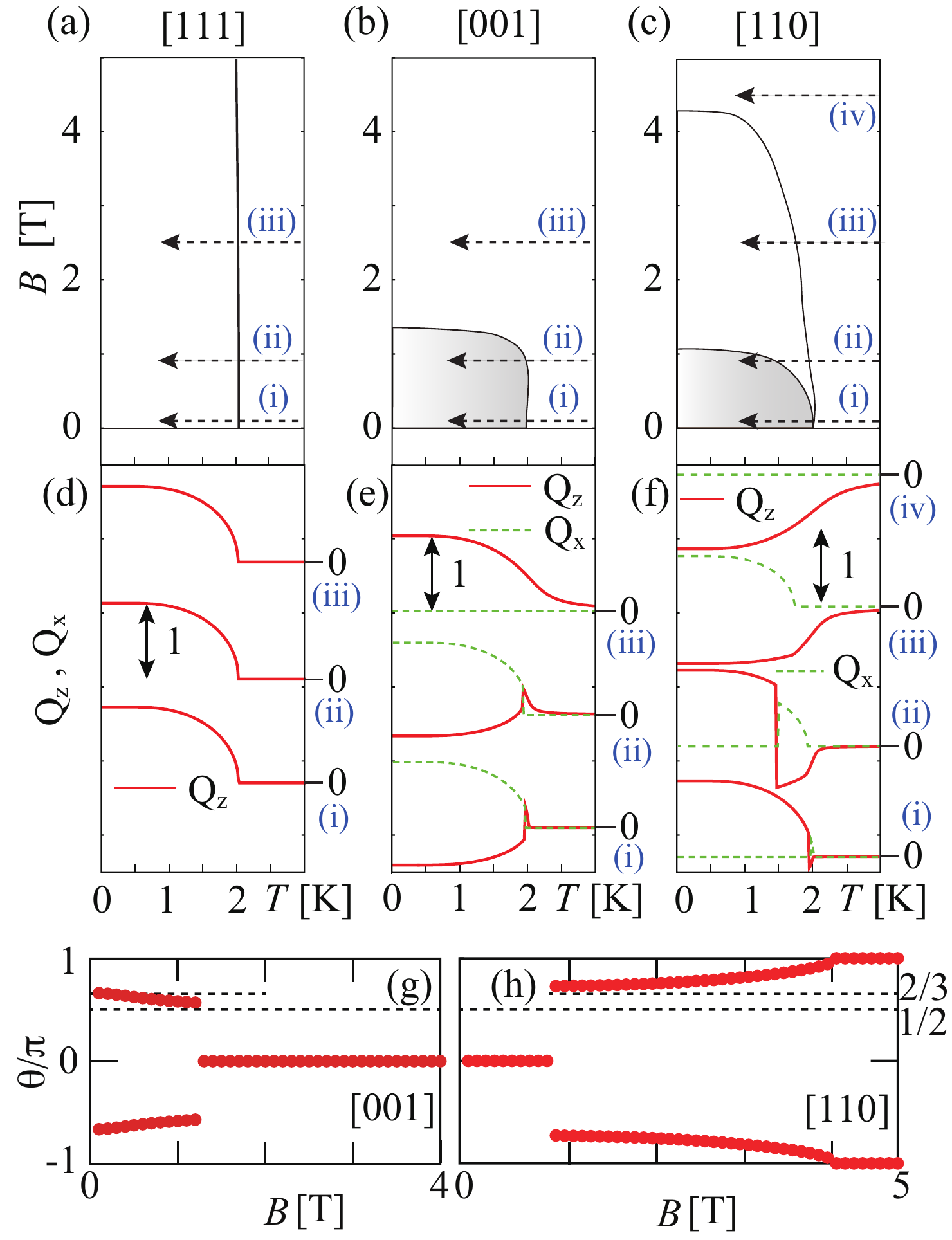}
\caption{\label{fig:phase_diagram1} (Color online) $B$-$T$ phase diagram for (a) ${\bm B}\parallel [001]$, (b) [110], and (c) [111],
where the arrows indicate the magnetic fields where the temperature dependence of the order parameters for $\lambda'=0.08$ K/T$^{2}$ are plotted in 
(d) for ${\bm B}\parallel [001]$, (e) for [110], and (f) for [111]. The values of the magnetic fields are (i) 0.1 T, (ii) 0.9 T, (iii) 2.5 T, and (iv) 4.5 T. In (d)-(f), the origins for each curve are shown in the right side. The magnetic field dependence of the order-parameter angle $\theta$ for $T=0.32$ K (this is essentially the same as the zero temperature limit) are shown in (g) for ${\bm B}\parallel [001]$ and (h) for $[110]$. The dotted lines represent $\theta=\pi/2$ and $2\pi/3$.}
\end{figure}

Figures \ref{fig:phase_diagram0} and \ref{fig:phase_diagram1} show $B$ - $T$ phase diagrams for 
$\lambda'=0$ and $\lambda'=0.08$ K/T$^{2}$, respectively. In Figs. \ref{fig:phase_diagram0}(a)-(e), the temperature dependence of 
the quadrupole moments for $\lambda'=0$ are also drawn for every 0.4 T from $H=0.1$ to 6.1 T with the offset set to the value of the magnetic field. 
When $\lambda'=0$, there is a crossover for ${\bm B}\parallel [001]$ as shown in Fig. \ref{fig:phase_diagram0}(b), while 
there is phase transitions for  ${\bm B}\parallel [110]$ and $[111]$. See Figs.~\ref{fig:phase_diagram0}(a), \ref{fig:phase_diagram0}(d), and \ref{fig:phase_diagram0}(e). For ${\bm B}\parallel [111]$, there are three equivalent domains $3J_x^2-\bm{J}^2$, $3J_y^2-\bm{J}^2$, and $3J_z^2-\bm{J}^2$. In Fig. \ref{fig:phase_diagram0}(a), the result for the domain $3J_z^2-\bm{J}^2\propto Q_z$ is shown.
	When the field is parallell to $[110]$, the two domains $3J_x^2-\bm{J}^2$ and $3J_y^2-\bm{J}^2$ are favored owing the the Zeeman coupling [Eq. (\ref{eq:FZ})]. The results shown in Fig. \ref{fig:phase_diagram0}(d) and \ref{fig:phase_diagram0}(e) are those for $3J_x^2-\bm{J}^2\sim (-Q_z+\sqrt{3}Q_x)/2$. These are trivial results as discussed in Sect. \ref{sec:level4_1}. 
Owing to the small factor $\eta T_c/E_{\Gamma_1}\sim 0.004$, the discontinuity in the order parameters at the transition is tiny and hardly seen in this scale.

In contrast, 
when $\lambda'=0.08$ K/T$^{2}$, nontrivial low-field phases emerge for ${\bm B}\parallel [001]$ [Fig.~ \ref{fig:phase_diagram1}(b)] and $[110]$ [Fig.~\ref{fig:phase_diagram1}(c)], while 
for ${\bm B}\parallel [111]$ [Fig.~\ref{fig:phase_diagram1}(a)], the phase diagram is independent of $\lambda'$, since the $\lambda'$ term vanishes in this field direction.
As is evident from Figs. \ref{fig:phase_diagram1}(b) and \ref{fig:phase_diagram1}(c), the critical field for 
${\bm B}\parallel [001]$ is higher than that for ${\bm B}\parallel [110]$, which is qualitatively consistent with the 
experimental results in Fig. \ref{fig:phase_diagram} and rough estimate done in Sect. \ref{sec:level4_1}. The directions of the order parameter for low fields are qualitatively the same as discussed in Sect. \ref{sec:level4_1}. See Figs.~\ref{fig:phase_diagram1}(d)-\ref{fig:phase_diagram1}(h). Namely, 
$\pi/2\le |\theta|\le  2\pi/3$ for ${\bm B} \parallel [001]$ [Fig.~\ref{fig:phase_diagram1}(g)] and 
$\theta=0$ for ${\bm B} \parallel [110]$ [Fig.~\ref{fig:phase_diagram1}(h)]. Thus, one can understand that the low-field phases are induced by the anisotropic interactions $\lambda'$ under finite fields and the present results are consistent with the experimental identification of the FQ order parameters as schematically shown in Fig.~\ref{fig:phase_diagram}. 
The actual temperature dependence of the order parameters are drawn in Figs. \ref{fig:phase_diagram1}(d)-\ref{fig:phase_diagram1}(f) for several magnetic fields (i)-(iv) as indicated by arrows in Figs. \ref{fig:phase_diagram1}(a)-\ref{fig:phase_diagram1}(c). The order parameter angle $\theta$ for $T=0.32$ K as a function of $|{\bm B}|$ is also shown in Figs. \ref{fig:phase_diagram1}(g) and \ref{fig:phase_diagram1}(h) for ${\bm B} \parallel [001]$ and [110], respectively, where values of $\theta$ for the two domains are plotted and they are essentially the same as those for $T=0$ K. 

For ${\bm B}\parallel [110]$ above the low-field phase, there is a phase similar to that exists 
for $\lambda'=0$. Let us call this intermediate-field phase. The shape of the intermediate-field phase is similar to that for $\lambda'=0$ in Figs. \ref{fig:phase_diagram0}(d) or \ref{fig:phase_diagram0}(e) 
and the critical field $\sim 4$ T is also very close to them. The magnetic-field dependence of $\theta$ for the intermediate-field phase is gradual and $\theta$ eventually reaches $\theta=\pm \pi$ at high fields. Although it is possible to destabilize the intermediate-field phase with keeping the low-field phase present by tuning the phenomenological parameters such as $\eta, c_2, $ and $c_4$, it seems that such parameter regime needs fine tuning of them. Since it is hard to detect small difference in the angle $\theta$ in the experiments, the presence/absence of the intermediate phase is not clear at this stage.  

Figure \ref{fig:susceptibilityFIT} shows comparison of the magnetic susceptibility between the present 
theory and the experiment.
The absence of the up turn as decreasing $T$ for the low-field phase $|{\bm B}|=1$ T is well reproduced for the results 
with $\lambda'=0.08$ in the theoretical calculations, while the data for $\lambda'=0$ fails to reproduce this. 
For $|{\bm B}|=4$ T, the two results for $\lambda'=0$ and 0.08 K/T$^{2}$ are almost the same and can capture the experimental temperature dependence.
Although there is a phase transition for $\lambda'=0.08$ K/T$^{2}$ for ${\bm B}\parallel [110]$ at this field, the effects of the transition is not visible in $M/B$. This is because $|{\bm B}|=4$ T is very close to the critical field. 
 
In addition to the temperature dependence of the magnetic susceptibility, the field dependence is examined 
in Fig. \ref{fig:MagnetizationFIT} for $T=0.32$ K. For $\lambda'=0$, the magnetization monotonically increases 
for both field directions ${\bm B} \parallel [001]$ and [110] as shown in Fig. \ref{fig:MagnetizationFIT}(a). 
For  [110] direction, there is a small anomaly at the phase boundary $B\sim 4$ T but not clearly visible in this scale. 
In Fig. \ref{fig:MagnetizationFIT}(b), the low-field suppression of $M$ for 
$\lambda'=0.08$ K/T$^{2}$ is well reproduced for ${\bm B}\parallel [001]$, 
while there is no qualitative change for higher fields compared with the data 
for $\lambda'=0$. Since the direction of the order parameter suddenly changes 
at the phase boundary between the low- and intermediate-field 
phases for $\lambda'=0.08$ K/T$^{2}$, the phase transition 
there is first order. Reflecting this, the theoretical data in 
Fig. \ref{fig:MagnetizationFIT}(b) show discontinuous jump around $|{\bm B}|\sim 1$ T. 
In contrast, the experimental data show no discontinuity and the magnetization curves 
are smooth functions of the field. As shown in Fig. \ref{fig:mag001}(c), this is 
owing to the phase coexistence of the low- and high-field phases. Thus, the calculated 
magnetization curve can capture the essential field dependence of the 
magnetization, i.e., the $|{\bm B}|$ dependence becomes steeper as the field increases. 
Note that this behavior is never explained by the data for $\lambda'=0$. 

\begin{figure}[t!]
\begin{center}
    \includegraphics[width=0.43\textwidth]{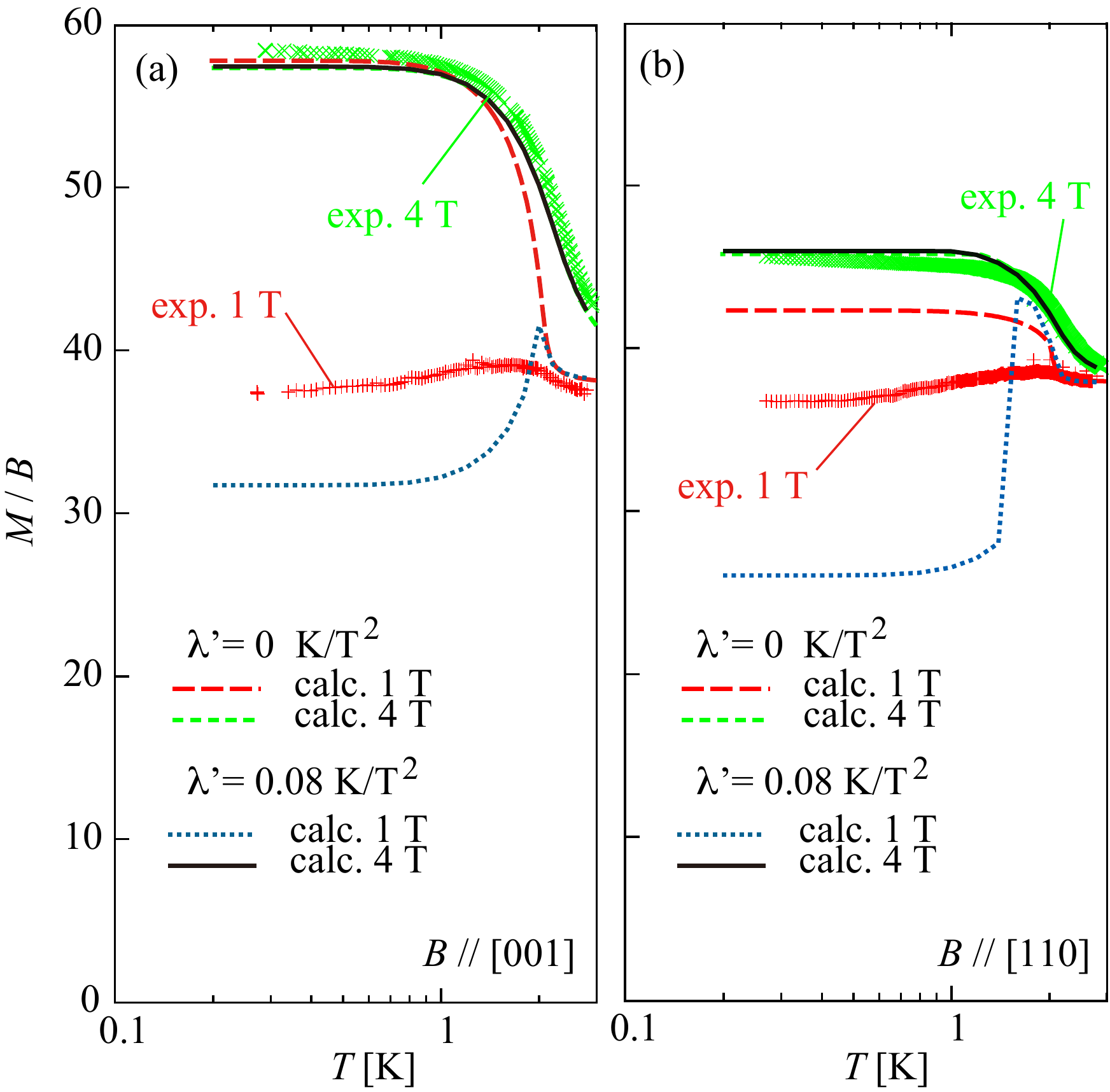}
\end{center}
\vspace{-0.5cm}
\caption{(Color online) \label{fig:susceptibilityFIT} Magnetic susceptibility of PrTi$_2$Al$_{20}$ as a function of temperature for (a) ${\bm B}\parallel [001]$ and (b) ${\bm B}\parallel [110]$.
The symbols represent the experimental data and the lines
 are the results by the CEF model (\ref{HcefMF}) with $\lambda'=0$ and $0.08$ K/T$^{2}$, which corresponds to Figs. \ref{fig:phase_diagram0} and \ref{fig:phase_diagram1}, respectively. }
\end{figure}

\begin{figure}[t!]
\begin{center}
    \includegraphics[width=0.48\textwidth]{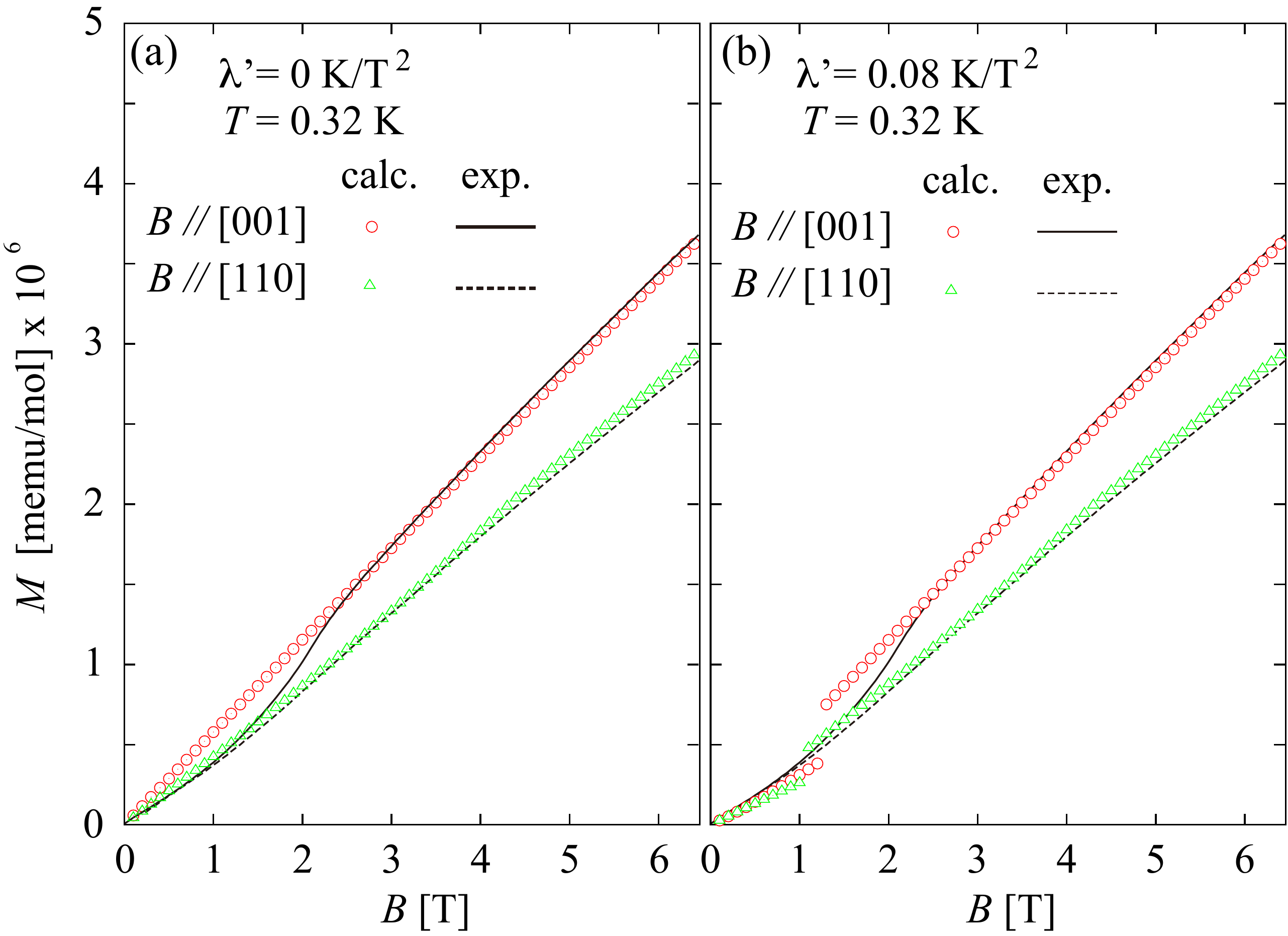}
\end{center}
\vspace{-0.5cm}
\caption{(Color online) \label{fig:MagnetizationFIT} Magnetization of PrTi$_2$Al$_{20}$ as a function of field for 
${\bm B} \parallel [001]$ (circle and red line) and [110] (triangle and green dotted line). 
The symbols represent the theoretical results by the CEF model (\ref{HcefMF}) for (a) $\lambda'=0$ K/T$^{2}$ and (b) $\lambda'=0.08$ K/T$^{2}$ and the 
lines are the experimental results.  }
\end{figure}

\section{\label{sec:level5}Concluding Remarks}
In this paper, we have thoroughly investigated ferro-quadrupole (FQ) order in PrTi$_{2}$Al$_{20}$ with the non-magnetic $\Gamma_{3}$ ground state by $^{27}$Al NMR and magnetization measurements and determined the magnetic field vs. temperature phase diagram for various directions of magnetic fields. The obtained phase diagram is highly anisotropic: while $Q_z \propto 3z^2-r^2$ type FQ order develops below $T_{Q}$ = 2.2~K for the field along [111] independent of the field strength, application of fields along [001] and [110] induces discontinuous switching of FQ order parameters at small field values less than a few tesla. 

Such anisotropic phase transitions are quite surprising because $\Gamma_{3}$ doublet has no dipole moment, therefore, does not couple directly to magnetic fields. Furthermore, the very small transition fields in the $\Gamma_{3}$ system suggest an extremely small energy scale involved in the transition. In fact, coexistence of two phases in a wide range of field observed for $\boldsymbol{B}_{\rm ext}$ \textbar{}\textbar{} {[}001{]} and {[}110{]} indicates inhomogeneous distribution of the small energy difference between the two phases. Such inhomogeneity could be easily generated, for example, by a small amount of disorder.

In order to understand the anisotropic phase diagram, we have constructed a Landau free energy with two-component order parameters, ${\bm Q}=(Q_z, Q_x) = Q (\cos \theta, \sin \theta )$.  At zero field, the crystalline electric field (CEF) is the only source of anisotropy, which gives a potential in ${\bm Q}$-space proportional to $-\cos 3\theta$ and selects the three stable orders of $Q_{z}$ type moments at $\theta$ = 0 and $\pm2\pi/3$. 

Application of magnetic fields generates additional anisotropy in two distinct routes. The first route is the Zeeman coupling to the field-induced dipole moment. Since $\Gamma_{3}$ is non-magnetic, the leading-order effect is due to the van Vleck paramagnetism, $F_{Z} = -\mu\left(2B_{z}^{2}-B_{x}^{2}-B_{y}^{2}\right)Q_{z}$ with $\mu>0$ [Eq.~(\ref{eq:FZ})], which can be also obtained by degenerate perturbation theory. (Here, we left only the term relevant to our experiments where the field is in the (1$\overline{1}$0) plane.) The second route is the anisotropy in the quadrupole-quadrupole interaction induced by the field. By symmetry, the lowest order term is restricted to the form $F_{I} = \lambda'\left(2B_{z}^{2}-B_{x}^{2}-B_{y}^{2}\right)\left(\langle Q_{z}\rangle Q_{z} - \langle Q_{x}\rangle Q_{x}\right)/2$ [Eq.~(\ref{eq:FI})] in a mean-field approximation. It is important to note that the form of $F_{I}$ is based only on symmetry argument and we do not know its microscopic mechanism yet, in contrast to the fully understood Zeemen coupling. Although interactions involving higher-rank multipoles might be able to generate similar effective anisotropy, examination of such possibilities is left for future studies.

Since both $F_{Z}$ and $F_{I}$ have the same field dependence, which is a remarkable property of the non-magnetic $\Gamma_{3}$ systems distinct form magnetic ones, the stable phase is selected by one of them with larger prefactor. Comparison between this phenomenological analysis and the experimentally determined phase diagram in Fig.~\ref{fig:phase_diagram} tells us that the interaction term $F_{I}$ with positive $\lambda'$ wins at low fields, selecting $Q_{x}$ order ($\theta = \pm\pi/2 \sim \pm2\pi/3$) for $\boldsymbol{B}_{\rm ext}$ \textbar{}\textbar{} {[}001{]} and $Q_{z}$ order with positive $\langle Q_{z} \rangle$ ($\theta$ = 0) for $\boldsymbol{B}_{\rm ext}$ \textbar{}\textbar{} {[}110{]}. On the other hand, the Zeeman term $F_{Z}$ wins at high fields, selecting $Q_{z}$ order with positive $\langle Q_{z} \rangle$ ($\theta$ = 0) for $\boldsymbol{B}_{\rm ext}$ \textbar{}\textbar{} {[}001{]} and $Q_{z}$ order with negative $\langle Q_{z} \rangle$ ($\theta = -\pi \sim \pm2\pi/3$)  for $\boldsymbol{B}_{\rm ext}$ \textbar{}\textbar{} {[}110{]}. For $\boldsymbol{B}_{\rm ext}$ \textbar{}\textbar{} {[}111{]}, since both $F_{Z}$ and $F_{I}$ vanish, the order should be the same as zero field.

Having summarized the main conclusion of our work, important open questions for better understanding of the quadrupole physics in PrTi$_{2}$Al$_{20}$ are the following: \\
(1) What is the mechanism for the field induced anisotropic interaction $F_{I}$? \\
(2) Why does the anisotropic interaction get suppressed relative to the Zeeman interaction by a very weak field of the order of one tesla? 

To answer the first question, we would have to calculate the quadrupole-quadrupole interaction, starting from the microscopic Hamiltonian of $c$-$f$ hybridization and the band structure of PrTi$_{2}$Al$_{20}$, then examine how it depends on magnetic fields. Although theoretical investigations of microscopic multipole interactions have been reported in several materials\cite{Shiba_1999,Takimoto_2006,Harima_2008,Kubo_2017}, effects of magnetic fields have not been discussed. This is clearly beyond the scope of this paper, however, should be an important future issue. 

The second question appears even more puzzling. The discontinuous jump of quadrupole order parameters must be caused by a sudden change in the quadrupole-quadrupole interaction, which in turn requires either changes in the Fermi surfaces and conduction-band structure or changes in the $c$-$f$ hybridization. However, the energy scale associated with the conducion bands and $c$-$f$ hybridization are usually quite large and it is highly unlikely that a few tesla of fields can make significant change, even though the band-structure calculation of LaTi$_{2}$Al$_{20}$ shows presence of small Fermi surfaces. 

A key observation to understand mechanism of the field induced transitions may be the remarkably different behavior of the NMR Knight shift and the susceptibility in the low-field FQ phase for $\boldsymbol{B}_{\rm ext}$ \textbar{}\textbar{} {[}001{]} as shown in Figs.~\ref{fig:Kx001} and \ref{fig:K_chi}. Since the susceptibility remains nearly unchanged by the FQ order, the rapid decrease of the Knight shift below $T_{Q}$ provides direct evidence that FQ order causes substantial reduction of the $c$-$f$ hybridization. The change in $c$-$f$ hybridization should then modify the quadrupole-quadrupole interaction, which in turn would influence the FQ order. Such a feedback effect may play an important role concerning stability of the FQ order. 

Similar violation of $K$-$\chi$ proportionality has been observed in magnetic heavy fermion materials, which is likely to be associated with Kondo screening and formation of heavy quasi-particles\cite{Curro_2009,Curro_2004}. It has been also proposed that magnetic RKKY interaction in heave fermion materials is renormalized by Kondo screening and becomes temperature dependent\cite{Broholm_1987}. Compared with the subtle many body effects in magnetic Kondo lattice systems, our case of quadrupole order involves change degrees of freedom, which can make direct influence on the $c$-$f$ hybridization. Thus, the physical mechanism appears to be more straightforward.    

Regarding the field-induced discontinuous transitions we observed, even if the direct effect of magnetic field on the conduction band structure is very weak, a small change in quadrupole moment may be amplified through the feedback effect, which eventually could lead to the switching of the order parameter. Since local charge distribution of the 4$f$ electrons is modified by FQ order, it is physically natural that such a modification of the charge density distribution results in a change in $c$-$f$ hybridization. This may be a new aspect of multipole physics of 4$f$ electron systems, which has not been recognized so far.

\begin{acknowledgment}
We would like to thank T. Arima, C. Broholm, H. Harima, K. Ishida, Y. B. Kim, T. J. Sato, Y. Shimura, S. Kitagawa, S. Kittaka, Y. Tokunaga, and H. Tsunetsugu for stimulating discussions. This work was financially supported by JSPS/MEXT Grants-in-Aids for Scientific Research (KAKENHI) Grant Numbers JP17H02918, JP16H04017, JP16H01079, JP18K03522, JP18H01164, JP16H02209, Grant Numbers JP18H04310, JP15H05883, JP15H05882, JP15H05883 (J-Physics) and JST/CREST Grant Number JPMJCR18T3. T. Taniguchi and M. Tsujimoto were supported by the JSPS Research Fellowship (JP17J08806 and JP16J08116, respectively) and by the Program for Leading Graduate School (MERIT). 
\end{acknowledgment}

\appendix

\section{Hyperfine fields from $T_{xyz}$ octupole}

\begin{table*}[t]
\begin{center}
\caption{\label{tab:Hf_oct}The magnetic hyperfine coupling vector $\mathbf{\boldsymbol{n}}$ at Al(3) sites describing the interaction between nuclear spins and $T_{xyz}$ octupole moments. The explicit values of its projection along the external field directions are also shown.
}
\begin{tabular}{cccccccc}
\hline 
 &\multicolumn{1}{c}{$$}&\multicolumn{6}{c}{$\mathbf{\boldsymbol{n}}\cdotp\boldsymbol{B}_{\rm ext}/B_{\rm ext}$}\\
\cline{3-8}
 site& $\mathbf{\boldsymbol{n}}$&site&$\boldsymbol{B}_{\rm ext}$\textbar{}\textbar{} {[}111{]} &site& $\boldsymbol{B}_{\rm ext}$\textbar{}\textbar{} {[}001{]} &site& $\boldsymbol{B}_{\rm ext}$\textbar{}\textbar{} {[}110{]}\\ 
\hline
A & $\left(a_{1},\,-a_{1},\,-a_{3}\right)$ & 3a & $-a_{3}/\sqrt{3}$ & 3$\alpha$ & $-a_{3}$ & 3q & 0\tabularnewline
B & $\left(-a_{1},\,a_{1},\,-a_{3}\right)$ & 3a & $-a_{3}/\sqrt{3}$ & 3$\alpha$ & $-a_{3}$ & 3q & 0\tabularnewline
C & $\left(-a_{3},\,a_{1},\,-a_{1}\right)$ & 3a & $-a_{3}/\sqrt{3}$ & 3$\beta$ & $-a_{1}$ & 3s & $\left(a_{1}-a_{3},\right)/\sqrt{2}$\tabularnewline
D & $\left(-a_{3},\,-a_{1},\,a_{1}\right)$ & 3a & $-a_{3}/\sqrt{3}$ & 3$\beta$ & $a_{1}$ & 3r & $-\left(a_{1}+a_{3}\right)/\sqrt{2}$\tabularnewline
E & $\left(-a_{1},\,-a_{3},\,a_{1}\right)$ & 3a & $-a_{3}/\sqrt{3}$ & 3$\beta$ & $a_{1}$ & 3r & $-\left(a_{1}+a_{3}\right)/\sqrt{2}$\tabularnewline
F & $\left(a_{1},\,-a_{3},\,-a_{1}\right)$ & 3a & $-a_{3}/\sqrt{3}$ & 3$\beta$ & $-a_{1}$ & 3s & $\left(a_{1}-a_{3}\right)/\sqrt{2}$\tabularnewline
G & $\left(-a_{1},\,-a_{1},\,a_{3}\right)$ & 3b & $\left(-2a_{1}+a_{3}\right)/\sqrt{3}$ & 3$\alpha$ & $a_{3}$ & 3p & $-\sqrt{2}a_{1}$\tabularnewline 
H & $\left(a_{3},\,-a_{1},\,-a_{1}\right)$ & 3b & $\left(-2a_{1}+a_{3}\right)/\sqrt{3}$ & 3$\beta$ & $-a_{1}$ & 3s & $\left(-a_{1}+a_{3}\right)/\sqrt{2}$\tabularnewline
I & $\left(-a_{1},\,a_{3},\,-a_{1}\right)$ & 3b & $\left(-2a_{1}+a_{3}\right)/\sqrt{3}$ & 3$\beta$ & $-a_{1}$ & 3s & $\left(-a_{1}+a_{3}\right)/\sqrt{2}$\tabularnewline
J & $\left(a_{1},\,a_{1},\,a_{3}\right)$ & 3c & $\left(2a_{1}+a_{3}\right)/\sqrt{3}$ & 3$\alpha$ & $a_{3}$ & 3p & $\sqrt{2}a_{1}$\tabularnewline
K & $\left(a_{3},\,a_{1},\,a_{1}\right)$ & 3c & $\left(2a_{1}+a_{3}\right)/\sqrt{3}$ & 3$\beta$ & $a_{1}$ & 3r & $\left(a_{1}+a_{3}\right)/\sqrt{2}$\tabularnewline
L & $\left(a_{1},\,a_{3},\,a_{1}\right)$ & 3c & $\left(2a_{1}+a_{3}\right)/\sqrt{3}$ & 3$\beta$ & $a_{1}$ & 3r & $\left(a_{1}+a_{3}\right)/\sqrt{2}$\tabularnewline
\hline 
\end{tabular}
\end{center}
\end{table*}

Here we derive symmetry allowed interaction between $T_{xyz}$ octupole and nuclear spins at Al(3) sites by following the method developed in Ref.~\citen{Sakai_1997}. We first consider the nuclear spin $\boldsymbol{I}$ at J site in Fig.~\ref{fig:op111}. Since both the nuclear spin and $T_{xyz}$ octupole are odd under time reversal, their interaction can be generally written as
\begin{eqnarray}
\text{\ensuremath{\mathscr{H}}}_{o}=-\gamma h \left(a_{1}I_{x}+a_{2}I_{y}+a_{3}I_{z}\right)T_{xyz}
\label{eq:Hhf_generally},
\end{eqnarray}
where $\boldsymbol{n} = (a_{1}, a_{2}, a_{3})$ is the hyperfine coupling vector. Since both the Pr and Al(3)-J sites are located on the $(1\overline{1}0)$ mirror plane, this interaction must be invariant under reflection with respect to the $(1\overline{1}0)$ plane. 

From the transformation properties of $\boldsymbol{I}$ and $T_{xyz}$ under reflection, 
\begin{eqnarray}
I_{x} & \rightarrow-I_{y},\\
I_{y} & \rightarrow-I_{x},\\
I_{z} & \rightarrow-I_{z},\\
T_{xyz} & \rightarrow-T_{xyz}
\label{eq:mirror}.
\end{eqnarray}
The transformed interaction $\text{\ensuremath{\mathscr{H}}}_{o}$ is given as,
\begin{eqnarray}
\text{\ensuremath{\mathscr{H}}}_{o}\rightarrow-\gamma\hbar\left(a_{1}I_{y}T_{xyz}+a_{2}I_{x}T_{xyz}+a_{3}I_{z}T_{xyz}\right)
\label{eq:Ho_mirror}.
\end{eqnarray}
The invariance of the interaction requires $a_{1}=a_{2}$ and we obtain
\begin{eqnarray}
\text{\ensuremath{\mathscr{H}}}_{o}=-\gamma\hbar a_{1}\left(I_{x}+I_{y}\right)T_{xyz}-\gamma\hbar a_{3}I_{z}T_{xyz}
\label{eq:Ho_J}.
\end{eqnarray}
The hyperfine field at J site is thus given as $\boldsymbol{B}_{\rm hf}^{(J)}=\boldsymbol{n}T_{xyz}$, where $\boldsymbol{n} = \left(a_{1}, a_{1}, a_{3}\right)$. 
Hyperfine fields at other sites can be obtained by appropriate coordinate transformation. 
The hyperfine coupling vector $\boldsymbol{n}$ and its projection along the external field directions at all Al(3) sites are listed in Tables~\ref{tab:Hf_oct}.

\newpage 

\newpage 


\bibliography{PrTi2Al20}

\begin{thebibliography}{10}

\bibitem{Kim_Balents_2014}
W.~Witczak-Krempa, G.~Chen, Y.~B. Kim, and L.~Balents: Annual Review of
  Condensed Matter Physics {\bfseries 5} (2014) 57.

\bibitem{Shaffer_Kim_2016}
R.~Schaffer, E.~K.-H. Lee, B.-J. Yang, and Y.~B. Kim: Reports on Progress in
  Physics {\bfseries 79} (2016) 094504.

\bibitem{Kuramoto_2009}
Y.~Kuramoto, H.~Kusunose, and A.~Kiss: Journal of the Physical Society of Japan
  {\bfseries 78} (2009) 072001.

\bibitem{Kusunose_2008}
H.~Kusunose: Journal of the Physical Society of Japan {\bfseries 77} (2008)
  064710.

\bibitem{Santini_2009}
P.~Santini, S.~Carretta, G.~Amoretti, R.~Caciuffo, N.~Magnani, and G.~H.
  Lander: Reviews of Modern Physics {\bfseries 81} (2009) 807.

\bibitem{Onimaru_2016}
T.~Onimaru and H.~Kusunose: Journal of the Physical Society of Japan {\bfseries
  85} (2016) 082002.

\bibitem{Cox_1987}
D.~Cox: Physical Review Letters {\bfseries 59} (1987) 1240.

\bibitem{Hoshino_2014}
S.~Hoshino and Y.~Kuramoto: Physical Review Letters {\bfseries 112} (2014)
  167204.

\bibitem{Tsuruta_2015}
A.~Tsuruta and K.~Miyake: Journal of the Physical Society of Japan {\bfseries
  84} (2015) 114714.

\bibitem{Matsubayashi_PRL}
K.~Matsubayashi, T.~Tanaka, A.~Sakai, S.~Nakatsuji, Y.~Kubo, and Y.~Uwatoko:
  Physical Review Letters {\bfseries 109} (2012) 187004.

\bibitem{Mannix_2005}
D.~Mannix, Y.~Tanaka, D.~Carbone, N.~Bernhoeft, and S.~Kunii: Physical Review
  Letters {\bfseries 95} (2005) 117206.

\bibitem{Wilkins_2006}
S.~B. Wilkins, R.~Caciuffo, C.~Detlefs, J.~Rebizant, E.~Colineau, F.~Wastin,
  and G.~H. Lander: Physical Review B {\bfseries 73} (2006) 060406.

\bibitem{Kuwahara_2007}
K.~Kuwahara, K.~Iwasa, M.~Kohgi, N.~Aso, M.~Sera, and F.~Iga: Journal of the
  Physical Society of Japan {\bfseries 76} (2007) 093702.

\bibitem{Onimaru_2005}
T.~Onimaru, T.~Sakakibara, N.~Aso, H.~Yoshizawa, H.~S. Suzuki, and T.~Takeuchi:
  Physical Review Letters {\bfseries 94} (2005) 197201.

\bibitem{Takigawa_1983}
M.~Takigawa, H.~Yasuoka, T.~Tanaka, and Y.~Ishizawa: Journal of the Physical
  Society of Japan {\bfseries 52} (1983) 728.

\bibitem{Sakai_1997}
O.~Sakai, R.~Shiina, H.~Shiba, and P.~Thalmeier: Journal of the Physical
  Society of Japan {\bfseries 66} (1997) 3005.

\bibitem{Shina_1998}
R.~Shiina, O.~Sakai, H.~Shiba, and P.~Thalmeier: Journal of the Physical
  Society of Japan {\bfseries 67} (1998) 941.

\bibitem{Tokunaga_2006}
Y.~Tokunaga, D.~Aoki, Y.~Homma, S.~Kambe, H.~Sakai, S.~Ikeda, T.~Fujimoto,
  R.~E. Walstedt, H.~Yasuoka, E.~Yamamoto, A.~Nakamura, and Y.~Shiokawa:
  Physical Review Letters {\bfseries 97} (2006) 257601.

\bibitem{Sakai_2005}
O.~Sakai, R.~Shiina, and H.~Shiba: Journal of the Physical Society of Japan
  {\bfseries 74} (2005) 457.

\bibitem{Kikuchi_2007}
J.~Kikuchi, M.~Takigawa, H.~Sugawara, and H.~Sato: Journal of the Physical
  Society of Japan {\bfseries 76} (2007) 043705.

\bibitem{Sakai_2007}
O.~Sakai, J.~Kikuchi, R.~Shiina, H.~Sato, H.~Sugawara, M.~Takigawa, and
  H.~Shiba: Journal of the Physical Society of Japan {\bfseries 76} (2007)
  024710.

\bibitem{Tayama_1997}
T.~Tayama, T.~Sakakibara, K.~Tenya, H.~Amitsuka, and S.~Kunii: Journal of the
  Physical Society of Japan {\bfseries 66} (1997) 2268.

\bibitem{Yamauchi_1999}
H.~Yamauchi, H.~Onodera, K.~Ohoyama, T.~Onimaru, M.~Kosaka, M.~Ohashi, and
  Y.~Yamaguchi: Journal of the Physical Society of Japan {\bfseries 68} (1999)
  2057.

\bibitem{Tanaka_2002}
H.~Adachi, H.~Kawata, M.~Mizumaki, T.~Akao, M.~Sato, N.~Ikeda, Y.~Tanaka, and
  H.~Miwa: Physical Review Letters {\bfseries 89} (2002) 206401.

\bibitem{Sato_2007}
H.~Sato, T.~Sakakibara, T.~Tayama, T.~Onimaru, H.~Sugawara, and H.~Sato:
  Journal of the Physical Society of Japan {\bfseries 76} (2007) 064701.

\bibitem{Iwasa_2005}
K.~Iwasa, L.~Hao, K.~Kuwahara, M.~Kohgi, S.~R. Saha, H.~Sugawara, Y.~Aoki,
  H.~Sato, T.~Tayama, and T.~Sakakibara: Physical Review B {\bfseries 72}
  (2005) 024414.

\bibitem{Takimoto_2006}
T.~Takimoto: Journal of the Physical Society of Japan {\bfseries 75} (2006)
  034714.

\bibitem{Harima_2008}
H.~Harima: Journal of the Physical Society of Japan {\bfseries 77} (2008) 114.

\bibitem{Kiss_2008}
A.~Kiss and Y.~Kuramoto: Journal of the Physical Society of Japan {\bfseries
  77} (2008) 034602.

\bibitem{Harima_2010}
H.~Harima, K.~Miyake, and J.~Flouquet: Journal of the Physical Society of Japan
  {\bfseries 79} (2010) 033705.

\bibitem{Kangas}
M.~J. Kangas, D.~C. Schmitt, A.~Sakai, S.~Nakatsuji, and J.~Y. Chan: Journal of
  Solid State Chemistry {\bfseries 196} (2012) 274.

\bibitem{Niemann}
S.~Niemann and W.~Jeitschko: Journal of Solid State Chemistry {\bfseries 114}
  (1995) 337.

\bibitem{Onimaru_2011}
T.~Onimaru, K.~T. Matsumoto, Y.~F. Inoue, K.~Umeo, T.~Sakakibara, Y.~Karaki,
  M.~Kubota, and T.~Takabatake: Physical Review Letters {\bfseries 106} (2011)
  177001.

\bibitem{Onimaru_2012}
T.~Onimaru, N.~Nagasawa, K.~T. Matsumoto, K.~Wakiya, K.~Umeo, S.~Kittaka,
  T.~Sakakibara, Y.~Matsushita, and T.~Takabatake: Physical Review B {\bfseries
  86} (2012) 184426.

\bibitem{Tsujimoto_2014}
M.~Tsujimoto, Y.~Matsumoto, T.~Tomita, A.~Sakai, and S.~Nakatsuji: Physical
  Review Letters {\bfseries 113} (2014) 267001.

\bibitem{Freyer_2018}
F.~Freyer, J.~Attig, S.~Lee, A.~Paramekanti, S.~Trebst, and Y.~B. Kim: Physical
  Review B {\bfseries 97} (2018) 115111.

\bibitem{Lee_2018}
S.~Lee, S.~Trebst, Y.~B. Kim, and A.~Paramekanti: Physical Review B {\bfseries
  98} (2018) 134447.

\bibitem{Y_B_Kim_2019}
A.~S. Patri, A.~Sakai, S.~Lee, A.~Paramekanti, S.~Nakatsuji, and Y.~B. Kim:
  arXiv  (2019) 1901.00012v1.

\bibitem{Sakai_2011}
A.~Sakai and S.~Nakatsuji: Journal of the Physical Society of Japan {\bfseries
  80} (2011) 063701.

\bibitem{Koseki}
M.~Koseki, Y.~Nakanishi, K.~Deto, G.~Koseki, R.~Kashiwazaki, F.~Shichinomiya,
  M.~Nakamura, M.~Yoshizawa, A.~Sakai, and S.~Nakatsuji: Journal of the
  Physical Society of Japan {\bfseries 80} (2011) SA049.

\bibitem{Ito}
T.~U. Ito, W.~Higemoto, K.~Ninomiya, H.~Luetkens, C.~Baines, A.~Sakai, and
  S.~Nakatsuji: Journal of the Physical Society of Japan {\bfseries 80} (2011)
  113703.

\bibitem{Sato_2012}
T.~J. Sato, S.~Ibuka, Y.~Nambu, T.~Yamazaki, T.~Hong, A.~Sakai, and
  S.~Nakatsuji: Physical Review B {\bfseries 86} (2012).

\bibitem{Sakai_super}
A.~Sakai, K.~Kuga, and S.~Nakatsuji: Journal of the Physical Society of Japan
  {\bfseries 81} (2012) 083702.

\bibitem{Matsunami}
M.~Matsunami, M.~Taguchi, A.~Chainani, R.~Eguchi, M.~Oura, A.~Sakai,
  S.~Nakatsuji, and S.~Shin: Physical Review B {\bfseries 84} (2011).

\bibitem{Machida}
Y.~Machida, T.~Yoshida, T.~Ikeura, K.~Izawa, A.~Nakama, R.~Higashinaka,
  Y.~Aoki, H.~Sato, A.~Sakai, S.~Nakatsuji, N.~Nagasawa, K.~Matsumoto,
  T.~Onimaru, and T.~Takabatake: Journal of Physics: Conference Series
  {\bfseries 592} (2015) 012025.

\bibitem{Kuwai}
T.~Kuwai, M.~Funane, K.~Tada, T.~Mizushima, and Y.~Isikawa: Journal of the
  Physical Society of Japan {\bfseries 82} (2013) 074705.

\bibitem{Nagashima}
S.~Nagashima, T.~Nishiwaki, A.~Otani, M.~Sakoda, E.~Matsuoka, H.~Harima, and
  H.~Sugawara: JPS Conference Proceedings {\bfseries 3} (2014) 011019.

\bibitem{Tokunaga}
Y.~Tokunaga, H.~Sakai, S.~Kambe, A.~Sakai, S.~Nakatsuji, and H.~Harima:
  Physical Review B {\bfseries 88} (2013) 085124.

\bibitem{Taniguchi_Proc}
T.~Taniguchi, M.~Yoshida, H.~Takeda, M.~Takigawa, M.~Tsujimoto, A.~Sakai,
  Y.~Matsumoto, and S.~Nakatsuji: Journal of Physics: Conference Series
  {\bfseries 683} (2016) 012016.

\bibitem{Taniguchi_JPSJ}
T.~Taniguchi, M.~Yoshida, H.~Takeda, M.~Takigawa, M.~Tsujimoto, A.~Sakai,
  Y.~Matsumoto, and S.~Nakatsuji: Journal of the Physical Society of Japan
  {\bfseries 85} (2016) 113703.

\bibitem{Matsubayashi_Proc}
K.~Matsubayashi, T.~Toshiki, S.~Junichirou, S.~Akito, N.~Satoru, K.~Kentaro,
  K.~Yasunori, and U.~Yoshiya: JPS Conference Proceedings {\bfseries 3} (2014)
  011077.

\bibitem{Sakai_2012_proc}
A.~Sakai and S.~Nakatsuji: Journal of Physics: Conference Series {\bfseries
  391} (2012) 012058.

\bibitem{Hattori_2014}
K.~Hattori and H.~Tsunetsugu: Journal of the Physical Society of Japan
  {\bfseries 83} (2014) 034709.

\bibitem{Abragam_1961}
A.~Abragam: {\em The Principles of Nuclear Magnetism} (Clarendon Press, 1961).

\bibitem{NMR_intro}
This formula is valid up to the first order in the nuclear quadrupole
  interaction with respect to the nuclear magnetic Zeeman energy. Values of the
  Knight shift reported in this paper, however, are obtained by numerical
  diagonalization of the total nuclear Hamiltonian, therefore, include all
  higher order effects.

\bibitem{Sakakibara}
T.~Sakakibara, H.~Mitamura, T.~Tayama, and H.~Amitsuka: Japanese Journal of
  Applied Physics {\bfseries 33} (1994) 5067.

\bibitem{Stevens}
K.~W.~H. Stevens: Proceedings of the Physical Society. Section A {\bfseries 65}
  (1952) 209.

\bibitem{mirror_note}
One should note that by applying a mirror operation to a system in a magnetic
  field parallel to the mirror plane, the field direction must be reversed.
  Therefore, precisely speaking, if two Al sites, say A and B, are related by
  (110) mirror, the NMR spectrum at site A with the field along $+z$ should be
  identical to the spectrum at site B with the field along $-z$. However, if
  there is no spontaneous hyperfine field at zero field, reversal of external
  magnetic field has no effect on the NMR spectrum. Thus we should have
  indential NMR spectra for the two sites in the same magnetic field.

\bibitem{Curro_2009}
N.~J. Curro: Reports on Progress in Physics {\bfseries 72} (2009) 026502.

\bibitem{Curro_2004}
N.~J. Curro, B.~L. Young, J.~Schmalian, and D.~Pines: Physical Review B
  {\bfseries 70} (2004) 235117.

\bibitem{Shiina_1997}
R.~Shiina, H.~Shiba, and P.~Thalmeier: Journal of the Physical Society of Japan
  {\bfseries 66} (1997) 1741.

\bibitem{Swatek_2018}
P.~Swatek, M.~Kleinert, P.~Wisniewski, and D.~Kaczorowski: Computational
  Materials Science {\bfseries 153} (2018) 461.

\bibitem{Shiba_1999}
H.~Shiba, O.~Sakai, and R.~Shiina: Journal of the Physical Society of Japan
  {\bfseries 68} (1999) 1988.

\bibitem{Kubo_2017}
K.~Kubo and T.~Hotta: Physical Review B {\bfseries 95} (2017) 054425.

\bibitem{Broholm_1987}
C.~Broholm, J.~K. Kjems, G.~Aeppli, Z.~Fisk, J.~L. Smith, S.~M. Shapiro,
  G.~Shirane, and H.~R. Ott: Physical Review Letters {\bfseries 58} (1987) 917.

\end{thebibliography}

\end{document}